%% file: arxiv_version.tex
\documentclass[10pt,journal,compsoc]{IEEEtran}

\usepackage{amsmath,amsfonts,amssymb}
\usepackage{mathtools}
\usepackage{amsthm}
\usepackage{subcaption}
\usepackage{alltt, xspace, times, epsfig, url}
\usepackage{gensymb,xfrac,array,booktabs,bm,xcolor}
\usepackage[linesnumbered,lined,ruled,noend]{algorithm2e}

\def\multiset#1#2{\ensuremath{\left(\kern-.3em\left(\genfrac{}{}{0pt}{}{#1}{#2}\right)\kern-.3em\right)}}

\ifCLASSOPTIONcompsoc
  \usepackage[nocompress]{cite}
\else
  \usepackage{cite}
\fi

\newtheorem{definition}{Definition}

\begin{document}

\title{Utility-Aware and Privacy-Preserving\\Mobile Query Services}

\author{Emre~Yigitoglu,
        Mehmet~Emre~Gursoy,~\IEEEmembership{Student~Member,~IEEE,}
        and~Ling~Liu,~\IEEEmembership{Fellow,~IEEE}
\IEEEcompsocitemizethanks{\IEEEcompsocthanksitem Emre Yigitoglu, Mehmet Emre Gursoy, and Ling Liu are with the School of Computer Science, Georgia Institute of Technology, Atlanta,
GA, 30332.\protect\\
E-mail: \{eyigitoglu,memregursoy\}@gatech.edu, ling.liu@cc.gatech.edu}
\thanks{Manuscript received X; revised Y.}}


\IEEEtitleabstractindextext{%
\begin{abstract}
Location-based queries enable fundamental services for mobile road network travelers. While the benefits of location-based services (LBS) are numerous, exposure of mobile travelers' location information to untrusted LBS providers may lead to privacy breaches. In this paper, we propose \textsc{StarCloak}, a utility-aware and attack-resilient approach to building a privacy-preserving query system for mobile users traveling on road networks. \textsc{StarCloak} has several desirable properties. First, \textsc{StarCloak} supports user-defined $k$-user anonymity and $l$-segment indistinguishability, along with user-specified spatial and temporal utility constraints, for utility-aware and personalized location privacy. Second, unlike conventional solutions which are indifferent to underlying road network structure, \textsc{StarCloak} uses the concept of stars and proposes cloaking graphs for effective location cloaking on road networks. Third, \textsc{StarCloak} achieves strong attack-resilience against replay and query injection-based attacks through randomized star selection and pruning. Finally, to enable scalable query processing with high throughput, \textsc{StarCloak} makes cost-aware star selection decisions by considering query evaluation and network communication costs. We evaluate \textsc{StarCloak} on two real-world road network datasets under various privacy and utility constraints. Results show that \textsc{StarCloak} achieves improved query success rate and throughput, reduced anonymization time and network usage, and higher attack-resilience in comparison to \textsc{XStar}, its most relevant competitor.
\end{abstract}

}

\maketitle

\IEEEdisplaynontitleabstractindextext
\IEEEpeerreviewmaketitle

\IEEEraisesectionheading{\section{Introduction}\label{sec:introduction}}

\newlength{\textfloatsepsave} \setlength{\textfloatsepsave}{\textfloatsep}

\IEEEPARstart{T}{he} growth of location-based services (LBSs) is fueled by ubiquitous wireless connectivity, universal presence of smart mobile devices with multi-modal sensing capability, and increased investments from industry and government on the Internet of Things. Juniper Research \cite{juniper} forecasted the LBS market to reach \$43.3 billion in revenue in 2019, rising from an estimated \$12.2 billion in 2014. \cite{pew} reports that 74\% of adult smartphone owners use their phones to get direction or information based on their current location. As more and more mobile travelers and vehicles are connected continuously and automatically, they are embraced by life-enriching location-based experiences and services, including but not limited to improved emergency assistance, real-time traffic alerts, and location recommendations. 

While there is ongoing research in answering queries and providing services for mobile users traveling on road networks \cite{lee2016road,miao2019efficiently,miao2018efficiently,luo2018toain}, users' location privacy poses an important concern. Unauthorized location exposure may cause vulnerability for abuse such as unwanted advertisement, stalking, and location spoofing. In addition, when private location data of a mobile user is linked to sensitive public locations such as health clinics, cancer treatment centers, nightclubs or religious organizations, such unauthorized linkage may cause ethical, professional, and social risks both to individuals and the society at large. As a result, it becomes imperative to protect road network travelers' location privacy as they interact with third-party LBS providers via service queries.

One viable approach to protecting location privacy of road network travelers is location anonymization through obfuscating or cloaking the mobile user's actual location. A practical anonymization framework should consider multiple aspects. First, the road network structure must be taken into account during anonymization, both for effective privacy protection and efficient query processing with anonymized locations. Second, the framework should support user-defined, personalized privacy goals such as $k$-user anonymity and $l$-segment distinguishability. Third, anonymization should incur as little utility loss as possible; in particular, if there are any utility constraints such as maximum spatial cloaking region size or maximum tolerable time delay in query response, the anonymization framework should satisfy these constraints. These enable the anonymization framework to be flexible in serving users with different privacy and utility needs. Fourth, the anonymization framework should be resilient to replay or query injection-based inference attacks. A sophisticated adversary who observes a cloaked region should not be able to infer the user's true location. Finally, the anonymization framework should have low communication and IO cost, i.e., anonymized cloaked locations should be compact enough to be sent through a wireless network without much network overhead, and they should be usable without increased processing effort.

To meet these goals, we propose and develop \textsc{StarCloak}, a utility-aware and attack-resilient approach to building a privacy-preserving location query system for mobile users traveling on \textit{road networks}. \textsc{StarCloak} relies on optimized data structures and algorithms for effectively and efficiently determining cloaked regions for incoming queries, such as the \textit{star}, \textit{star graph}, and \textit{cloaking graph} data structures. \textsc{StarCloak} maintains its internal data structures as new queries are processed, and generates candidate star-sets as cloaked regions when it identifies that certain users' queries can be successfully served. \textsc{StarCloak}'s candidate star-set pruner, which is implemented with high parallelism, enables pruning of candidate star-sets to generate low-cost cloaked subgraphs with improved attack-resilience via randomized pruning. In addition, we also propose two variants of \textsc{StarCloak}, namely spatially bounded \textsc{StarCloak} and hybrid \textsc{StarCloak}, for generating more compact cloaked regions with negligible sacrifice in query success rate and throughput. 

We evaluate \textsc{StarCloak} and its variants through extensive experiments on real-world Georgia and California road networks of different scales, under varying privacy and utility constraints. We also compare \textsc{StarCloak} with two baseline anonymization approaches (random sampling and network expansion) as well as \textsc{XStar} \cite{wang:vldb09}, which is the most relevant work to ours from the literature. Results show that \textsc{StarCloak} offers significantly improved query success rate and throughput. Furthermore, compared to \textsc{XStar}, \textsc{StarCloak} achieves substantially reduced anonymization time, network bandwidth usage, and improved resilience to inference attacks.

\vspace{-8pt}
\section{\textsc{StarCloak} Overview and Concepts} \label{sec:overview}

\textsc{StarCloak} can be viewed as a trusted location anonymization service. It forms a middle layer between mobile users and their untrusted LBS providers. Assume that user Alice issues a service query while she is moving on a road segment. Without \textsc{StarCloak}, Alice's device directly sends her query with her true current location to an untrusted LBS provider, which executes the query based on Alice's location and sends the results to Alice's device. However, if Alice is using \textsc{StarCloak}, \textsc{StarCloak} will first compute an anonymized location for Alice and replace her true location with the anonymized location transparently from Alice, before the query is sent to the untrusted LBS provider. 

\begin{figure}[t]
\centering
\includegraphics[width=.48\textwidth]{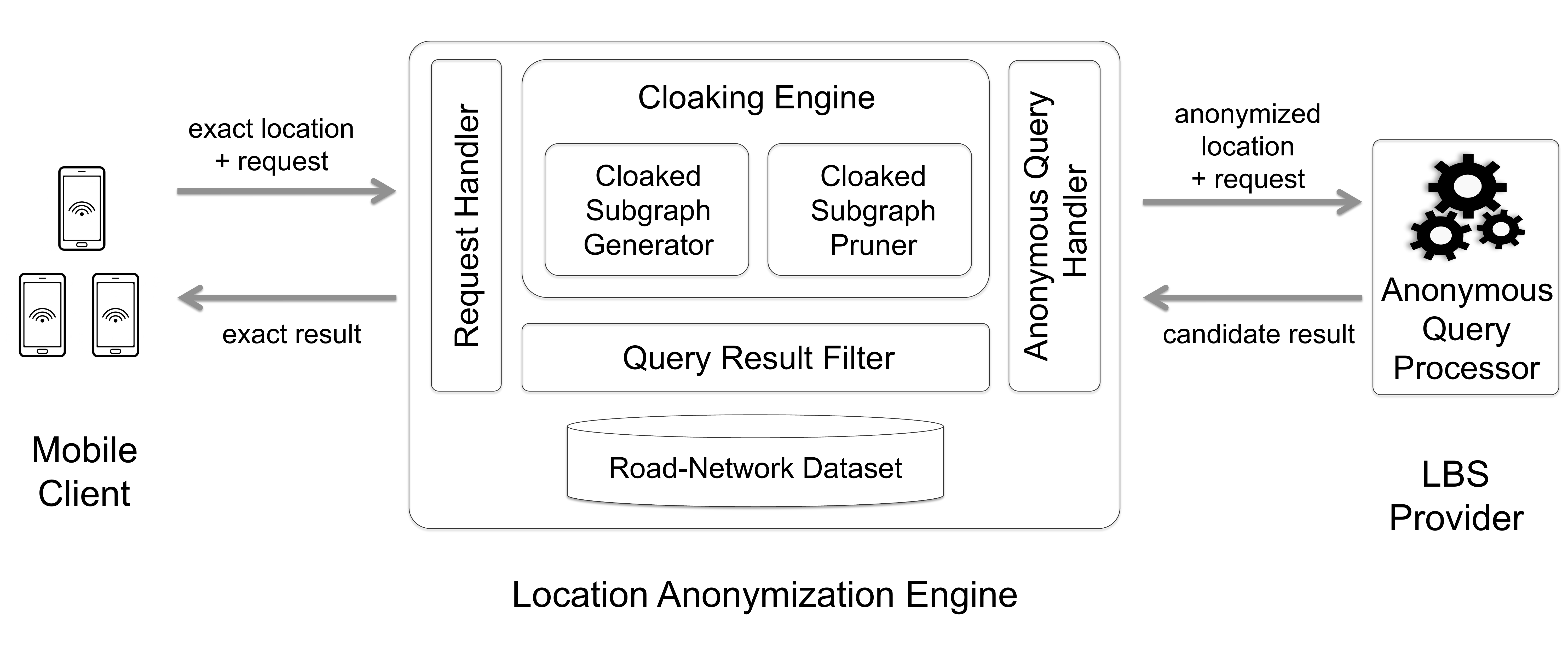}
\vspace{-6pt}
\caption{Overall architecture of \textsc{StarCloak}} 
\label{fig:architecture}
\vspace{-8pt}
\end{figure}

Figure \ref{fig:architecture} illustrates the reference architecture. Let $q$ denote the original query of mobile user $u$. When $u$ issues query $q$ with their true location, the location and query are intercepted by the location anonymization engine. The engine transforms $u$'s true location to a cloaked location $S$ while meeting the personalized privacy and utility profile of $u$. Next, the engine relays the anonymized location and query to the LBS provider. The LBS provider computes a candidate result, and the candidate result is received by the location anonymization engine. Since the cloaked location often has lower resolution than the actual location to meet privacy goals, the candidate result received by the anonymization engine may contain false positives. The anonymization engine performs post-processing of results to filter false positives. Finally, the anonymization engine delivers the exact query answer to $u$.


This section presents an overview of \textsc{StarCloak} and describes its privacy, utility, attack, and cost models. \textsc{StarCloak} assumes that mobile users travel on spatially constrained road networks or walk paths. Thus, we first introduce the basic models for road networks, location privacy, inference attacks, and query costs, followed by our problem statement combining these models.

\vspace{-6pt}
\subsection{Road Network Model} \label{sec:roadnetwork}


We represent a road network as an undirected graph $G = \langle V_G, E_G \rangle$ with the node set $V_G$ denoting road junctions and edge set $E_G$ denoting road segments, respectively. Each road segment connects a pair of junctions. Figure~\ref{fig:query_inject} illustrates a road network. We use $d_G(v)$ to denote the degree of a node $v$ with respect to $G$, $d_G(v)=|\{w | (v,w)\in E_G\}|$. We call $v$ an \textit{intersection node} if $d_G(v) \geq 3$. For example, in Figure~\ref{fig:query_inject}, $v_5$ is an intersection node.

An anonymized location in the road network can be represented as a subgraph. \textit{Border nodes} are nodes that connect a subgraph $S$ to the remainder of the main graph $G$.

\begin{definition}[Subgraph] \label{def:subgraph}
$S$ is a subgraph of road network $G$, denoted by $S = \langle V_S,E_S \rangle$, if and only if $V_S \subset V_G$ and $E_S \subset E_G$.
\end{definition}

\begin{definition}[Border Node] \label{def:bordernode}
Let $S$ denote a subgraph of $G$. The set of border nodes of $S$, denoted $BV(S)$, are nodes in both $S$ and $G$ but have edges that are in $E_G$ but not in $E_S$. Formally:
\[
BV(S) = \{ v ~|~ \exists w \in (V_G \setminus V_S) ~\text{s.t.}~ (v,w) \in E_G \}
\]
\end{definition}

Equivalently, border nodes are those nodes $v$ in $S$ that satisfy the condition: $d_G(v) > d_S(v)$. As an example, we can construct a subgraph $S$ in Figure \ref{fig:query_inject} as: $V_S = \{v_2, v_4, v_5, v_6, v_7, v_{10} \}$ and $E_S = \{(v_4,v_5), (v_5, v_{10}), (v_5,v_6), (v_6,v_7), (v_2,v_6) \}$. Then, it holds that: $BV(S) = \{v_2, v_4, v_7, v_{10} \}$. A sequence of edges $(v_0, v_1), \ldots, (v_i, v_{i+1}), \ldots, (v_{L-1},v_{L})$, where all $v_i$ are unique and satisfy the conditions $d_G(v_0) \geq 3$, $d_G(v_L) \geq 3$, and $d_G(v_i) = 2$ for $0 < i < L$, constitute a \textit{segment} denoted by $\overline{v_0v_L}$.




\subsection{Utility-Aware Location Privacy Model} \label{sec:privacymodel}

\textsc{StarCloak} enforces location privacy for mobile travelers while considering privacy and utility metrics simultaneously. It supports personalized location $k$-user anonymity and $l$-segment indistinguishability, such that instead of using a system-supplied fixed $k$ or $l$ for all users and queries \cite{gruteser2003anonymous}, it achieves high versatility via user-specified privacy needs and specifications \cite{clique-cloak}. In addition, we introduce two utility metrics to capture location utility constraints: maximum spatial and temporal cloaking resolutions. These utility metrics constrain and regulate \textsc{StarCloak} so that it performs anonymization while meeting the spatial and temporal tolerances. 

\textsc{StarCloak} performs location anonymization via cloaking, which is a process that transforms the user's exact location into a cloaked region with lower resolution to satisfy user-defined privacy requirements. The goal is to choose the cloaked region with as little utility loss and query costs as possible. We start by formalizing the privacy notions: $k$-user anonymity and $l$-segment indistinguishability. $k$-user anonymity protects user $u$'s location by ``hiding $u$ in a crowd", i.e., enforcing at least $k-1$ other users in the vicinity of $u$ report the same cloaked location. We observe that $k$-anonymity is not sufficient to prevent the linkage of user $u$ with a sensitive public location or road segment, since the cloaked $k$-anonymized region may lack sufficient segment diversity, e.g., it may contain only a single road segment. This motivates the proposal of $l$-segment indistinguishability.


\begin{definition}[$k$-user anonymity]
An anonymized location $S$ (subgraph of road network $G$) is said to satisfy $k$-user anonymity, if at least $k$ active users report $S$.  
\end{definition}

\begin{definition}[$l$-segment indistinguishability]
An anonymized location $S$ is said to satisfy $l$-segment indistinguishability, if it contains at least $l$ different road segments and any one segment could be plausibly associated with a user reporting $S$.
\end{definition}

In \textsc{StarCloak}, a query $q$ is allowed to specify a custom privacy requirement as $(\delta^q_k, \delta^q_l)$, such that $\delta^q_k \geq 1$ is the desired $k$-user anonymity level and $\delta^q_l \geq 1$ is the desired $l$-segment indistinguishability level.

A trivial approach to achieve maximum protection could be to assign the whole road network $G$ as the anonymized location. However, this approach clearly provides weak utility and low quality of service. Hence, we incorporate \textit{spatial} and \textit{temporal cloaking resolutions} as utility constraints. The spatial constraint $\sigma_s$ bounds the spatial resolution of the anonymized location. This is necessary so that anonymized locations are not arbitrarily large. The temporal constraint $\sigma_t$ bounds the maximum time delay resulting from anonymization. This is necessary so that the query-issuing user receives a response in timely manner.

\begin{definition}[Query profile]
For user $u$ with query $q$, we denote by $(\delta^q_k, \delta^q_l, \sigma^q_s, \sigma^q_t)$ the complete service profile of $q$, where $\delta^q_k, \delta^q_l$ are the privacy parameters and $\sigma^q_s, \sigma^q_t$ are the utility parameters.
\end{definition}

We expect \textsc{StarCloak} to operate in an environment with diverse user and query profiles and diverse road conditions. It is sometimes possible, e.g., due to low traffic density or few active users in the system at night time, that desired $\delta^q_k$-anonymity level is unachievable under strict $\sigma^q_s$ and $\sigma^q_t$ constraints for some query $q$. In such cases where $q$ cannot be serviced, it is discarded (dropped).

\subsection{Inference Attack Models} \label{sec:attackmodels}

An adversary may run sophisticated inference attacks with the goal of identifying probabilities of each segment $s$ in anonymized $S$ to be the user's actual segment. From an attack-resilience perspective, the ideal case is when the association of the mobile user with the segments in $S$ follows a uniform distribution (with equal probability $1/|S|$). In order to formalize an adversary's association power, we use the notion of \textit{linkability} \cite{wang:vldb09}.

\begin{definition}[Linkability]
For user $u$ with anonymized location $S$, the linkability of $u$ with a specific segment $s^* \in S$ is the probability that adversary associates $u$ with $s^*$ based on adversarial background knowledge $K_{ad}$, denoted as: $\text{link}[u \leftarrow s^* | 
S, K_{ad}]$.
\end{definition}

The background knowledge considered here includes knowledge of the location anonymization algorithm, underlying road network structure, and estimation of overall query cost (Sec.~\ref{sec:costmodel}). 

In a general replay attack, the adversary observes the anony\-mized location as set of segments $S$ and attempts to perform reverse-engineering with understanding of the anonymization algorithm. Specifically, the adversary re-runs the anonymization algorithm, denoted $\mathcal{A}(\cdot)$, for each segment $s \in S$ that could potentially be the mobile user's actual location. The similarity between $S$ and the algorithm's output $S'$ generated by $\mathcal{A}$, is used to estimate the likelihood of $s$ having generated $S$:
\begin{equation} \label{eq:like}
\textit{like}[S | u \leftarrow s, K_{ad}] = \frac{|S'\cap S|}{|S|}
\end{equation}

We propose that this can be improved in two ways to create a \textit{correlation-based replay attack}. First, the general replay attack only takes into account the placement of a single user in $S$; however, in utility-preserving $k$-anonymity algorithms, one user's location is cloaked together with other active users in the vicinity. Then, the placement of the remaining $k-1$ users in $S$ should also play a role in the likelihood calculation. Note that there are a total of $\multiset{S}{k-1}$ different possible placements of $k-1$ queries in $S$. We denote by $m_i$ each placement, such that $1 \leq i \leq \multiset{S}{k-1}$. Second, the adversary may have statistical knowledge of mobile users' distribution on the road network. For example, the adversary may know the traffic density distribution of the city during rush hour, which enables the adversary to predict that there is higher probability that the user is actually located on a dense segment rather than a sparse segment. We denote by $\text{Pr}[s]$ and $\text{Pr}[m_i]$ the probability of user $u$ being located on segment $s$ and remaining $k-1$ users being located as in the placement of $m_i$ according to the background distribution knowledge. Combining the two improvements, we compute $\textit{like}^{c}[S | u \leftarrow s, K_{ad}]$ as:
{ \small
\begin{equation*} 
\textit{like}^{c}[S | u \leftarrow s, K_{ad}] = \sum_{i = 1}^{\multiset{S}{k-1}} \text{Pr}[s] \cdot \text{Pr}[m_i] \cdot \textit{like}[S|u \leftarrow s, m_i, K_{ad}]
\end{equation*} 
}
Then, linkability can be calculated as:
\begin{equation} \label{eq:link}
\textit{link}[u \leftarrow s^* | S, K_{ad}] = \frac{\textit{like}^{c}[S | u \leftarrow s^*, K_{ad}]}{\sum_{s \in S} \textit{like}^{c}[S | u \leftarrow s, K_{ad}]}
\end{equation}

\begin{figure}
\includegraphics[width=.25\textwidth]{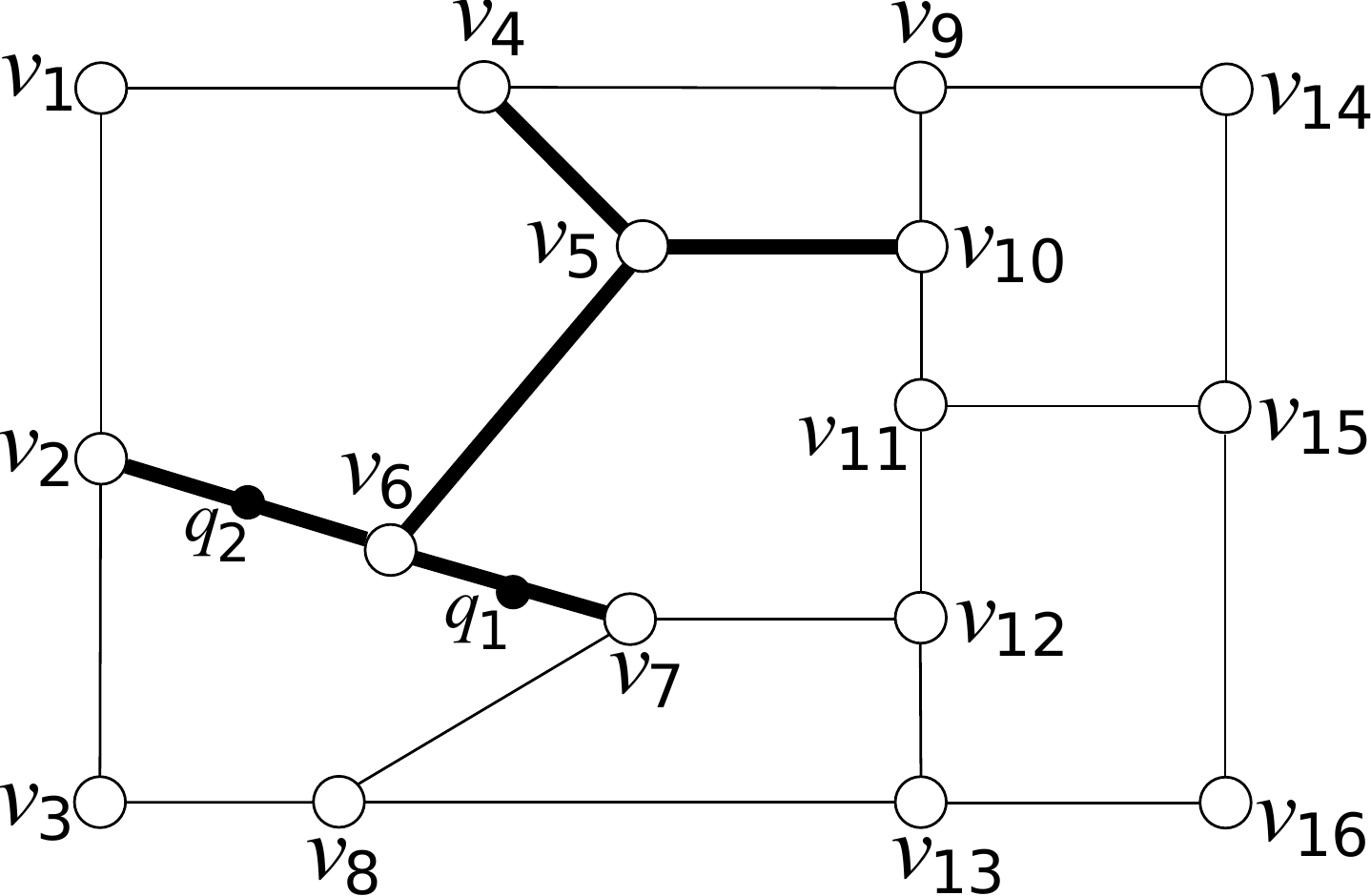}
\centering
\vspace{-4pt}
\caption{Road network and query injection example}
\label{fig:query_inject}
\vspace{-8pt}
\end{figure}

In replay attacks, the assumption is that the adversary is an observer only. Next, we also consider active adversaries who can inject queries into the system, i.e., execute a query injection attack. We expect anonymization algorithms with strong minimality (tightness) to be more vulnerable to the query injection attack. Consider an anonymization algorithm which cloaks segments into the same anonymized location if and only if they include at least one active query and through the shortest paths between the active queries. This knowledge can be exploited by an inference attack. For example, consider the anonymized location $S$ that consists of the bold lines in Figure \ref{fig:query_inject}. Suppose that $q_1$ and $q_2$ are the queries injected by the adversary with privacy profiles $(\delta_k,\delta_l)=(3,3)$. Then, by the minimality property, the adversary can infer that the third (actual) query was issued from either segment $\overline{v_4v_5}$ or segment $\overline{v_5v_{10}}$. To capture the effects of query injection attack, we slightly modify the $\textit{like}^c$ calculation. We assign zero to the likelihood value of placement $m_i$ if the segments corresponding to this placement conflict with the injected queries' locations.

\vspace{-2pt}
\subsection{Query Cost Model} \label{sec:costmodel}

An important challenge in finding an optimal anonymized location $S$ to a query $q$ is to minimize the \textit{cost} of the query when executed with the anonymized location. We study two types of cost: cost of query \textit{evaluation} and cost of \textit{communication}.

\textbf{Cost of Anonymized Query Evaluation:} Most query processing approaches for road networks are based on two types of fundamental operations: edge-based and node-based. The edge-based operation takes a query $q$ and an edge $e$ as input and returns a set of objects on $e$ denoted $\mathcal{O}_e(q,e)$ which satisfy the query condition. For segment $s$ potentially composed of a sequence of edges, we have: $\mathcal{O}_s(q,s) = \cup_{e \in s} \mathcal{O}_e(q,e)$. We denote by $\mathcal{C}_s$ the average computation cost of evaluating the query on a segment. While $\mathcal{C}_s$ depends on a variety of factors, in our current system, we set $\mathcal{C}_s$ statically according to the underlying spatial index implementation (e.g., look-up table, R-Tree). The node-based operation takes a query $q$ and a node $v$ as input and returns a set of objects in the vicinity of $v$ denoted $\mathcal{O}_v(q,v)$ which satisfy the query condition. The computation cost of evaluating a node-based query is denoted by $\mathcal{C}_v$.

\begin{figure}
\centering
	\includegraphics[width=.25\textwidth]{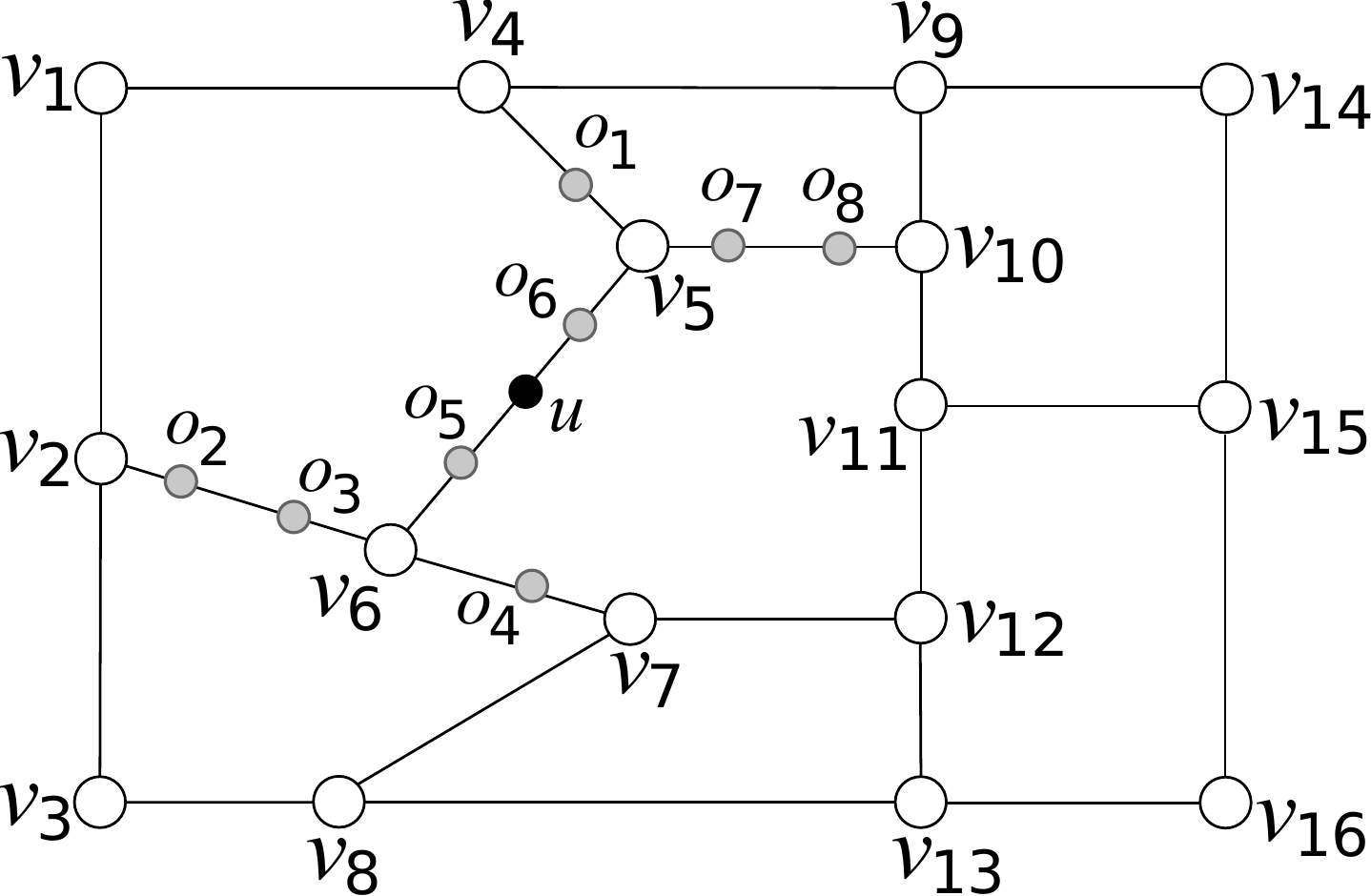}
	\vspace{-4pt}
	\caption{Illustration of query processing on a road network \label{fig:query_process}}
	\vspace{-10pt}
\end{figure}

Let $q$ denote a query issued at some position while traveling on segment $s$, and let $v^s_b$ and $v^s_e$ denote the two ends of $s$. The query result $\mathcal{R}(q,s)$ satisfies the following:
\begin{equation}
\mathcal{R}(q,s) \subseteq \mathcal{O}_s(q,s) \cup \mathcal{O}_v(q,v^s_b) \cup \mathcal{O}_v(q,v^s_e)
\end{equation}
We give an example in Figure \ref{fig:query_process}. A 3-nearest neighbor query is issued by a user $u$ located on segment $\overline{v_5 v_6}$. The exact answer to this query is $\mathcal{R}(q,s) = \{o_5, o_6, o_7\}$, which is indeed a subset of the union of ${\cal O}_{s}(q, \overline{v_5v_6})$ = $\{o_5, o_6\}$, 
${\cal O}_{v}(q, v_5)$ = $\{o_1, o_6, o_7\}$ and ${\cal O}_{v}(q, v_6)$ = $\{o_3, o_4, o_5\}$. We extend this model from a single segment $s$ to anonymized locations which potentially consist of a set of segments $S$ by employing the concept of border nodes (see Definition \ref{def:bordernode}). Concretely, the result of query $q$ with $S$ as its anonymized location satisfies:
\begin{equation} \label{eq:5}
\mathcal{R}(q,S) \subseteq \left(\cup_{s \in S}{\cal O}_{s}(q,s)\right)
\cup \left(\cup_{v \in BV(S)}{\cal O}_{v}(q, v)\right)
\end{equation}

Finally, the evaluation cost of $q$ with anonymized location $S$, denoted by $cost_{eval}(q,S)$ can be estimated as:
\begin{equation} \label{eq:evaluation}
cost_{eval}(q,S) = \mathcal{C}_{s} \cdot |S| + {\cal C}_{v} \cdot |BV(S)| 
\end{equation}
where $|BV|$ denotes the number of border nodes in $S$ and $|S|$ denotes the number of segments in $S$. 

\textbf{Cost of Communication:} We presented the architecture and communication phases of \textsc{StarCloak} in Figure \ref{fig:architecture}. We focus specifically on the cost that is added by a location anonymization service such as \textsc{StarCloak}. For query $q$, the communication cost in mobile client's exact request sent and the exact result it receives do not change depending on whether an anonymization engine is used or not, since a service request takes a fixed encoded format and the size of the exact answer is fixed. With respect to the messages exchanged between the location anonymization engine and the LBS provider, we measure communication cost as the length of the sent and received messages, and use $\Arrowvert x \Arrowvert$ to denote the encoded length of object $x$. For the message sent from the location anonymization engine to the LBS provider, the query remains intact while the location information is anonymized by cloaking it to a set of segments $S$. Therefore, the communication cost here is $\Arrowvert q \Arrowvert + \Arrowvert S \Arrowvert$. The message sent from the LBS provider to the location anonymization engine contains the candidate result $\mathcal{R}(q,S)$; hence, the communication cost here is $\Arrowvert {\cal R}(q, S) \Arrowvert$. As discussed above, a query $q$ usually has fixed length. Also, for given location privacy requirements, the number of segments in $S$ tends to be fairly stable. As such, we conclude that $\Arrowvert {\cal R}(q, S) \Arrowvert$ is the dominant and most ``optimizable" communication cost.

For query $q$, let $res\_size$ denote the average exact result size of $q$, e.g., if $q$ is the popular $k$-NN query, then $res\_size=k$. Following Equation \ref{eq:5}, given a query $q$ and anonymized location $S$ represented as a set of segments, the size of the candidate result $\mathcal{R}(q,S)$ can be estimated as:
\begin{equation}
|\mathcal{R}(q,S)| \leq res\_size \cdot |BV(S)| + \sum_{s \in S}\sum_{e \in s}|{\cal O}_{e}(q,e)| 
\end{equation}
Then, denoting by $\rho_o$ the average number of objects on an edge and $\mathcal{C}_o$ the cost of sending/receiving an object $o$ over the wireless channel (e.g., sending unique identifier of $o$), the total communication cost for $q$ with anonymized location $S$ is:
\begin{equation*} \label{eq:communication}
cost_{comm}(q,S) = \mathcal{C}_o \cdot \left[ res\_size \cdot |BV(S)| + \rho_o \cdot \sum_{s \in S}\sum_{e \in s}|e| \right]
\end{equation*} 

\textbf{Overall Cost:} It is desirable to combine $cost_{eval}$ and $cost_{comm}$ to find an estimation of the overall cost. In \textsc{StarCloak}, we consider a linear combination scheme:
\begin{equation*} \label{eq:overallcost}
cost(q,S) = \beta \cdot cost_{comm}(q,S) + (1-\beta) \cdot cost_{eval}(q,S)    
\end{equation*}
where $\beta$ is the parameter tuning the trade-off between evaluation cost (mainly CPU computation on server side) and the communication cost (mainly bandwidth of wireless channel).

\vspace{-4pt}
\subsection{Problem Statement}

Given a road network represented as a graph $G$ with mobile users traveling on the road network while issuing queries, where each user $u$'s query $q$ is associated with its profile $(\delta^q_k, \delta^q_l, \sigma^q_s, \sigma^q_t)$, the principles and objectives of \textsc{StarCloak} are:
\vspace{-4pt}
\begin{itemize}
    \item It transforms $u$'s true location to an anonymized (cloaked) location $S$, where $S$ is a subgraph of $G$.
    \item $S$ satisfies the privacy requirements of $q$ in terms of $\delta^q_k$-user anonymity and $\delta^q_l$-segment indistinguishability.
    \item $S$ satisfies the utility constraints of $q$, i.e., spatial size of $S$ is no larger than $\sigma^q_s$ and the temporal delay caused by location anonymization is no more than $\sigma^q_t$. 
    \item $S$ achieves high attack-resilience; measured in terms of low linkability and high segment entropy.
    \item Anonymized location $S$ yields low $cost(q,S)$.
\end{itemize}

\vspace{-6pt}

\section{\textsc{StarCloak} Algorithms}

This section explains the \textsc{StarCloak} constructs and algorithms in detail. We first describe the concept of cloaking \textit{star}, \textit{star graph}, and relevant data structures used in implementing \textsc{StarCloak} efficiently in Section \ref{sec:xs-step1}. We explain how an incoming query $q$ is pre-processed by \textsc{StarCloak} and added to the appropriate data structures in Section \ref{sec:preprocessing}. The overview of the main \textsc{StarCloak} algorithm is presented in Section \ref{sec:nutshell}. The main algorithm relies on several methods, such as selecting a star, updating the cloaking graph (adding and removing queries from the cloaking graph), candidate star-set selection, and star-set pruning. These methods are described in Sections \ref{sec:selectstar}, \ref{sec:cloaking_graph}, \ref{sec:candidatestarsetselection}, and \ref{sec:prune}, respectively.

\vspace{-8pt}

\subsection{Star Concept and \textsc{StarCloak} Data Structures} \label{sec:xs-step1}

Unlike conventional solutions which are indifferent to underlying road network structure, use a random waypoint mobility model, and rely on a rectangular or circular region as the basic unit of location cloaking; \textsc{StarCloak} introduces a \textit{star} as the basic unit of location cloaking. Each star is defined by a vertex with its adjacency segment list in $G$. 

\begin{definition}[Star] 
Let $G = \langle V_G, E_G \rangle$ denote the road network of interest. We define a star $\Phi_i$ anchored at vertex $v_i\in V_G$ as a subgraph of $G$, denoted by $\Phi_i = \langle V_{\Phi}^{i}, E_{\Phi}^i \rangle$, and $V_{\Phi}^{i}=\{v_i\}$ and $E_{\Phi}^{i} = \{ w | w\neq v_i, w\in V_G, (v_i, w)\in E_G\}$. 
\end{definition}

\begin{figure}[t]
\centering
\includegraphics[width=.45\textwidth]{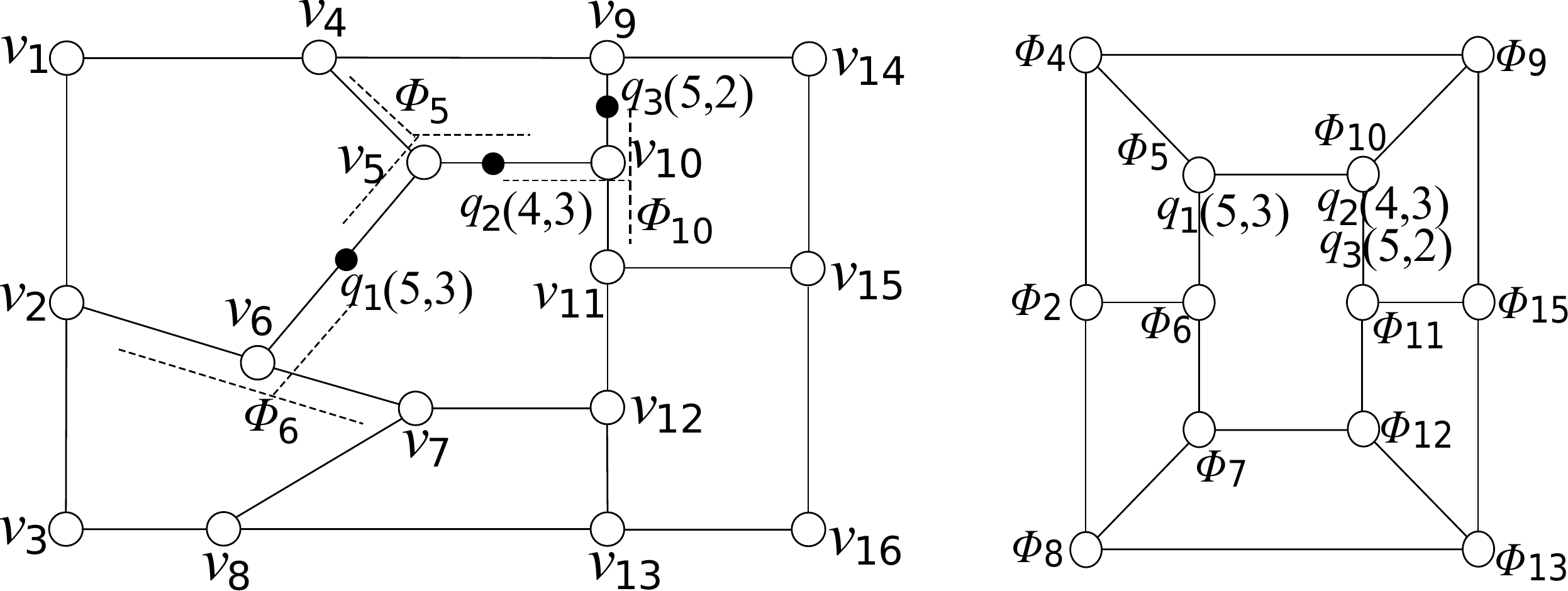}
\vspace{-2pt}
\caption{Illustration of the \textsc{Star} concept} 
\label{fig:star_model}
\vspace{-8pt}
\end{figure}

Accordingly, every node $v_i$ with $d_G(v_i) \geq 3$ is associated with a unique star $\Phi_i$, which consists of vertex $v_i$ and all of its adjacent road segments, that is, those segments with  $v_i$ as one of two end nodes. For example, in the left plot of Figure~\ref{fig:star_model}, star $\Phi_{5}$ is composed of node $v_5$ and segments $\{\overline{v_5v_4}$, $\overline{v_5 v_6}$, 
$\overline{v_5 v_{10}}\}$. 


The road network can then be transformed into a \textit{star graph}, as shown on the right of Figure \ref{fig:star_model}. Each vertex in the star graph is a star in $G$, and two vertices are adjacent in the star graph if and only if their corresponding stars in $G$ share a segment. All edges in the star graph are of unit length. The \textit{hop distance} between two stars $\Phi_i$ and $\Phi_j$ in a road network $G$ is measured by the number of hops in the shortest path between $\Phi_i$ and $\Phi_j$. For example, in Figure \ref{fig:star_model}, the hop distance between $\Phi_{6}$ and $\Phi_{10}$ is 2, since their shortest path in the star graph is $\Phi_{6} \rightarrow \Phi_{5} \rightarrow \Phi_{10}$.

In addition to the star concept, \textsc{StarCloak} uses some important data structures for improved effectiveness and efficiency.

\textit{\textbf{Query Queue, $\mathcal{Q}$}}: A first-in-first-out (FIFO) queue that records the incoming queries which must be anonymized before they are relayed to the respective LBS provider. Incoming queries are inserted into the queue from the tail. The anonymization engine pops each query from $\mathcal{Q}$ to find a suitable cloaked subgraph $S$. 

\textit{\textbf{Expiration Heap, $H$}}: A max-min heap that maintains the queries in the order of their expiration time computed according to query arrival time and temporal delay constraint $\sigma^q_t$. Anonymization engine checks $H$ to identify queries that are close to their expiration time in order to prioritize certain queries or to identify queries that have been expired and should be removed from $\mathcal{Q}$. 

\textit{\textbf{Cloaking Graph, $G_C$}}: An undirected graph dynamically constructed in-memory, for recording the set of queries associated to a star based on their similarities with respect to their privacy requirements, their spatial proximity, and their expiration deadlines. The cloaking graph will be explained further in Section \ref{sec:cloaking_graph}. 

\textit{\textbf{Star-Map, $M_{S}$ and Query-Map, $M_{Q}$}}: We create one hash map to index stars called a Star-Map, and similarly, one hash map to index queries associated to a node in the cloaking graph called Query-Map, for fast star and query look-up.

\textit{\textbf{Candidate Star-Set Queue, $\mathcal{Q}_C$}}: A FIFO-based queue structure that records generated candidate cloaking star-sets. The pruning unit in \textsc{StarCloak} pops star-sets from $\mathcal{Q}_C$ and applies an effective and randomized pruning algorithm to generate the final cloaked star regions.

\subsection{Incoming Query Pre-processing} \label{sec:preprocessing}

Let $q$ denote an incoming location service query. \textsc{StarCloak} pre-processes $q$ to generate the internal representation of $q$ by performing the following sequence of tasks. First, a unique identifier is assigned to $q$ using a secure hash function with user ID and query issue time, i.e., $hash(q.u || q.t)$. Second, using the latitude and longitude values of $q$'s focal location, the spatial index and road network graph, the road segment of $q$ is determined. Third, $q$ is inserted to queue $\mathcal{Q}$. Fourth, $q$ is inserted to the expiration heap $H$ with query expiration time as key and query identifier as value. Query expiration time is the sum of query issue time and user-specified temporal delay constraint: $q.t_{exp} = q.t + \sigma^q_t$.

\vspace{-6pt}

\subsection{Main \textsc{StarCloak} Procedure} \label{sec:nutshell}

\setlength{\textfloatsep}{6pt}
\begin{algorithm}[t]
\DontPrintSemicolon
\LinesNumbered
\KwIn{${\cal Q}$:~query queue,  ~$H$:~expiration heap, \\ \qquad ~~ $G_{C}$:~cloaking graph}
\KwOut{${\cal Q_C}$:~candidate star-set queue}
\While{${\cal Q} \ne \emptyset $}{
	$L_u \gets \emptyset$\;
	\While{$true$}{
		$q_e \gets$ first entry of $H$\;
		\If{$q_e$ \text{is expired}}
		{
			$v_u \gets$ \textit{removeQueryFromCloakingGraph($q_e$)}\;
			\If{$v_u \ne \varnothing $}{
				Add $v_u$ to $L_u$\;
			}
			Pop $q_e$ from $H$\;
		}
		\Else{
			\textbf{break}\;	
		}	
	}
	\ForEach{$v_u \in L_u$}{
		$C \gets$ \textit{SearchStarSet($v_u$)}\;
		\If{$C \ne \emptyset$}{
			Add $C$ to $\mathcal{Q}_C$ for pruning\;
		}
	}
	$q_i \gets$ first entry of ${\cal Q}$\;
	$\Phi_i \gets$ \textit{SelectStar($q_i$)}\;
	$v_u \gets$ \textit{addQuerytoCloakingGraph($q_i$, $\Phi_i$)}\;
	$C \gets$ \textit{SearchStarSet($v_u$)}\;
	\If{$C \ne \emptyset$}{
		Add $C$ to $\mathcal{Q}_C$ for pruning\;
	}
}
\caption{Main \textsc{StarCloak} Algorithm}
\label{alg:starcloak}
\end{algorithm}

Before going into the details of each step, we summarize the main \textsc{StarCloak} procedure in Algorithm \ref{alg:starcloak}. The location anonymization engine continuously pops queries from the query queue $\mathcal{Q}$ and processes them to find anonymized location $S$ fitting the desired privacy and utility requirements. Prior to processing a new query $q$, \textsc{StarCloak} removes expired queries from the system. When an expired query denoted $q_e$ is removed from a cloaking graph node, it is possible to find cloaked subgraph for the remaining queries in the node. All updated nodes are stored in the list $L_u$ after expired queries are removed, and \textsc{StarCloak} attempts to find anonymized locations for these nodes before processing new queries (lines 12-15). Next, it pops a new query from query queue $\mathcal{Q}$ and selects the star to assign to it (lines 16-17). Thereafter, \textsc{StarCloak} updates the cloaking graph with the new query $q_i$ and searches a possible cloaked location for the updated cloaking graph node (lines 18-21). Finally, in the pruning phase, \textsc{StarCloak} randomly selects and removes non-active stars from the candidate star-sets. In the next sections, we present the details of these operations.

\vspace{-4pt}
\subsection{Select Star} \label{sec:selectstar}

\textsc{StarCloak} engine performs anonymization by scanning through the FIFO query queue $\mathcal{Q}$. All segments that are associated with active queries are marked as \textit{active}. The anonymization engine first selects a star to assign queries on the active segment as the initial cloaking star. Each segment has two end nodes and if both nodes are intersection nodes, i.e., $d_G(v_s) \geq 3$ and $d_G(v_t) \geq 3$, \textsc{StarCloak} needs to determine to which star this active segment should be assigned: $\Phi_{s}$ or $\Phi_{t}$. For example, in Figure \ref{fig:star_model} when $q_1$ arrives and segment $\overline{v_5 v_6}$ becomes active, one of the two possible stars $\Phi_{5}$ or $\Phi_{6}$ will be determined as the initial cloaking star. When a star $\Phi$ is ``selected" and segment $s$ is assigned to $\Phi$, we denote this by $s \leftarrow \Phi$. 

In \textsc{StarCloak}, we use a cost-aware star selection strategy, taking into account the cost model described in Section \ref{sec:costmodel}. Let $q$ be the query, $AS$ denote the set of currently active segments on the road network $G$, and $\phi$ be the set of selected stars. Then, the minimization of the overall cost can be formally stated as:
\begin{align} \label{nphard}
\min_{\phi} & \sum_{\Phi \in \phi} cost(q, \Phi) \\
\textrm{s.t.} & ~~ \forall s \in AS, \exists \Phi \in \phi, s \leftarrow \Phi    \nonumber
\end{align}
This optimization problem aims at finding an assignment between stars and segments such that the stars cover all segments with active queries, while having the minimum total cost. 
It can be shown that the optimization problem in Expression \ref{nphard} is NP-Hard. The proof follows from a reduction from the weighted Vertex-Cover problem, which is a well-known NP-Complete problem. Specifically, if for all stars $\Phi$ in the star graph we set $cost(q, \Phi)=1$, i.e., all stars have identical cost, then the problem is equivalent to the classical Vertex-Cover problem. Motivated by the hardness of finding a globally optimal solution to our optimization problem, we propose a randomized algorithm called Select Star, which finds approximate solutions with high assignment quality and attack-resilience. The intuition is, for each query which has two endpoints as viable stars, the algorithm probabilistically selects one of the two stars with probability inversely proportional to their $cost$.

The technical description of our Select Star algorithm is given in Algorithm \ref{alg:selectstar}. The algorithm works as follows. Let $q$ be an incoming query with travel segment $s$, and let $\Phi_{a}$ and $\Phi_{b}$ be the two stars on the two endpoints of segment $s$. For simplicity, we assume both endpoints are stars; if not, then $s$ is trivially assigned to the endpoint which is a star. If only one of $\Phi_{a}$ or $\Phi_{b}$ is currently active, Select Star assigns $s$ to the active star. If both $\Phi_{a}$ and $\Phi_{b}$ are active, then $s$ is assigned to $\Phi_{a}$ with probability $cost(q, \Phi_{b})/[cost(q, \Phi_{a}) + cost(q, \Phi_{b})]$, or $\Phi_{b}$ otherwise. If neither star is active, then the same probabilistic assignment to either $\Phi_{a}$ or $\Phi_{b}$ is carried out, additionally, the assigned star is marked as active for next iterations. This assignment has the desirable property that the outcome of our randomized Select Star algorithm is not far from an optimal assignment. More formally, denoting by $cost^{opt}$ the cost achieved by the optimal assignment, and denoting by $cost^{rnd}$ the cost achieved by our Select Star algorithm, it holds in expectation that: $\mathbb{E}\left[cost^{rnd}\right]$ $\leq$ $2 \cdot cost^{opt}$. 

\IncMargin{1.2em}
\begin{algorithm}[h]
\DontPrintSemicolon
	\LinesNumbered
	\Indm
		\KwIn{$q$:~new query, $I_{\Phi}$:~active star index }
		\KwOut{${\Phi}$:~selected star}
	\Indp
	
	$s$ $\leftarrow$ the segment containing $q$\;
	\tcp{$\Phi_a$ and $\Phi_b$ are two stars that share segment $s$}
	\If{$s$ is already assigned to $\Phi_a$ $(\textit{resp. } \Phi_b)$}{
			\textbf{return} $\Phi_a$ $(\textit{resp. } \Phi_b)$\;
	}
	\ElseIf {$\{\Phi_a, \Phi_b\} \cap I_{\Phi}$ = 
		$\{\Phi_{a}, \Phi_{b}\}$}{
		\tcp{both active, neither covers s}
		Assign $s$ to $\Phi_{a}$ w.p. $\frac{\textit{cost}(q,\Phi_{b})}{\textit{cost}(q,\Phi_{a}) 
			+ \textit{cost}(q,\Phi_{b})}$ or $\Phi_{b}$ otherwise\;
	}
	\ElseIf{$\{\Phi_{a}, \Phi_{b}\} \cap I_{\Phi}$ = 
		$\Phi_{a}$  $(\textit{resp. } \Phi_{b})$}{
		\tcp{only one star is active}
		Assign $s$ to $\Phi_{a}$ (resp. $\Phi_{b}$)\;
	}
	\ElseIf{$\{\Phi_{a}, \Phi_{b}\} \cap I_{\Phi}$ = $\emptyset$}{      
	\tcp{neither star is selected yet}
		\If{${d}_G(v_a^s)$ = 1 or ${d}_G(v_b^s)$ = 1}{
			\tcp{only one end is a star}
			Add $\Phi_{b}$ (resp.~$\Phi_{a}$) to $I_{\Phi}$\;
			Assign $s$ to $\Phi_{b}$ (resp.~$\Phi_{a}$)\;
		}
		\Else{
			Assign $s$ to $\Phi_{a}$ w.p. $\frac{\textit{cost}(q, \Phi_{b})}
			{\textit{cost}(q, \Phi_{a}) + \textit{cost}(q, \Phi_{b})}$ or  $\Phi_{b}$ otherwise\;
			Add $\Phi_{a}$ (or  $\Phi_{b}$) to $I_{\Phi}$\;
		}
	}
    \textbf{return} $\Phi_a$ $(\textit{or } \Phi_b)$ according to chosen assignment\;
	\caption{Select Star Algorithm}
	\label{alg:selectstar}
\end{algorithm}

\setlength{\textfloatsep}{\textfloatsepsave}

\subsection{Cloaking Graph Update} \label{sec:cloaking_graph}

We use the cloaking graph data structure (previously introduced and denoted by $G_C$) to group nearby queries and efficiently index other query groups that can be cloaked together for easy access. The cloaking graph $G_C(V_C,E_C)$ is an undirected graph, where $V_C$ is the set of vertices each representing a set of requests grouped by the star they are assigned to and their profiles (similarities in privacy and utility requirements). $E_C$ is the set of edges; there is an edge $e=(v_i,v_j) \in E_C$ between $v_i$ and $v_j$ iff queries associated with both vertices can be cloaked together based on $k$-user anonymity, $l$-segment indistinguishability, and spatial tolerance. Each vertex $v$ in $V_C$ stores the following information.

The \textbf{corresponding star $v.\Phi$}: for each active star, there is at least one vertex in $V_C$. The \textbf{query set $v.Q$} stores the queries assigned to $v.\Phi$. We compute the \textbf{combined privacy of utility requirements $\langle \delta_k^v, \delta_l^v, \sigma^v_s \rangle$} of the queries in $v.Q$ as:
\begin{equation} \label{eq:updreqs}
    \langle \delta_k^v, \delta_l^v, \sigma^v_s \rangle := \langle \max_{q \in v.Q} \delta_k^q, \max_{q \in v.Q} \delta_l^q, \min_{q \in v.Q} \sigma^q_s \rangle
\end{equation}
We denote by \textbf{covered star-set $v.\varTheta$} the set that contains the identifiers of stars which are within $\sigma^v_s$ distance from $v.\Phi$. The \textbf{segment count $v.sc$} denotes the number of segments associated with stars in star-set $v.\varTheta$. Finally, \textbf{adjacency list $v.N$} is stored with $v$, where being neighbors indicates that requests in corresponding nodes can be cloaked together. Two cloaking nodes $v_i$ and $v_j$ are considered to be neighbors iff: (i) stars associated with each node are an element of the star-set of the other node, i.e., $v_i.\Phi \in v_j.\varTheta$ and $v_j.\Phi \in v_i.\varTheta$, and (ii) the number of segments that cover both nodes is enough to satisfy their $l$-segment indistinguishability requirements, i.e., $|v_i.\varTheta \cap v_j.\varTheta| \geq max \{ \delta_l^{v_i}, \delta_l^{v_j} \}$.

\textsc{StarCloak} performs two types of updates on the cloaking graph: add query to cloaking graph, remove query from cloaking graph. The function for adding queries to the cloaking graph is given in Algorithm \ref{alg:insert}. When the function is called to insert a new query, it checks all cloaking graph vertices associated with the corresponding star to add the new query (lines 4-14). If there is no possible vertex to add, a new vertex is created (lines 15-16). The new query can be added to an existing vertex only if its privacy profile does not conflict with the profile of the existing node. A conflict occurs when new spatial tolerance is not able to satisfy the new $l$-segment indistinguishability requirement. Thus, we need to perform the checks under lines 4-14 to avoid any conflicts. 

\setlength{\textfloatsep}{6pt}
\begin{algorithm}[t]
\DontPrintSemicolon
\Indm
	\KwIn{$q$:~new query, $\Phi_n$: assigned star}
	\KwOut{$v_u$:~updated cloaking graph node}
\Indp
$v_u \gets \varnothing$\;
$N \gets M_{S}.get(\Phi_n)$\;
\If{$N \ne \emptyset$}{
	\ForAll{$v_i \in N$}{
		\If{$\sigma^q_s < \sigma^{v_i}_s$}{
			$sc \gets$ \# of segments within $\sigma_s^q$ from $\Phi_n$\;
			\If{$sc \ge max\{ \delta_l^q, \delta_l^{v_i} \}$ }{
				Add $q$ to the node $v_i$\;
				$v_u \gets v_i$\;
				\textbf{break}\;
			}
		}
		\ElseIf{$v_i.sc \ge \delta_l^{q}$}{
			Add $q$ to the node $v_i$\;
			$v_u \gets v_i$\;
			\textbf{break}\;
		}
	}
}
\If {$v_u = \varnothing$}{
	$v_u \gets $ Create new node for query $q$\;
}
\textbf{return} $v_u$\; 
\caption{Add Query to Cloaking Graph}
\label{alg:insert}
\end{algorithm}

The function for removing queries from the cloaking graph is given in Algorithm \ref{alg:remove}. Say that $q_e$ is the expired query that should be removed. We first perform a look-up from $M_Q$ to find the cloaking graph node $v_u$ associated with $q_e$. If $|v_u.Q| > 1$, i.e., $v_u$ contains other queries as well, its information is updated based on remaining queries after deletion of $q_e$. The update is performed according to Equation \ref{eq:updreqs} to re-compute $\delta_k^{v_u}$, $\delta_l^{v_u}$, and $\sigma_s^{v_u}$. If the updated $v_u$ now has either $\delta_l^{v_u} < \delta_l^{q_e}$ or $\sigma_s^{v_u} > \sigma_s^{q_e}$, then ${v_u}.\varTheta$, ${v_u}.sc$ and $v_u.N$ are also re-computed. Note that the latter is only necessary if segment indistinguishability or spatial tolerance requirements are relaxed. On the other hand, if $|v_u.Q| = 1$, i.e., $q_e$ was the only query associated with $v_u$, then $v_u$ is removed from $G_C$, and $M_S$ is updated. The return value of the function is $v_u$, which is an input for the next step (candidate star-set selection). 

\begin{algorithm}[!h]
\DontPrintSemicolon
\Indm
\KwIn{$q_e$:~expired query}
\KwOut{$v_u$:~updated cloaking graph node}
\Indp
$v_u \gets M_{Q}.get(q_e)$\;
\If{$|v_u.Q| > 1$}{
	Update $\delta_k^{v_u}$, $\delta_l^{v_u}$, and $\sigma_s^{v_u}$ \;
	\If{$\delta_l^{v_u} < \delta_l^{q_e}$ or $\sigma_s^{v_u} > \sigma_s^{q_e}$}{
			Update ${v_u}.\varTheta$, ${v_u}.sc$ and $v_u.N$\; 
	}
	
}
\Else{
	Remove $v_u$ form $G_C$\;
	$v_u \gets null$\;
	Update $M_S$\;
}
Remove $q_e$ from $M_{Q}$\;
\textbf{return} $v_u$\; 
\caption{Remove Query from Cloaking Graph}
\label{alg:remove}
\end{algorithm}


\vspace{-4pt}

\subsection{Candidate Star-Set Selection} \label{sec:candidatestarsetselection}

\setlength{\textfloatsep}{6pt}
\begin{algorithm} [t]
\DontPrintSemicolon
\Indm
\KwIn{$v_u$: updated vertex} 
\KwOut{$\vartheta$: candidate star-set} 
\Indp
$NS \gets v_u$\;
\If{$(\vartheta \gets checkReqs(NS)) \ne \emptyset$}{
	\Return $\vartheta$\;
}
$Q_{Comb} \gets \emptyset$ \;
\ForAll{$v \in v_u.N$}{
	$NS \gets v_u \cup v$\;
	\If{$( \vartheta \gets checkReqs(NS)) \ne \emptyset$}{
		\Return $\vartheta$\;
	}
	$Q_{Temp} \gets Q_{Comb}$\;
	\ForAll{$C \in Q_{Comb}$}{
		\If{$\forall v_c \in C, v_c \in v.N$}{
			$NS \gets v_u \cup v \cup C$\;
			\If{$( \vartheta \gets checkReqs(NS)) \ne \emptyset$}{
				\Return $\vartheta$\;
			}
			\Else{
				$Q_{Temp} \gets Q_{Temp} \cup C \cup v$ \;
			}
		}
	}		
	$Q_{Comb} \gets Q_{Temp}$
}
\caption{Search Candidate Star-Set}
\label{alg:select}
\end{algorithm}

The goal of this step is to discover a set of stars, called \textit{candidate star-set}, which constitutes a possible anonymized sub-graph for certain queries. In order to find such star-set, \textsc{StarCloak} searches over the cloaking graph and identifies a set of nodes, denoted by $NS$, that satisfy the privacy requirements of all queries associated with each node. Formally, let $\vartheta$ denote a candidate star-set, and let $seg(\vartheta)$ be a function that returns all segments associated with input stars. $NS$ meets $k$-user anonymity and $l$-segment indistinguishability if and only if:
\begin{equation} \label{eq:privcheck}
\forall v \in NS: ~~ \Big[ \delta_k^v \leq \sum_{\hat{v} \in NS} |\hat{v}.Q| \Big] ~~\land~~ \Big[ \delta_l^v \leq |seg(\bigcap_{\hat{v} \in NS} \hat{v}.\varTheta)| \Big]
\end{equation}
Such $NS$ forms candidate star-set $\vartheta$ with all stars shared within the covered star-set of each node in $NS$:
\begin{equation} \label{eq:starset}
\vartheta = \bigcap_{v \in NS} v.\varTheta
\end{equation}
We assume the existence of a procedure named \textit{checkReqs(NS)}, which takes as input a set of nodes $NS$, performs the privacy check given in Equation \ref{eq:privcheck}, and returns either the $\vartheta$ built in Equation \ref{eq:starset} if $NS$ passes the privacy check or an empty set $\emptyset$ otherwise. We use the \textit{checkReqs} procedure in Algorithm \ref{alg:select} for candidate star-set selection.

Algorithm \ref{alg:select} specifies the technical details of candidate star-set selection process. Searching over the cloaking graph for finding a candidate star-set starts with the updated vertex $v_u$. If the number of queries assigned to this vertex is fewer than the $k$-user anonymity requirement of the vertex, then the algorithm continues the search process over the neighboring nodes, ordered by the hop distance between their associated star and the star of the starting vertex. For each neighbor node, it applies \textit{checkReqs} to $v_u$ and the neighbor node combined (lines 6-8). If a candidate star-set still cannot be found, neighbor node is evaluated with all possible node combinations generated with the previously processed neighbor nodes (lines 10-16). On line 10, we denote by $C$ a clique in $Q_{Comb}$, and line 11 checks if the clique satisfies the $l$-segment indistinguishability requirement. The possible node combinations are tracked by variable $Q_{Comb}$, which is enlarged in each iteration so that newly visited nodes are added (lines 16-17). The output of the algorithm is $\vartheta$, a candidate a star-set.

\setlength{\textfloatsep}{\textfloatsepsave}

\vspace{-4pt}
\subsection{Star-Set Pruning} \label{sec:prune}

Final component of \textsc{StarCloak} is star-set pruning: pruning of extra segments from candidate star-set. As specified in the main \textsc{StarCloak} procedure in Algorithm \ref{alg:starcloak}, the candidate star-sets found by Algorithm \ref{alg:select} are added to the candidate star-set queue denoted $\mathcal{Q}_C$, and then they are pruned by the star-set pruning component. Star-set pruning plays an important role in the generation of low cost cloaked subgraphs and improved attack-resilience by randomizing the star selection from outer to center of the candidate star-set. Note that star-set pruning is highly parallelizable, i.e., it is possible to implement one or more pruning processes running in parallel (each popping from $\mathcal{Q}_C$) while another process in the anonymization engine performs remaining tasks and adds to $\mathcal{Q}_C$.

The function for pruning star-sets is given in Algorithm \ref{alg:prune}. Pruning starts by popping a candidate star-set from queue $\mathcal{Q}_C$. Let $\vartheta$ denote the popped star-set. We find the set of boundary stars $BS$ of $\vartheta$, which are the stars that have at least one neighbor star not in $\vartheta$, as well as the set of active stars $AS$ of $\vartheta$ which cannot be removed from the star-set. Let $l^{max}$ denote the maximum $l$-segment indistinguishability requirement in the star-set. We run multiple iterations, and within each iteration, the following are performed. First, a random star denoted $\Phi_r$ is selected from $BS \setminus FS$. If $\vartheta$ still satisfies $l^{max}$-segment indistinguishability after removing $\Phi_r$ from $\vartheta$; then $\Phi_r$ is removed from $\vartheta$, $BS$ is updated by removing $\Phi_r$ from $BS$, and we proceed to the next iteration. However, if $l^{max}$-segment indistinguishability is violated after removing $\Phi_r$ from $\vartheta$, then the pruning stops here and the current $\vartheta$ (without removing $\Phi_r$) is produced as the final output of the pruning process.

\begin{algorithm}[t]
\LinesNumbered
\DontPrintSemicolon
\Indm
\KwIn{$\mathcal{Q}_c$:~candidate star-set queue}
\Indp
\While{$true$}{
	\If{$\mathcal{Q}_c \ne \emptyset $}{
		$\vartheta \gets$ Pop the first entry of $\mathcal{Q}_c $\;
		$BS \gets$ Find boundary stars of $\vartheta$\;
		$FS \gets$ Select active stars\;
		$l^{max} \gets$ Find max segment requirement\;
		\While{$true$}{
			$\Phi_r \gets$ Randomly select star from $BS$\;
			\If{$\Phi_r \notin FS$}{
				$\vartheta \gets \vartheta \setminus \Phi_r $\;
				\If { $\#$ of segments in $\vartheta > l^{max}$ }{
					Update $BS$\;
				}
				\Else{
					$\vartheta \gets \vartheta \cup \Phi_r $\;
					Output $\vartheta$\;
					\textbf{break}
				}
			}
		}
	}
}
\caption{Prune Candidate Star-Set}
\label{alg:prune}
\end{algorithm}

\begin{figure}
\centering
\includegraphics[width=.48\textwidth]{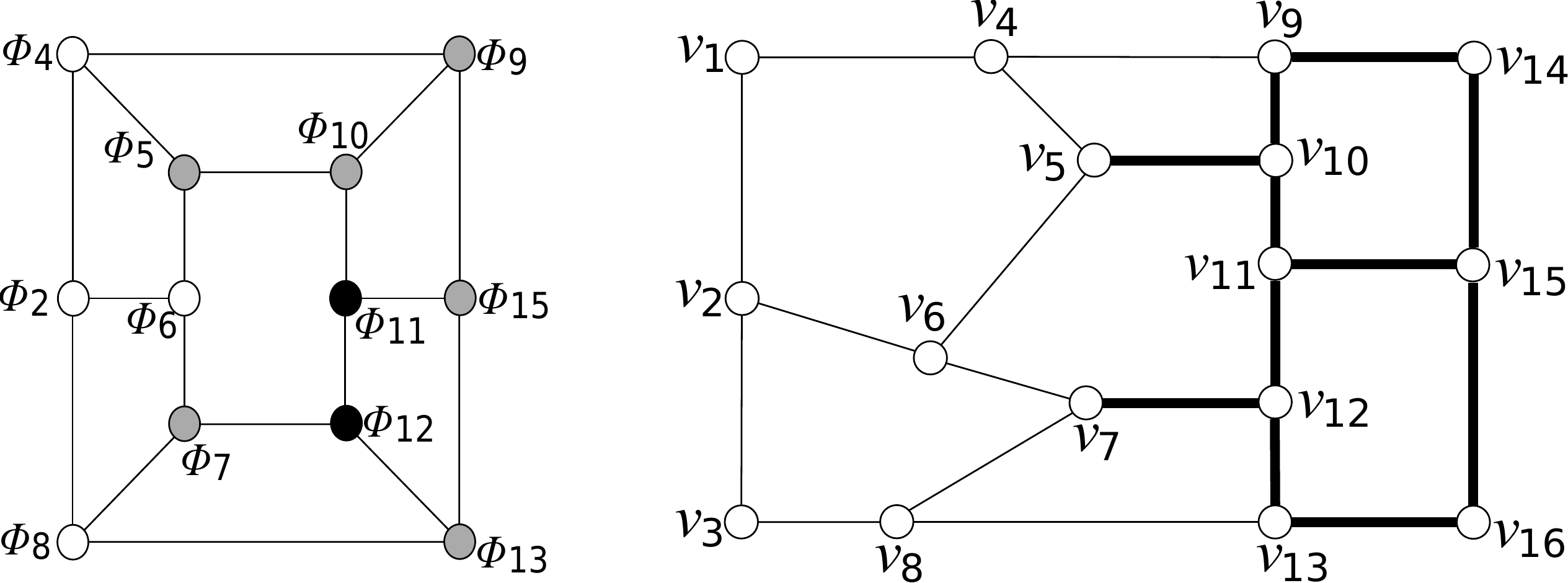}
\caption{Star-set pruning example} 
\label{fig:prun}
\vspace{-8pt}
\end{figure}

We give an example of star-set pruning in Figure \ref{fig:prun}. The candidate star-set is shown on the left with the black and grey circles. Black circles depict active nodes that cannot be removed because of their association with active queries. Suppose that the maximum segment requirement is 9. First, one generates boundary star list $\{ \Phi_5, \Phi_7, \Phi_9, \Phi_{13} \}$, i.e., gray circles that are connected with the white circles. Then, one of the boundary stars is selected randomly, say $\Phi_5$ for sake of example. After removing $\Phi_5$, remaining stars still meet the segment requirement. Boundary star set is updated with the new star $\Phi_{10}$ and another star is selected randomly from the set. Suppose $\Phi_7$ is selected for pruning; indistinguishability requirement is still satisfied with the remaining stars. Note that there are no new boundary stars because $\Phi_{12}$ is an active star. One selects another star from the current boundary star set which is $\{ \Phi_5, \Phi_9, \Phi_{10}, \Phi_{13} \}$. Assume $\Phi_9$ is selected for removal; then boundary star set is updated with new star $\Phi_{15}$. However, removing any of the current stars now violates the queries' segment indistinguishability requirement, thus we have to add back the selected star to the star list. Associated segments constitute the cloaked subgraph. We show the resulting cloaked subgraph with bold lines on the right side of Figure \ref{fig:prun}.

\vspace{-4pt}

\section{Variants and Optimizations}  \label{sec:xs-query}

In this section, we introduce two variants of \textsc{StarCloak} for finding cloaking regions with lower cost, query processing time, and network bandwidth usage without sacrificing privacy. 

\vspace{-4pt}
\subsection{Spatially Bounded \textsc{StarCloak}}

Basic \textsc{StarCloak} generates cloak regions whenever it finds a star-set that satisfies all queries' privacy requirements. However, this approach may cause cloak regions that consist of stars that are far from each other and scattered across the part of the road network within $\sigma_s$. An example scenario is given in supplementary material. We propose spatially bounded \textsc{StarCloak} to generate more compact cloaked subgraphs. The essence of this optimization is to sacrifice anonymization time in favor of lower query processing and communication costs. We define a system parameter $\lambda \geq 1$ called the \textbf{compactness factor}, that controls the maximum hop distance between selected vertices in the candidate star-set. To generate more compact cloaked subgraphs, we make some modifications to the candidate star-set selection algorithm. First, we group neighbors by their distance $d$ to the starting node. $\lfloor d / \lambda \rfloor$ determines the level of each group element. At each level, the algorithm only considers neighbor nodes which can be cloaked with the node combinations generated in the previous level. The algorithm searches level by level iteratively in top-down manner. Spatially bounded \textsc{StarCloak} enforces compactness by selecting active stars that, for each star in the star-set there is at least one other star which is no further than $2\lambda-1$ hop distance.

\begin{figure*}[!ht]
\begin{minipage}[!b]{0.25\linewidth}
    \includegraphics[width=\linewidth]{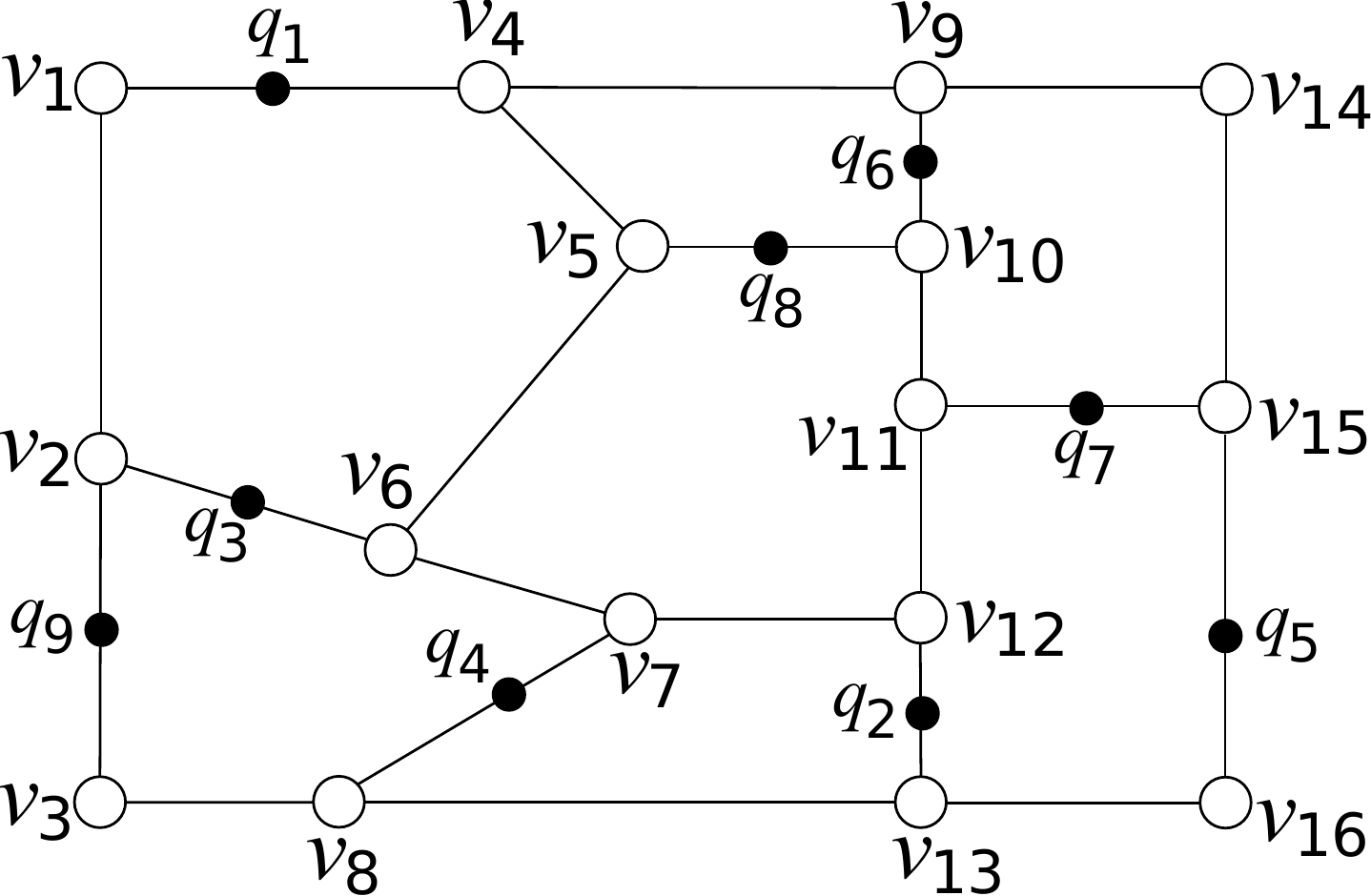}
    \subcaption{Queries on road network}
    \label{fig:opt1}
\end{minipage}%
\hspace{10pt}
\begin{minipage}[c]{0.15\linewidth}
\small{
    \begin{tabular}[b]{|c|c|}
			\hline
			{\bf query} & star  \\ 
			\hline
			\hline
			\textbf{$q_1$} & $\phi_4$ \\
			\textbf{$q_2$} & $\phi_{12}$ \\
			\textbf{$q_3$} & $\phi_6$ \\
			\textbf{$q_4$} & $\phi_8$  \\
			\textbf{$q_5$} & $\phi_{13}$ \\
			\textbf{$q_6$} & $\phi_9$ \\
			\textbf{$q_7$} & $\phi_{11}$\\
			\textbf{$q_8$} & $\phi_5$\\
			\textbf{$q_9$} & $\phi_8$ \\
			\hline
	\end{tabular}
}
	\subcaption{Star assignment}
	\label{fig:opt_s}
\end{minipage}%
\begin{minipage}[!b]{0.17\linewidth}
    \includegraphics[width=\linewidth]{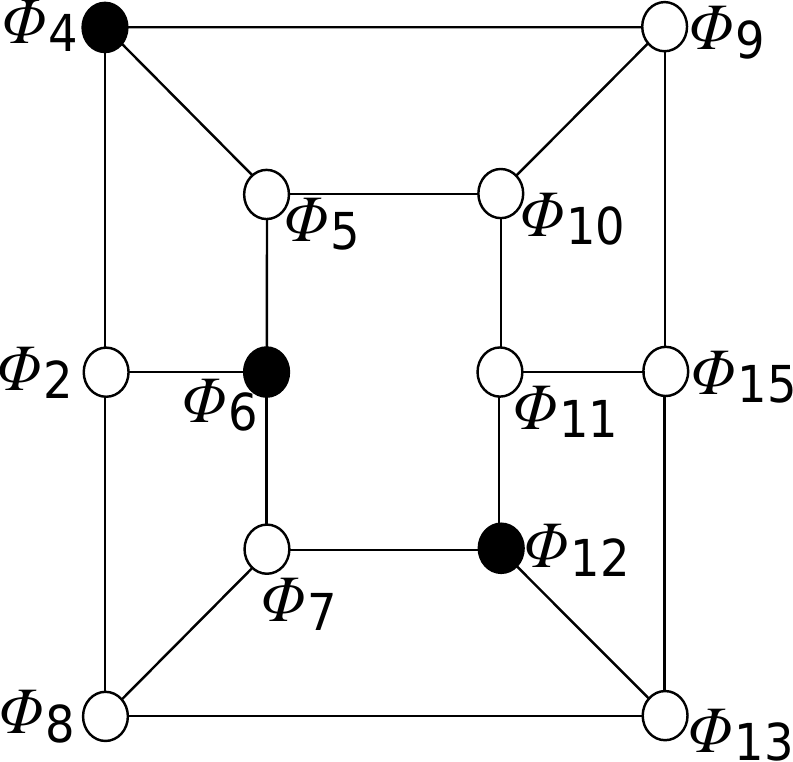}
    \subcaption{$q_1, q_2, q_3$}
    \label{fig:opt2}
\end{minipage}%
\hspace{5pt}
\begin{minipage}[!b]{0.17\linewidth}
    \includegraphics[width=\linewidth]{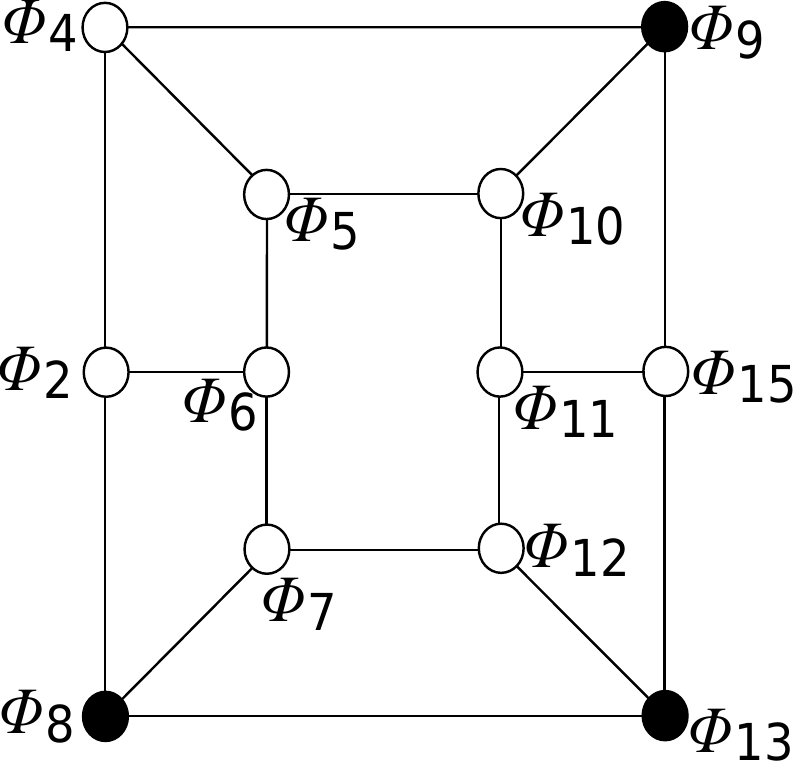}
    \subcaption{$q_4, q_5, q_6$}
    \label{fig:opt3}
\end{minipage}%
\hspace{5pt}
\begin{minipage}[!b]{0.17\linewidth}
    \includegraphics[width=\linewidth]{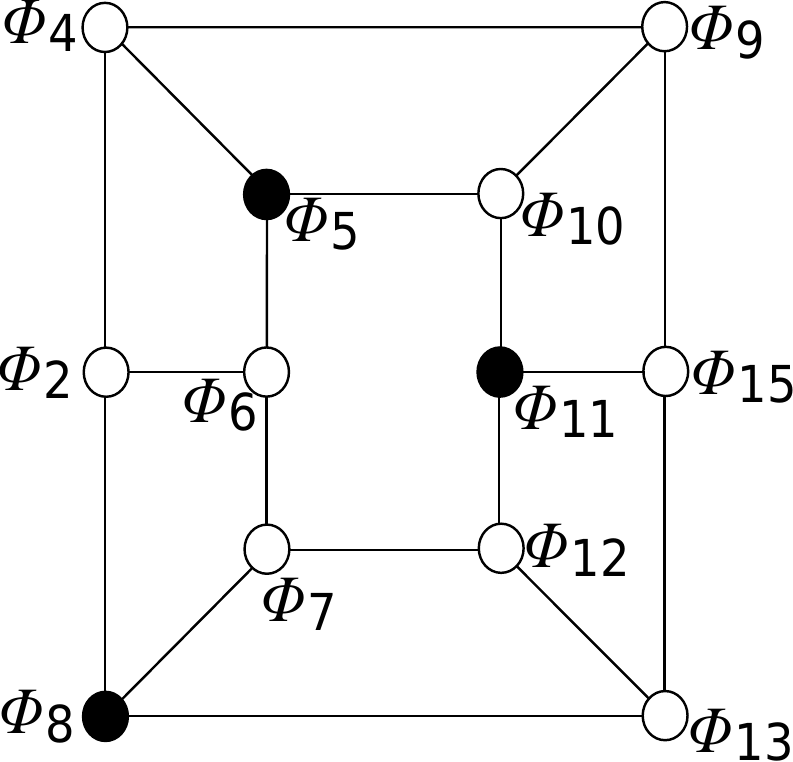}
    \subcaption{$q_7, q_8, q_9$}
    \label{fig:opt4}
\end{minipage}%
\caption{A spatially suboptimal cloaking output that can potentially be produced by basic \textsc{StarCloak}}
\label{fig:optEx}
\vspace{-8pt}
\end{figure*}

\begin{figure}[!h]
\begin{minipage}[!b]{0.155\textwidth}
    \includegraphics[width=\textwidth]{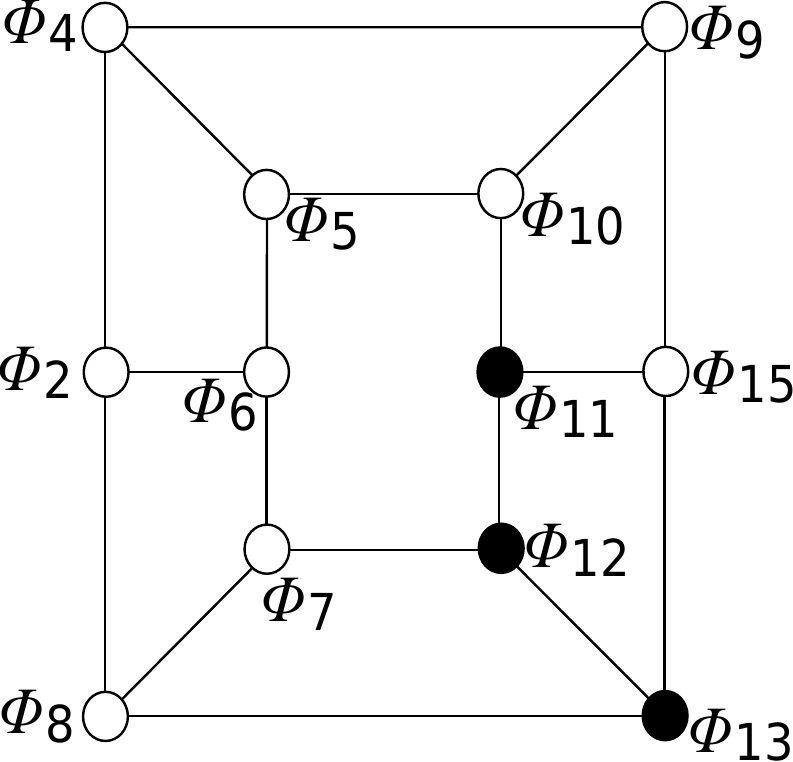}
    \vspace{-10pt}
    \subcaption{$q_2, q_5, q_7$}
    \label{fig:opt5}
\end{minipage}%
\hspace{2pt}
\begin{minipage}[!b]{0.155\textwidth}
    \includegraphics[width=\textwidth]{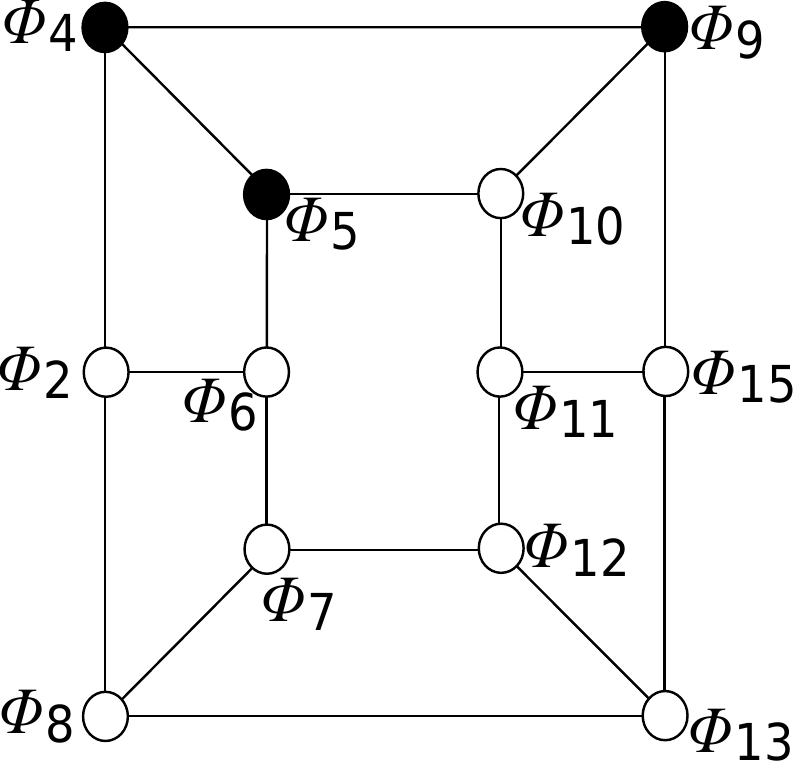}
    \vspace{-10pt}
    \subcaption{$q_1, q_6, q_8$}
    \label{fig:opt6}
\end{minipage}%
\hspace{2pt}
\begin{minipage}[!b]{0.155\textwidth}
    \includegraphics[width=\textwidth]{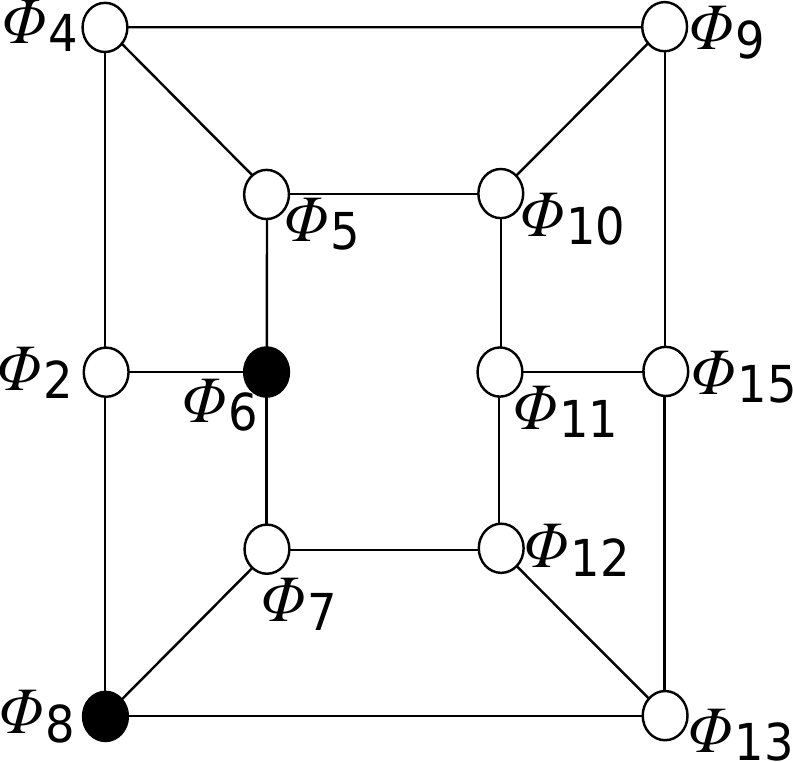}
    \vspace{-10pt}
    \subcaption{$q_3, q_4, q_9$}
    \label{fig:opt7}
\end{minipage}%
\caption{Improvement of spatially bounded \textsc{StarCloak} for the same queries from Figure \ref{fig:optEx}a and \ref{fig:optEx}b}
\label{fig:optEx2}
\vspace{-8pt}
\end{figure}

\textbf{Illustrative Example:} Figure \ref{fig:optEx} shows an example scenario with queries $q_1, q_2, \ldots, q_9$ distributed on the road-network as in Figure \ref{fig:opt1}. Suppose that queries are issued in the order of their ID, and all queries have 3-user anonymity requirement. (For simplicity, we assume the spatial tolerance is high enough to cover all stars in our small road network, and no $l$-segment indistinguishability requirement exists.) In Figure \ref{fig:opt_s} we give an example star assignment for the 9 queries. When we apply basic \textsc{StarCloak} for the given queries, selected stars will be as shown in Figures \ref{fig:opt2}, \ref{fig:opt3} and \ref{fig:opt4}. It can be observed that selected stars are spread all over the network. On the other hand, Figure \ref{fig:opt5}, \ref{fig:opt6} and \ref{fig:opt7} show a more compact star selection possibility for the same queries with different combinations.

Suppose that for the above example, we used spatially bounded \textsc{StarCloak} with compactness factor $\lambda=1$. During the processing of $q_3$, it would first check neighbor nodes which are 2 hop distance from the star $\Phi_6$. Since there is no neighbor in level 1, the anonymization engine would then accept new queries to process. While it is processing $q_7$, it first adds neighbor node associated with the star $\Phi_{12}$ which is in the 1 hop distance level. Two nodes do not satisfy $k$-user anonymity requirement yet, thus the anonymization engine continues to process neighbor nodes at the second level ($\Phi_9$ and $\Phi_{13}$). Since we use a FIFO-based query processing ordering to decrease waiting time, $\Phi_{13}$ is selected in the next iteration. Three nodes together meet all users' privacy requirement and can be removed from the cloaking graph.

\vspace{-4pt}
\subsection{Hybrid \textsc{StarCloak}}

The main difficulty in spatially bounded \textsc{StarCloak} is the choice of $\lambda$. At first sight, $\lambda$ can be determined by the query density of a general area. However, query density is often highly dynamic and changes street-by-street or star-by-star. Even neighboring segments may have different densities. Thus, the $\lambda$ determined based on query density of a general area may be undesirable for sparse sub-areas, and it is not possible to define an optimal compactness factor for each individual star at each time. To overcome this problem, we propose \textit{hybrid} \textsc{StarCloak}, which leverages advantages from both basic \textsc{StarCloak} and spatially bounded \textsc{StarCloak}. In hybrid \textsc{StarCloak}, we first try to generate cloaking regions with spatially bounded \textsc{StarCloak}, and then for queries which could not be cloaked yet and are close to their expiration time, we apply basic \textsc{StarCloak}. We use a \textbf{consideration factor} denoted by $\alpha$ as the system parameter to decide when to apply basic \textsc{StarCloak}. Hybrid \textsc{StarCloak} periodically checks the expiration heap $H$ to see if any query is closer than $\alpha$ to their expiration time.

To demonstrate the usefulness of hybrid \textsc{StarCloak}, we consider the example from Figure \ref{fig:optEx}. When we use spatially bounded \textsc{StarCloak} with compactness factor $\lambda =1$ for the example in Figure \ref{fig:optEx}, queries $q_3, q_4, q_9$ would remain in the system until new queries were issued in their neighborhood, since the distance between stars $\Phi_6$ and $\Phi_8$ is two hops. Assume now that these queries have an expiration time, then they may have to be dropped before the new queries arrive, even though there is a possible cloaked subgraph. With hybrid \textsc{StarCloak}, we would be able to apply basic \textsc{StarCloak} towards the queries' expiration time, and we would be able to cloak those stars together, thereby saving the queries from being dropped.

\vspace{-4pt}
\section{Experimental Evaluation} \label{sec:experiments}


\subsection{Experimental Setup} \label{sec:exp_setup}

We used two different road network datasets, California \footnote{\url{http://www.cs.utah.edu/~lifeifei/SpatialDataset.htm}} and Georgia \footnote{\url{https://www.census.gov/geo/maps-data/data/tiger-geodatabases.html}}, with varying sizes to observe the effect of map density on the efficiency and effectiveness of \textsc{StarCloak}. California road network contains only highways with 21,693 edges and 21,048 nodes. 87,635 points of interest from 62 different classes (e.g., hospital, school, etc.) are associated with the road network. Georgia is the larger road network dataset, which contains primary and secondary roads with 430,849 edges and 428,708 nodes. To simulate user movements, we used the Brinkhoff data generator for moving objects \footnote{\url{http://iapg.jade-hs.de/personen/brinkhoff/generator/}}. We assign the same number of moving objects (10,000) to each map, with the intention of simulating high user density and low user density conditions since the two maps have different scale. In each simulation, we define two classes of moving objects: vehicles with fast speed (such as passenger cars) and vehicles with slow speed (such as trucks). 

During the simulation, each vehicle generates $k$-NN queries with randomized probability with parameters specified as: 
(1) $k$ denotes the number of nearest points of interest requested; (2) $\delta_k$ and $\delta_l$ are the personalized privacy parameters; (3) $\sigma_s$ and $\sigma_t$ are the personalized spatial and temporal tolerance constraints; (4) $\gamma$ is the \textit{waiting time}, i.e., amount of time a vehicle waits until its previous query is either answered or dropped, before issuing another query. The values of each individual query are drawn independently from Gaussian distributions with default mean and standard deviation parameters listed in Table \ref{tab:parameter}. The values of parameters $\sigma_t$, $\alpha$, and $\gamma$ are in seconds. The compactness factor $\lambda$ and consideration factor $\alpha$ are only used in spatially bounded \textsc{StarCloak} and hybrid \textsc{StarCloak}. All algorithms are implemented in Java and tested on a Windows 7 platform with Intel(R) Core(TM) CPU (4.00 GHz) and 16GB memory.

\setlength{\textfloatsep}{6pt}
\begin{table}[t]
\centering
\caption{Default parameter settings used in our experiments}
\begin{tabular}{|c|c|c|c|c|c|c|c|c|c|} \hline
Parameter&$k$&$\delta_k$&$\delta_l$&$\sigma_s$&$\sigma_t$&$\gamma$&$\lambda$&$\alpha$\\ \hline \hline
Mean&$5$&$5$&$5$&$4$&$10$&$20$&$1$&$2$\\ \hline 
Deviation&$1$&$1.5$&$1.5$&$1$&$2$&$2$&$0$&$0$\\ 
\hline\end{tabular}\\
\label{tab:parameter}
\end{table}

\vspace{-4pt}
\subsection{Compared Approaches} \label{sec:compared_approaches}

In our evaluation, we compare multiple approaches. Random sampling and network expansion serve as two baseline anonymization approaches. \textsc{XStar} \cite{wang:vldb09} is the most relevant system to \textsc{StarCloak}. We also include three versions of \textsc{StarCloak} in our comparison: basic, spatially bounded, and hybrid.

\textbf{Random Sampling:} Given an incoming query with profile $(\delta^q_k, \delta^q_l, \sigma^q_s, \sigma^q_t)$, this approach iteratively samples segments randomly from the spatial region within $\sigma^q_s$ one-by-one, and adds them to the anonymized location. It terminates when $(\delta^q_k, \delta^q_l)$ privacy requirements are satisfied. The strength of random sampling is its high resilience to inference attacks. Its weakness is the high query processing cost due to random segment selection.

\textbf{Network Expansion:} For incoming query $q$, this approach starts from the actual segment of the query and incrementally adds a neighboring segment using Dijkstra's deterministic network expansion algorithm. The order of expansion is based on the distance between $q$'s focal position and neighboring segments' midpoints. The approach terminates when $(\delta^q_k, \delta^q_l)$ privacy requirements are satisfied. Network expansion results in a cloaked location connected as a densely compact subgraph. Its advantage is low query processing cost. Its main weakness is added vulnerability to attack since the expansion follows a deterministic best-first search.

\textbf{\textsc{XStar}:} The most related work to \textsc{StarCloak} in the literature is \textsc{XStar} \cite{wang:vldb09}, which performs road network anonymization under utility and privacy constraints. Our evaluation shows \textsc{StarCloak} is superior to \textsc{XStar} in aspects including reduced query processing and anonymization time, higher success rate, and higher attack-resilience.

\textbf{\textsc{StarCloak} and Variants:} In our result graphs, we denote by \textsc{StarCloak} the basic version of \textsc{StarCloak}. We include its two optimized variants, which are spatially bounded \textsc{StarCloak} and hybrid \textsc{StarCloak}, in our experimental comparison.

\vspace{-4pt}
\subsection{Evaluation Metrics}  \label{sec:metrics}

\begin{figure*}[!ht]
\centering
\begin{minipage}[!b]{0.25\textwidth} 
    \includegraphics[width=\textwidth]{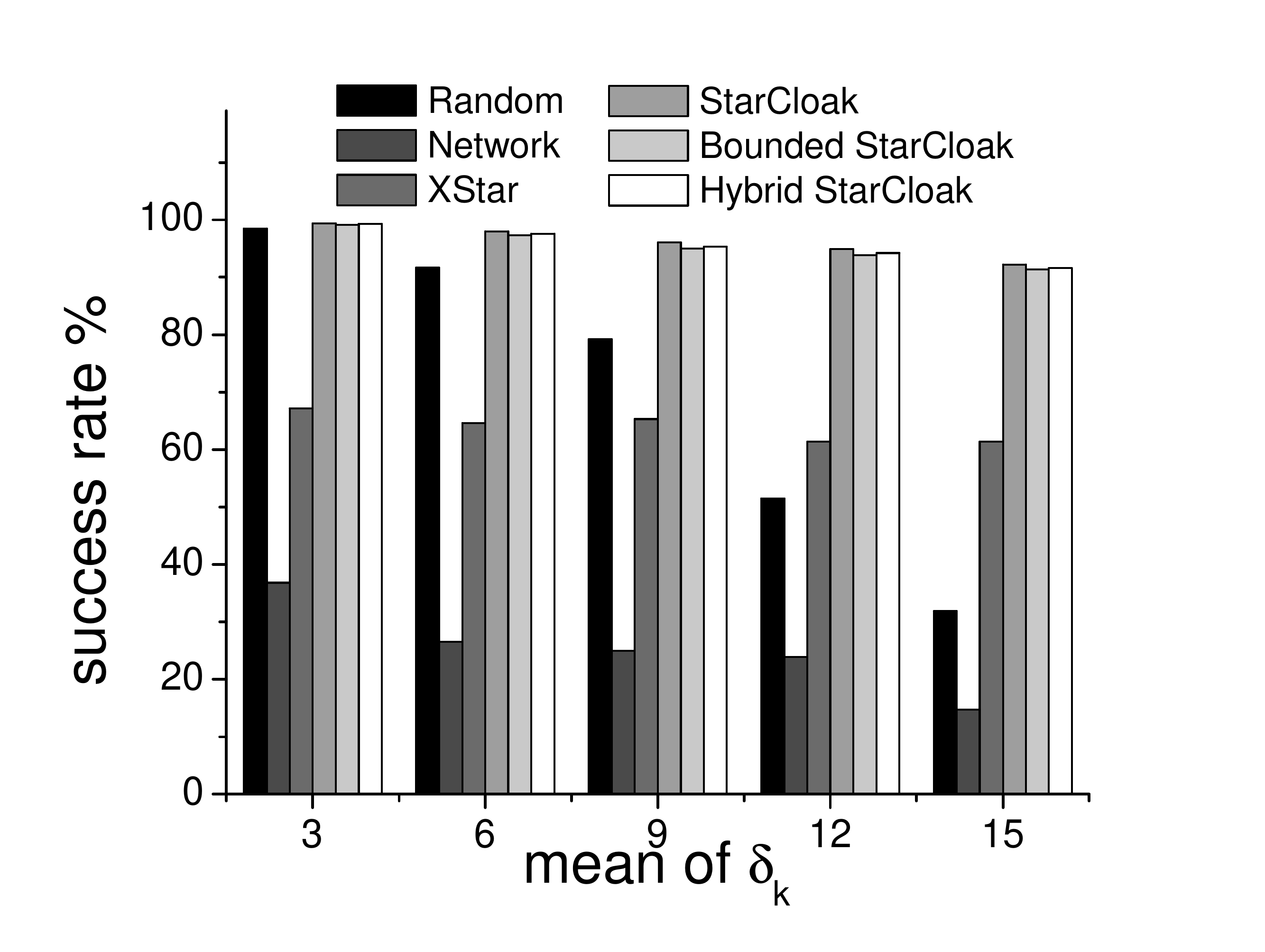}
\end{minipage}%
\hspace{-7mm}
\begin{minipage}[!b]{0.25\textwidth}
    \includegraphics[width=\textwidth]{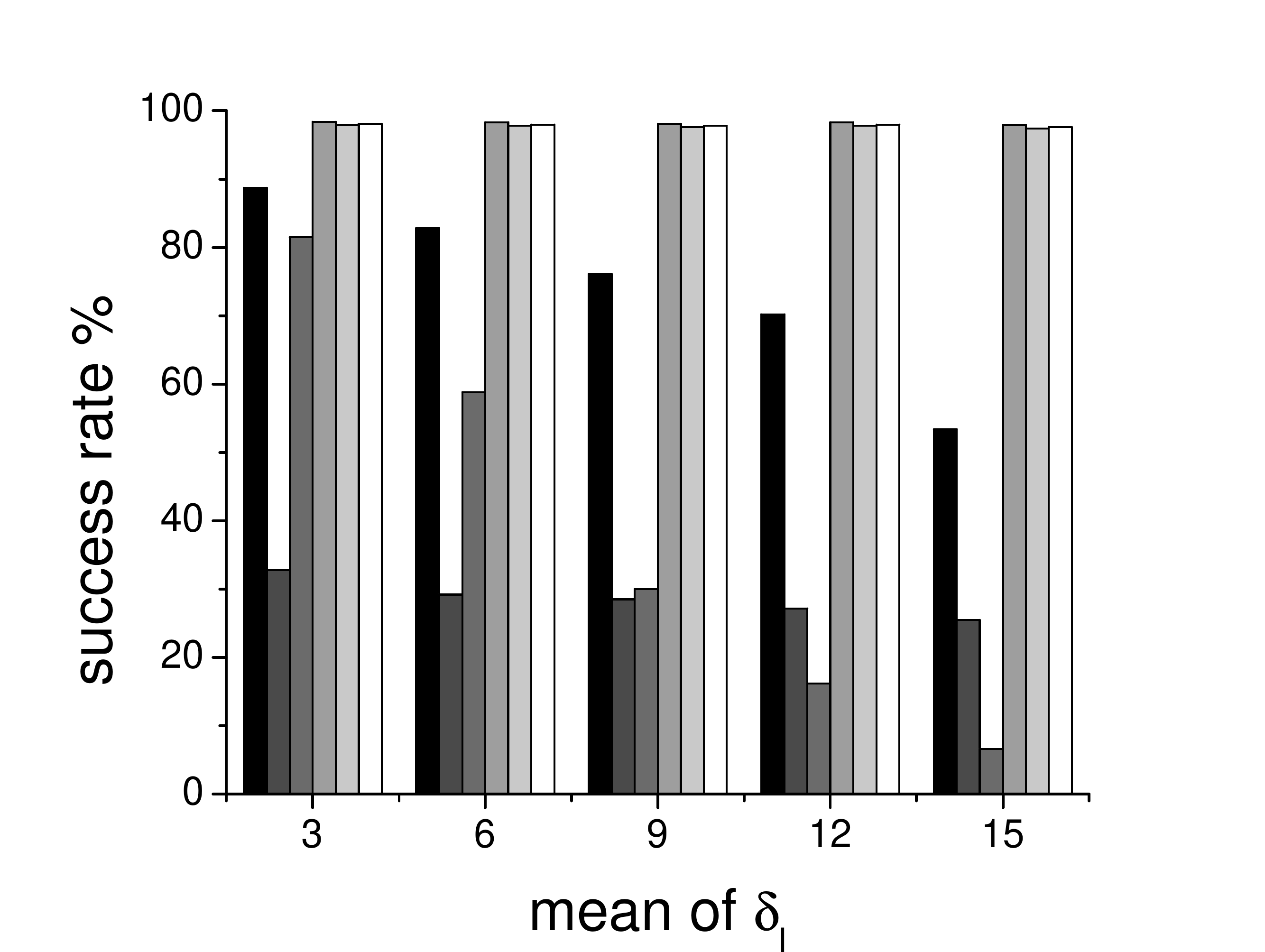}
\end{minipage}%
\hspace{-7mm}
\begin{minipage}[!b]{0.25\textwidth}
    \includegraphics[width=\textwidth]{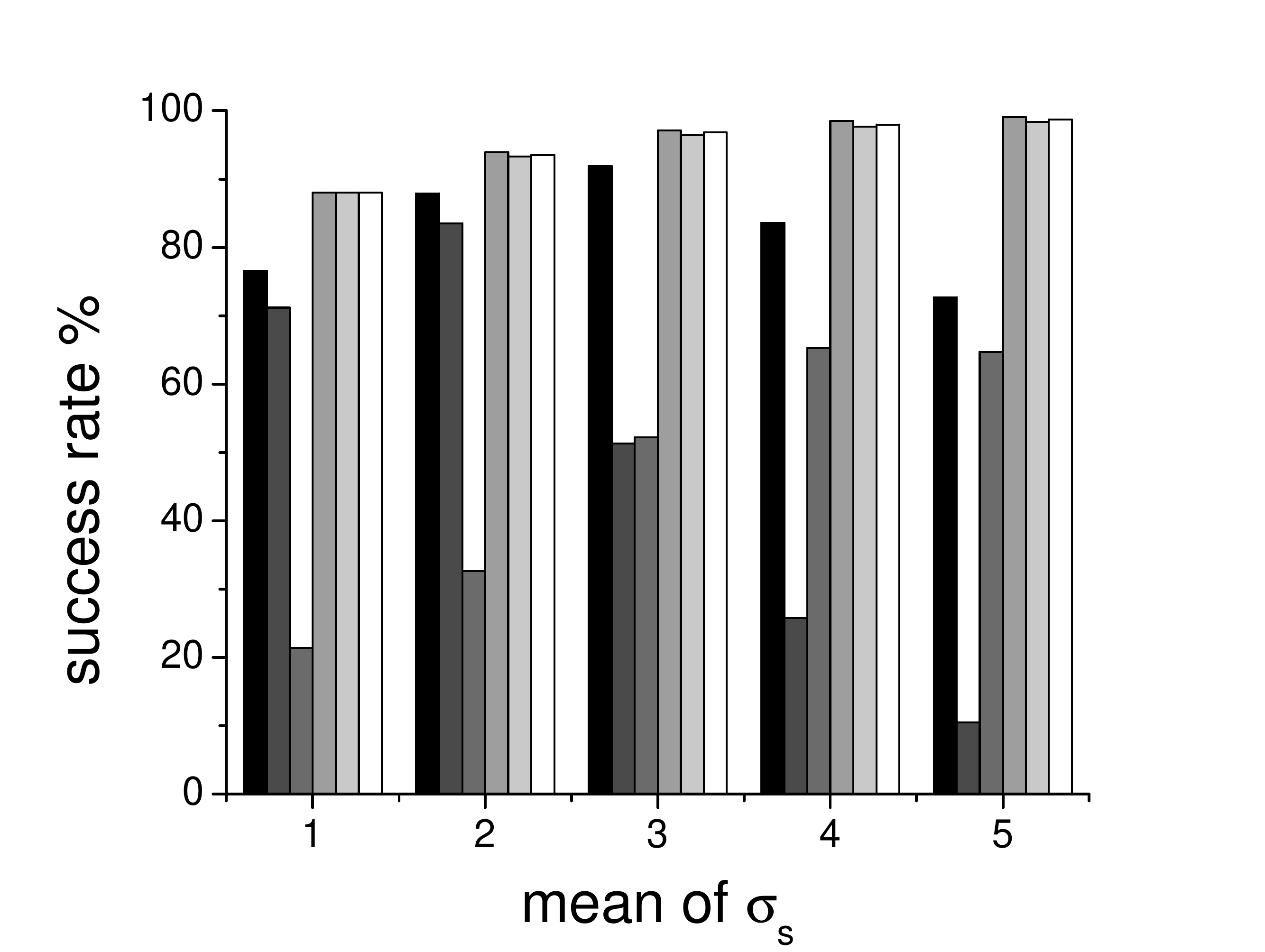}
\end{minipage}%
\hspace{-7mm}
\begin{minipage}[!b]{0.25\textwidth}
    \includegraphics[width=\textwidth]{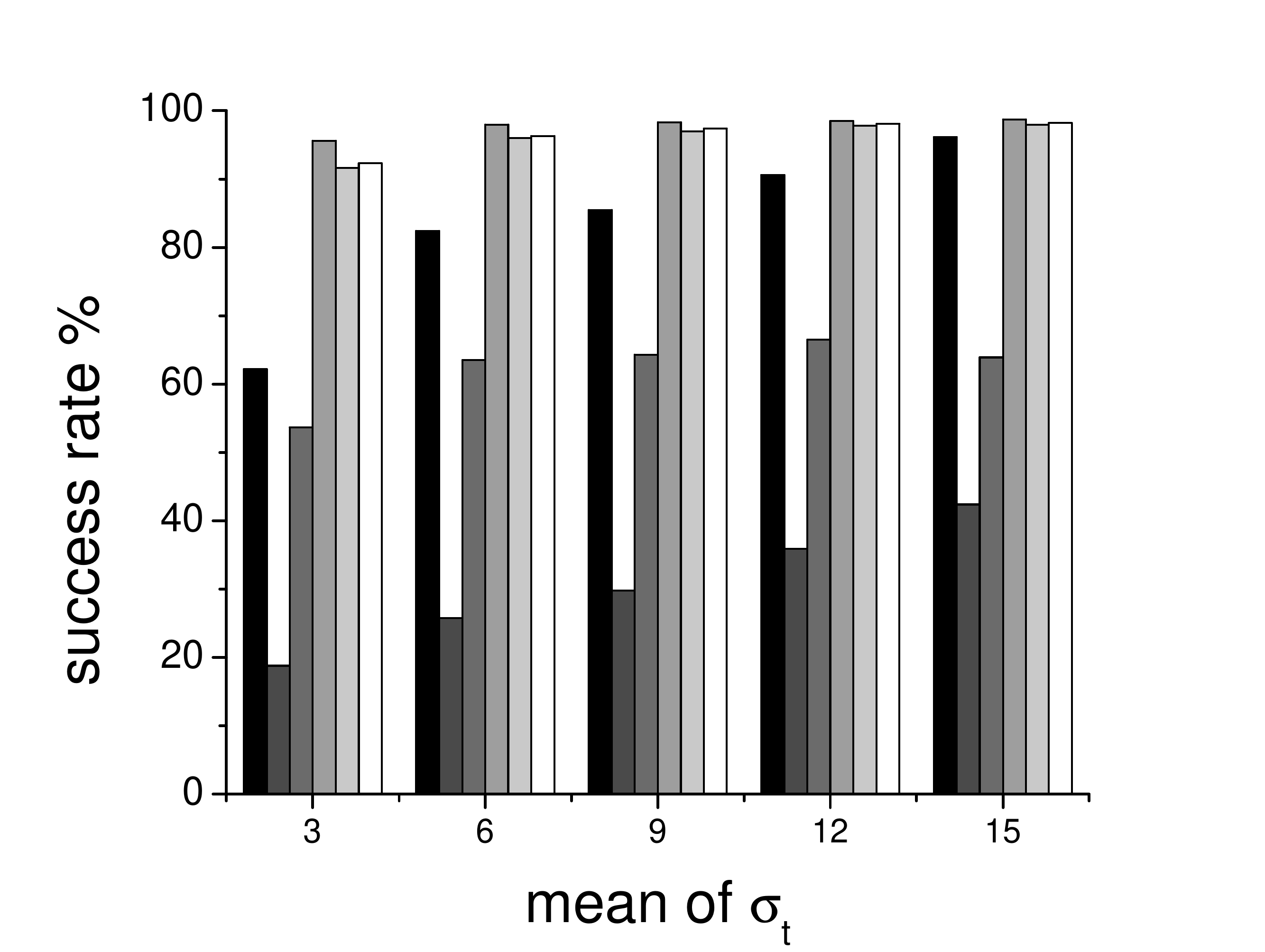}
\end{minipage}%
\\
\begin{minipage}[!b]{0.25\textwidth}
    \includegraphics[width=\textwidth]{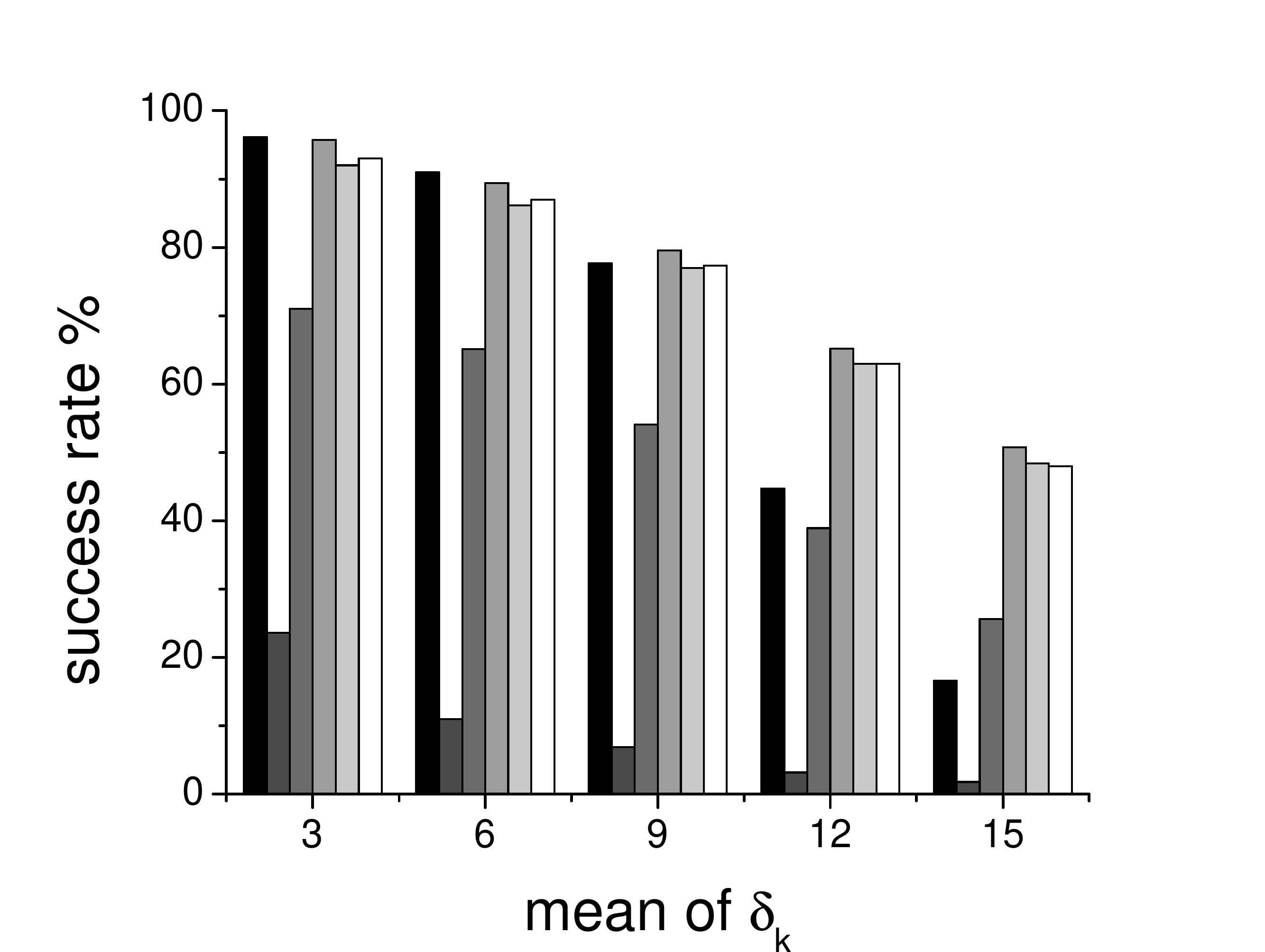}
\end{minipage}%
\hspace{-7mm}
\begin{minipage}[!b]{0.25\textwidth}
    \includegraphics[width=\textwidth]{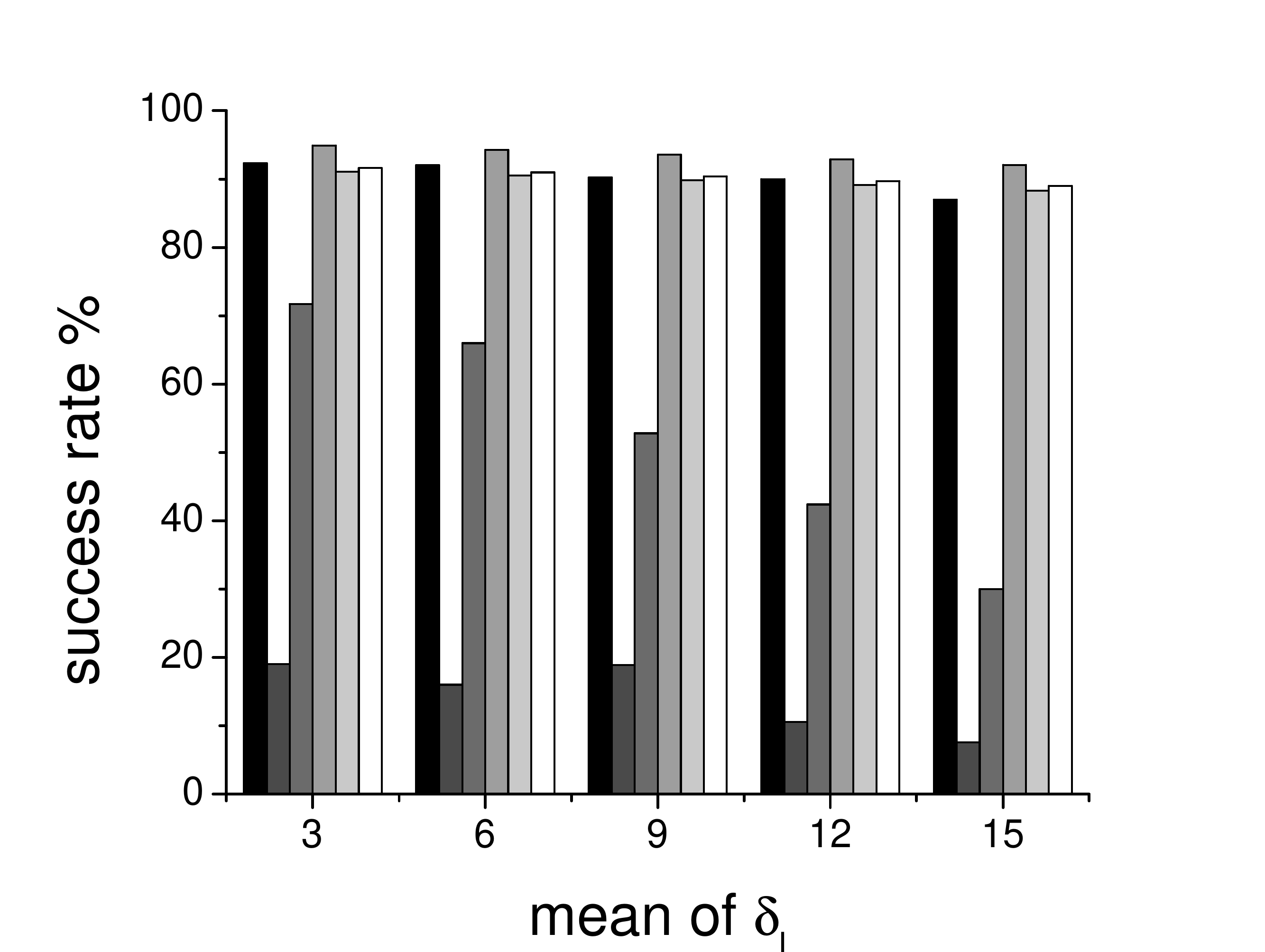}
\end{minipage}%
\hspace{-7mm}
\begin{minipage}[!b]{0.25\textwidth}
    \includegraphics[width=\textwidth]{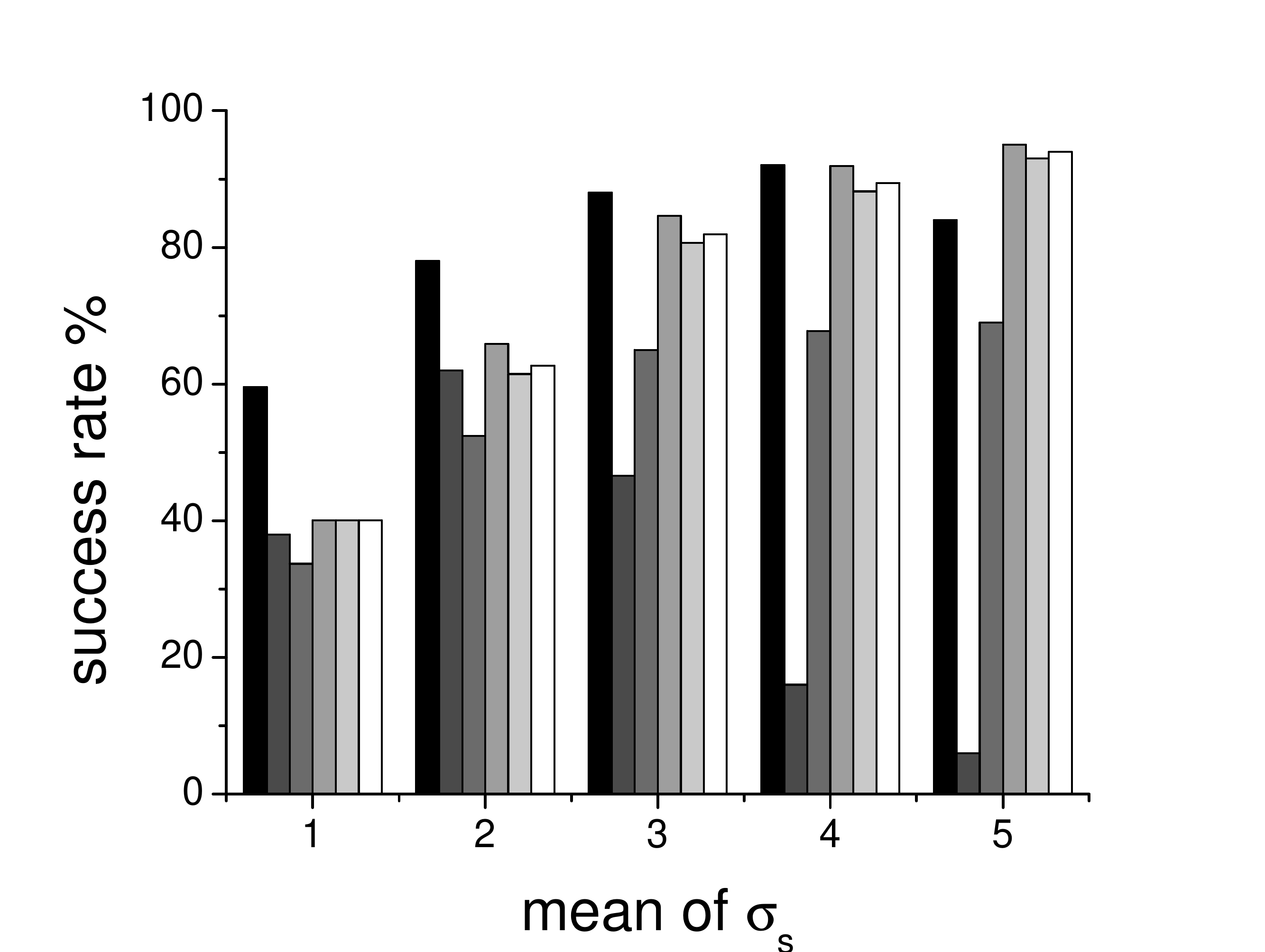}
\end{minipage}%
\hspace{-7mm}
\begin{minipage}[!b]{0.25\textwidth}
    \includegraphics[width=\textwidth]{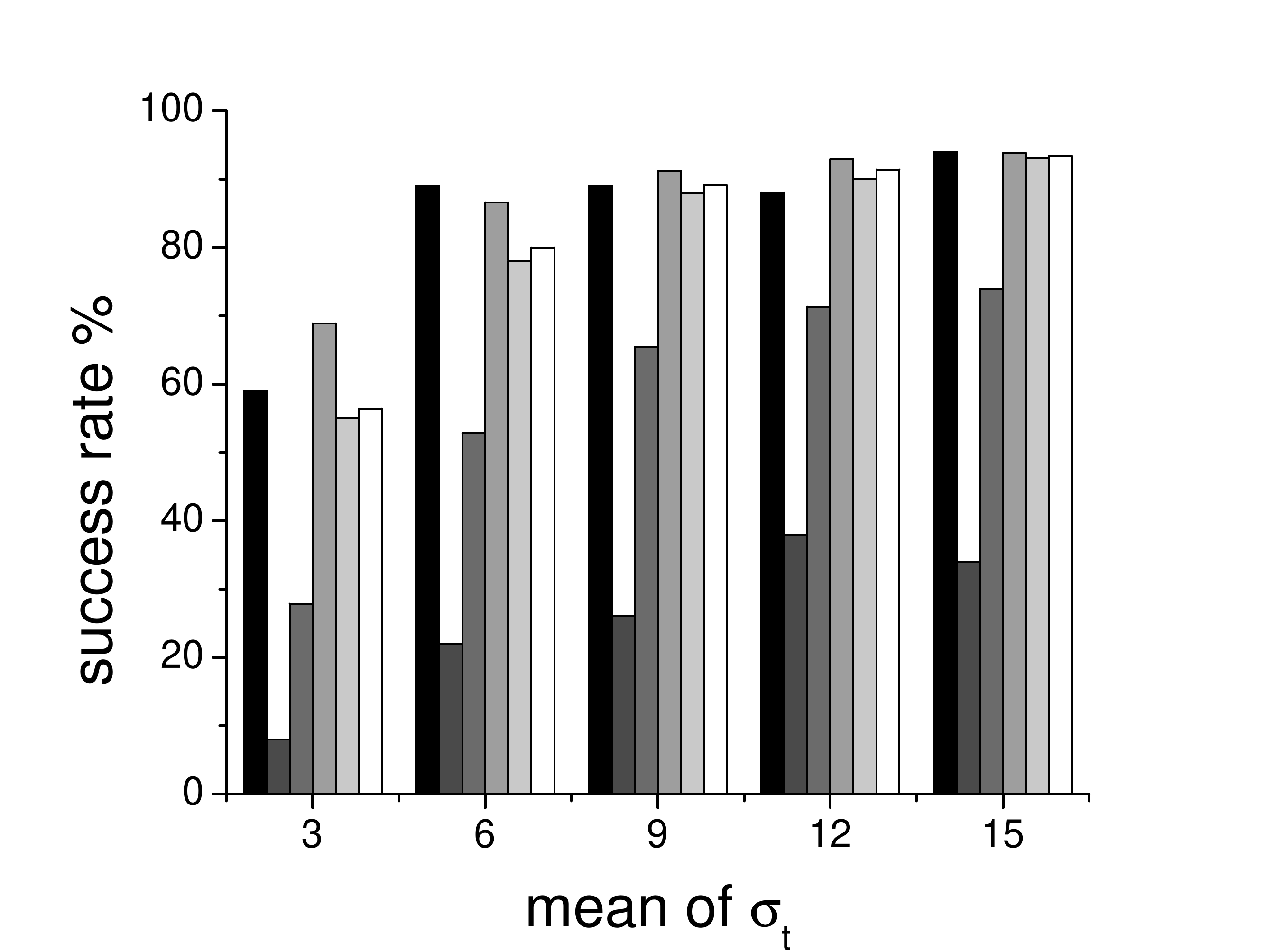}
\end{minipage}%
\vspace{-4pt}
\caption{Success rate for California map (four graphs in top row) and Georgia map (four graphs in bottom row)}
\label{fig:exp_sr_cal_ga} 
\vspace{-10pt}
\end{figure*}

\begin{figure*}[!ht]
\centering
\begin{minipage}[!b]{0.25\textwidth} 
    \includegraphics[width=\textwidth]{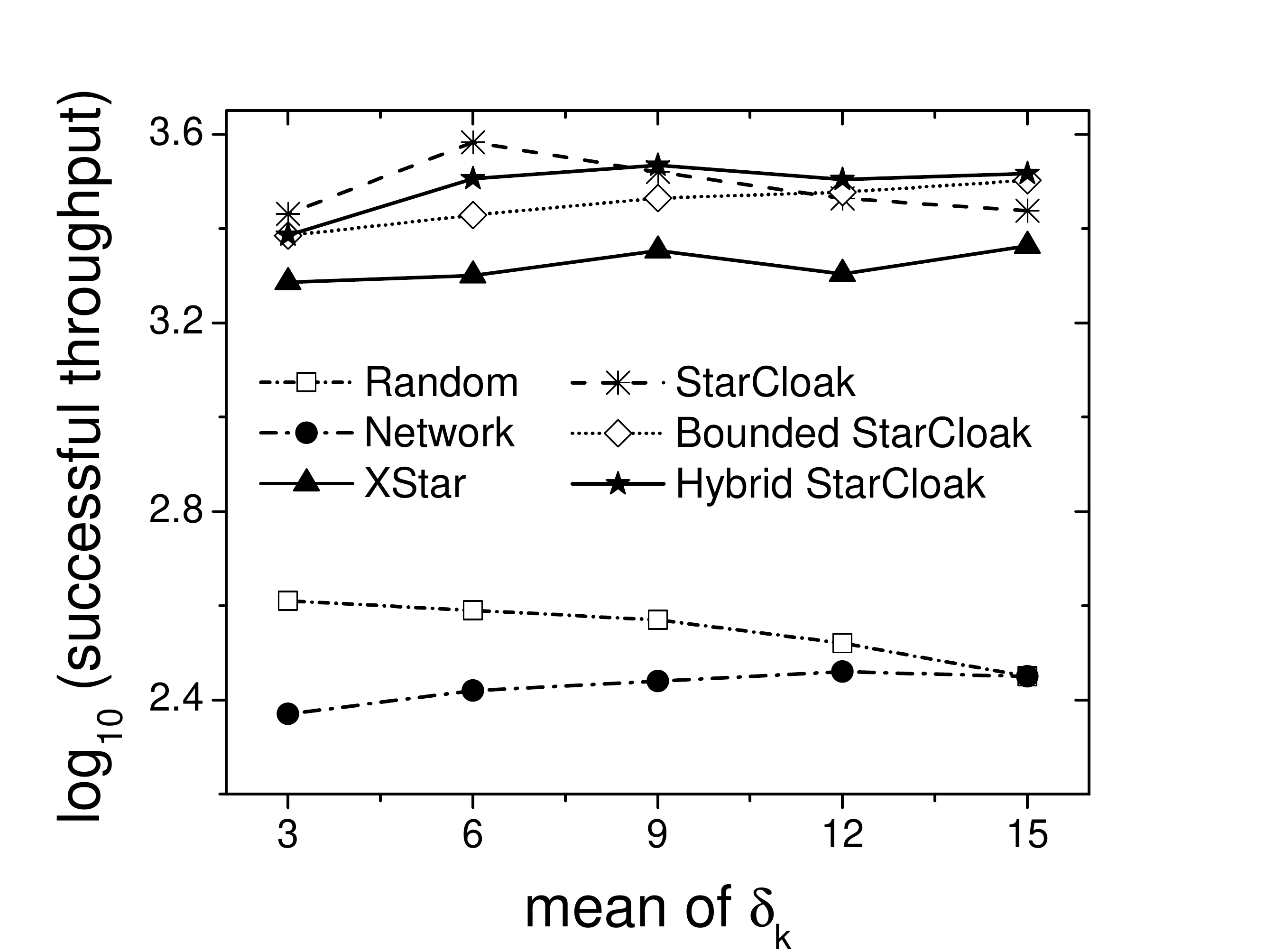}
\end{minipage}%
\hspace{-6mm}
\begin{minipage}[!b]{0.25\textwidth}
    \includegraphics[width=\textwidth]{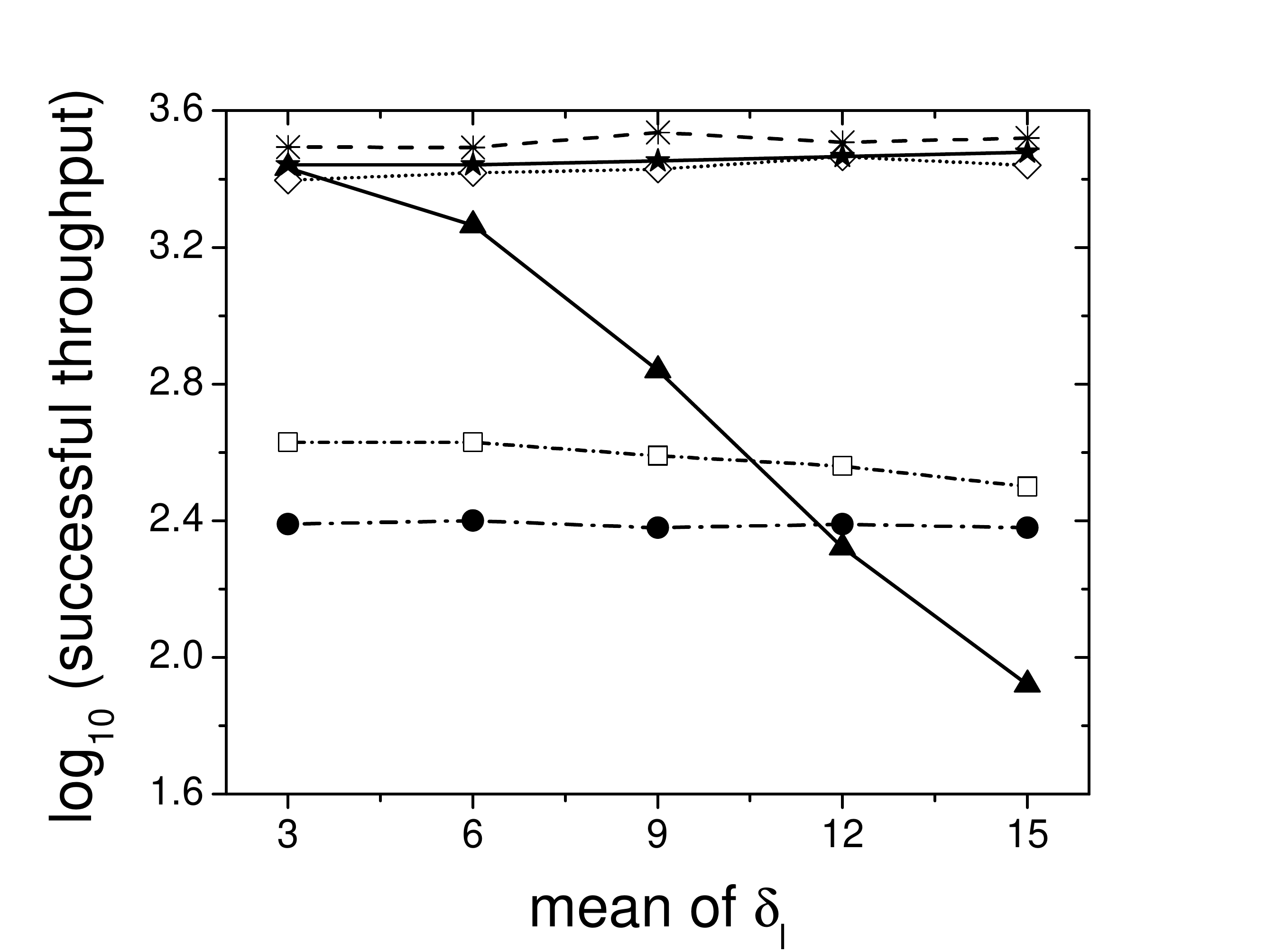}
\end{minipage}%
\hspace{-6mm}
\begin{minipage}[!b]{0.25\textwidth}
    \includegraphics[width=\textwidth]{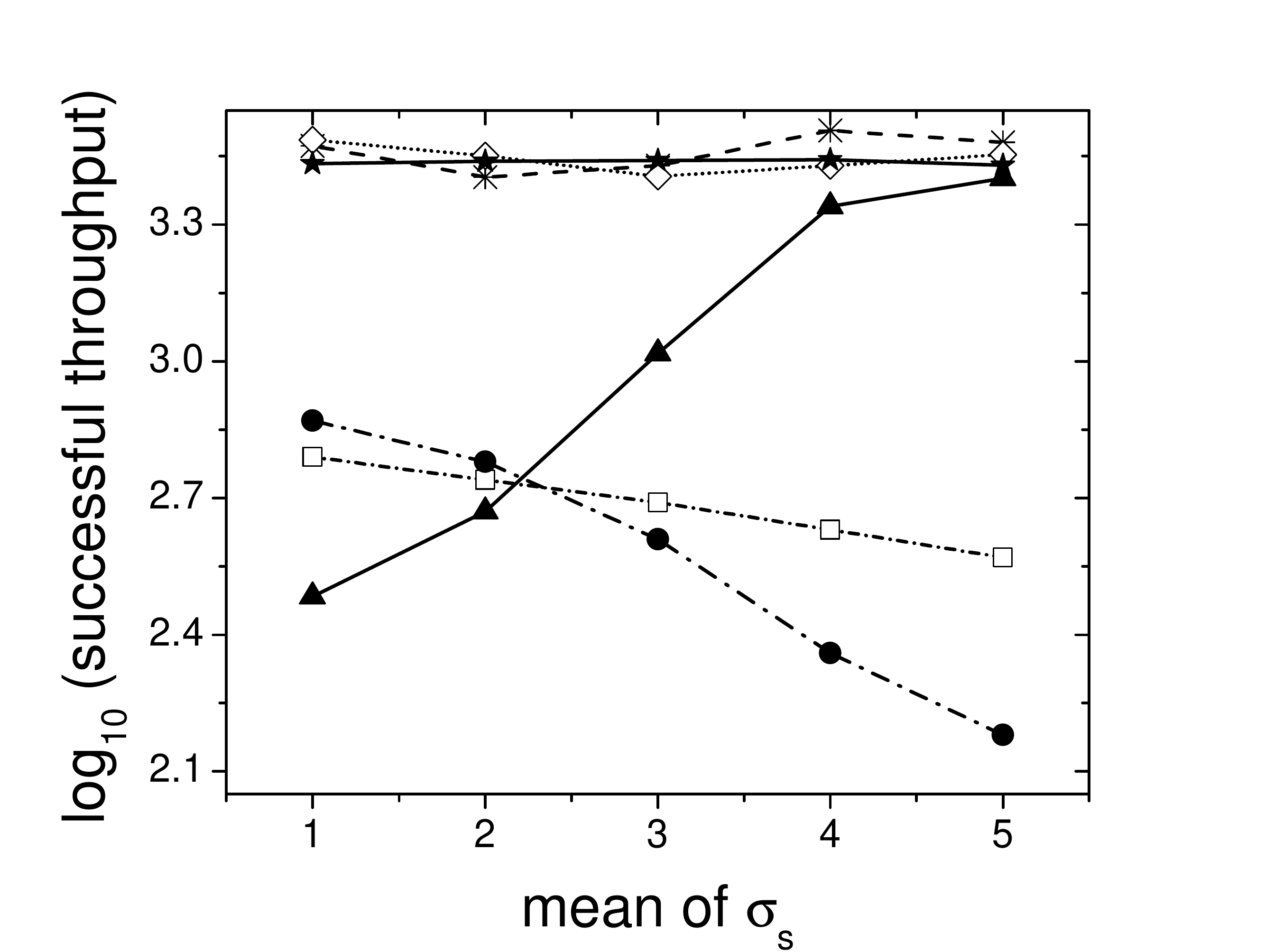}
\end{minipage}%
\hspace{-6mm}
\begin{minipage}[!b]{0.25\textwidth}
    \includegraphics[width=\textwidth]{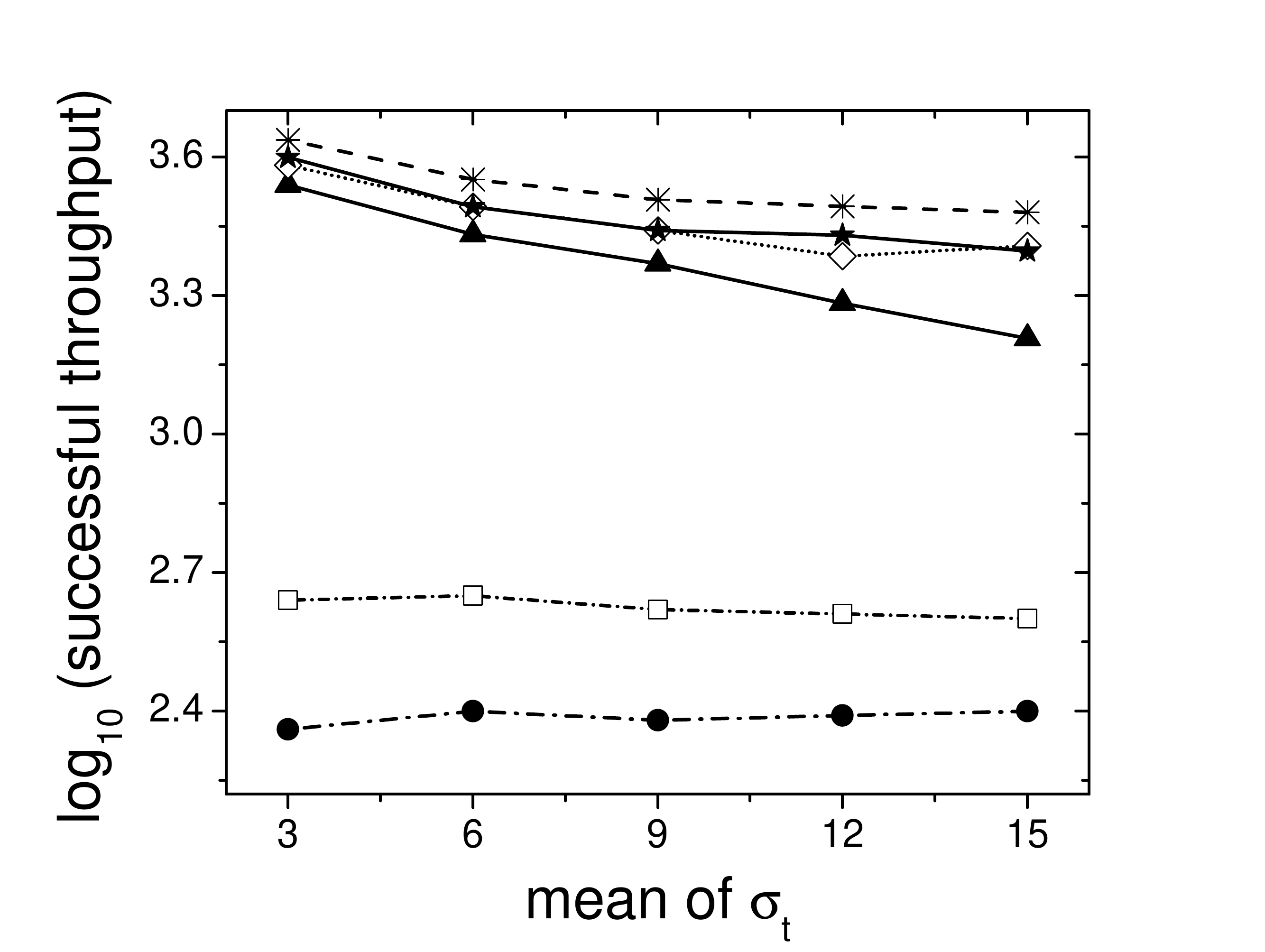}
\end{minipage}%
\\
\begin{minipage}[!b]{0.25\textwidth}
    \includegraphics[width=\textwidth]{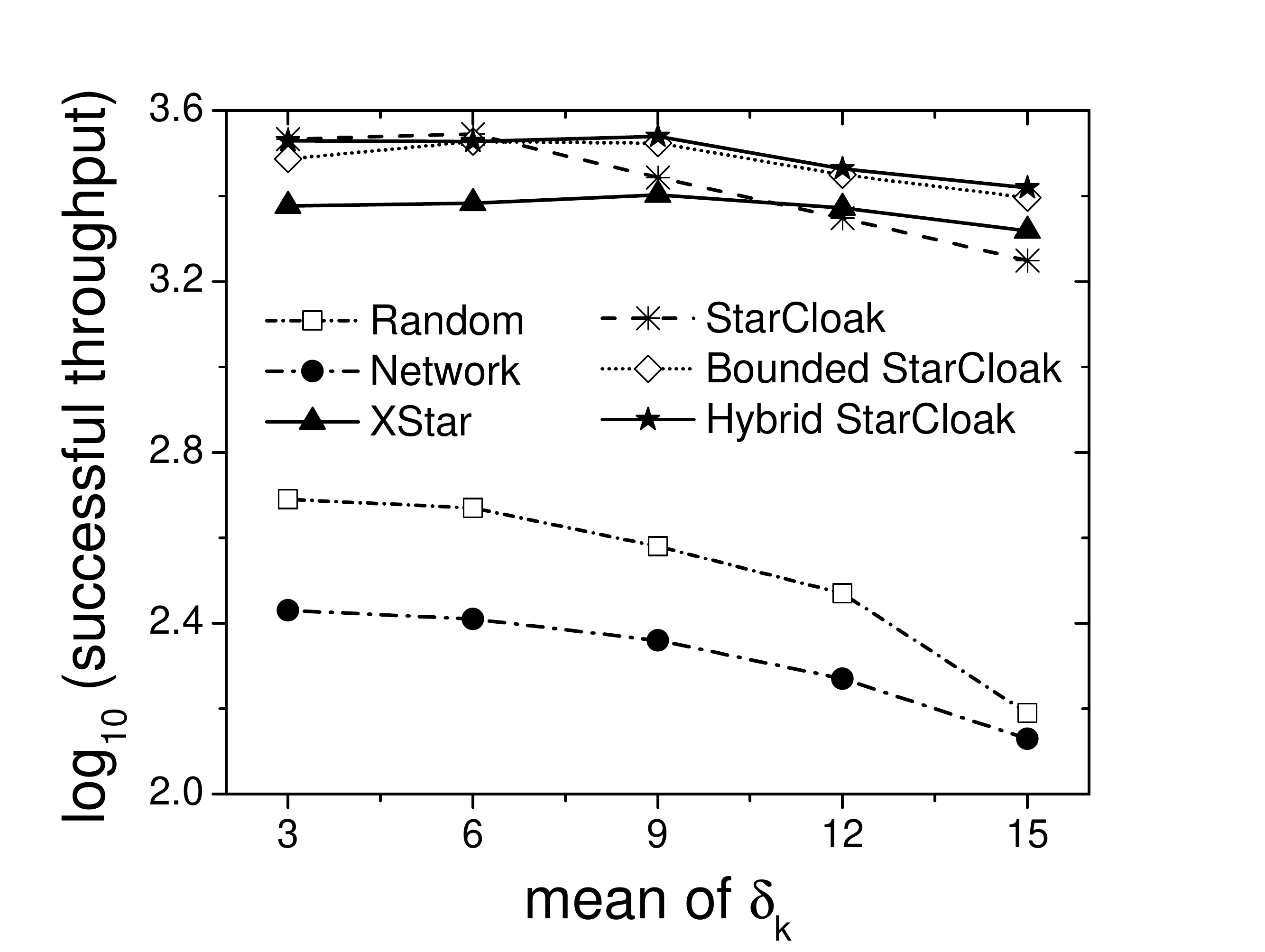}
\end{minipage}%
\hspace{-6mm}
\begin{minipage}[!b]{0.25\textwidth}
    \includegraphics[width=\textwidth]{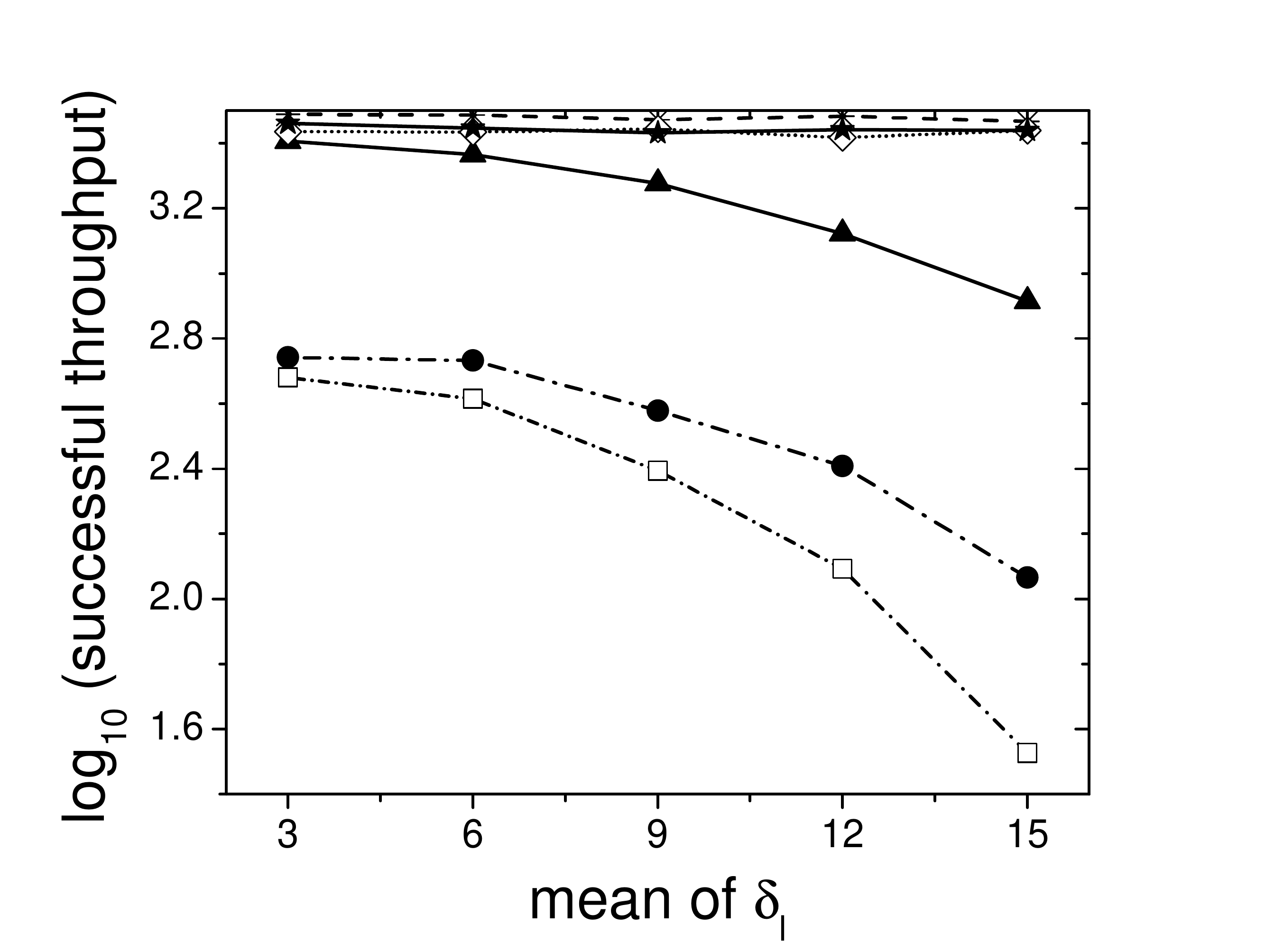}
\end{minipage}%
\hspace{-6mm}
\begin{minipage}[!b]{0.25\textwidth}
    \includegraphics[width=\textwidth]{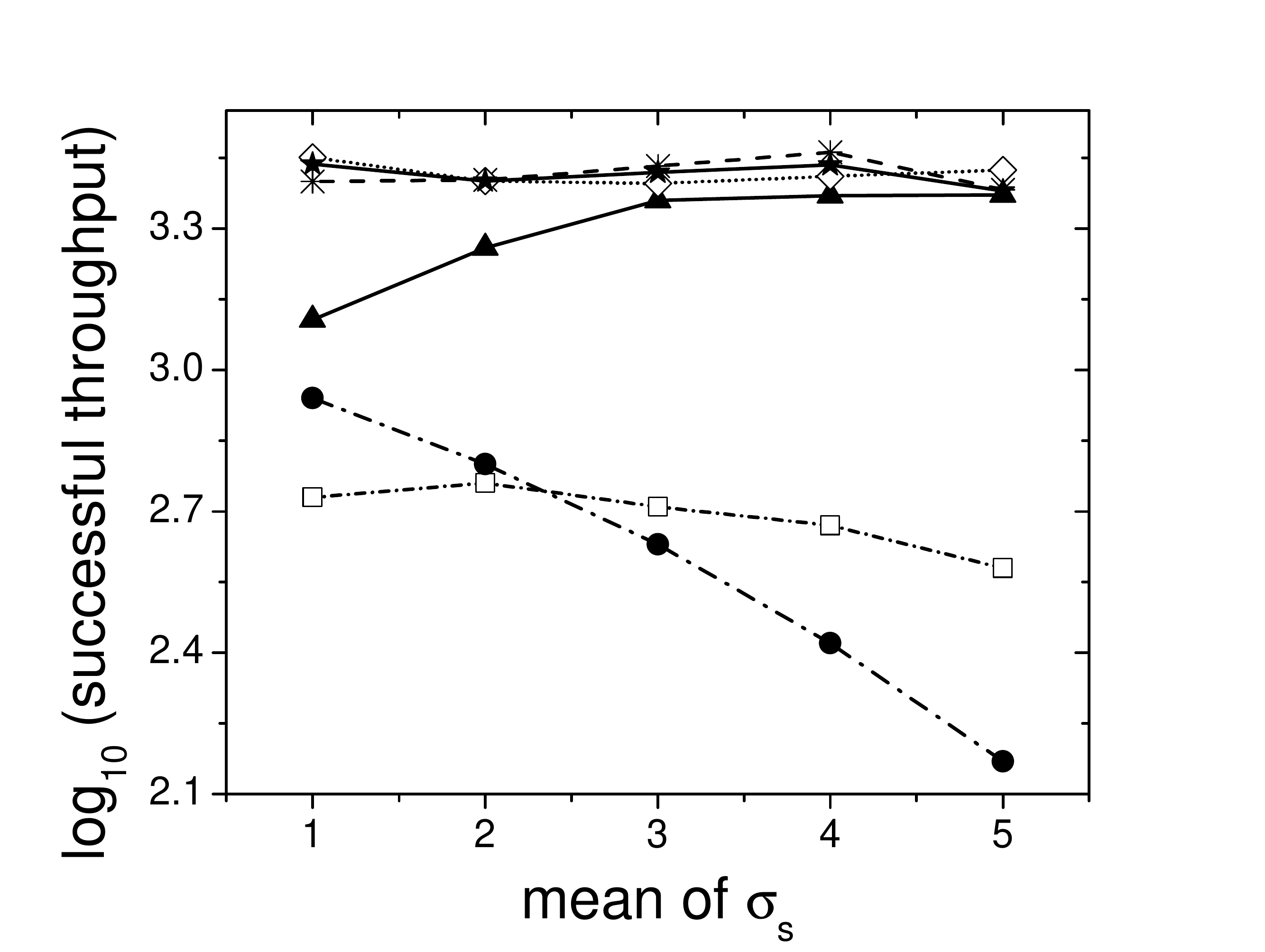}
\end{minipage}%
\hspace{-6mm}
\begin{minipage}[!b]{0.25\textwidth}
    \includegraphics[width=\textwidth]{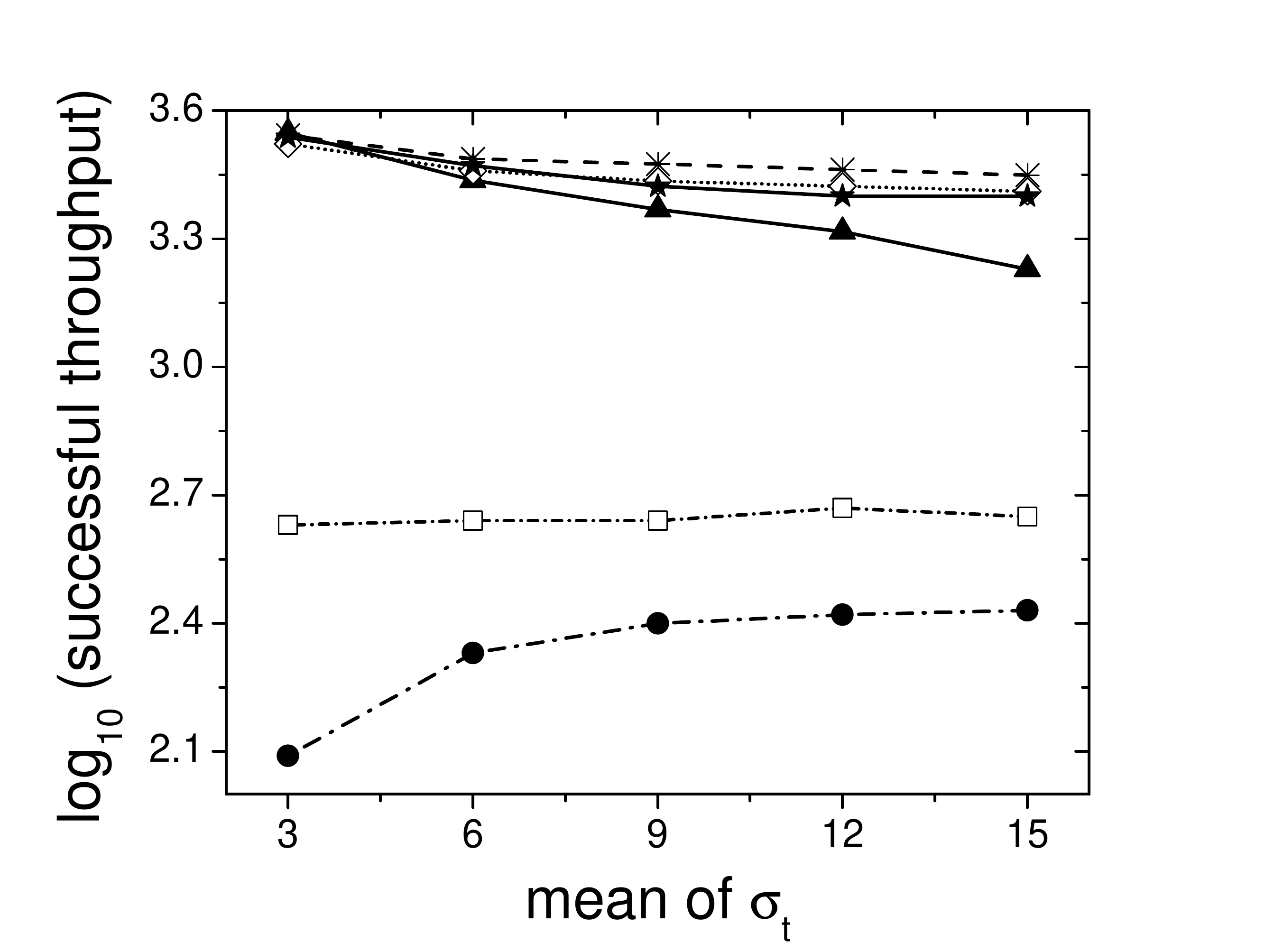}
\end{minipage}%
\vspace{-4pt}
\caption{Successful throughput for California map (four graphs in top row) and Georgia map (four graphs in bottom row)}
\label{exp:st_ca_ga} 
\vspace{-10pt}
\end{figure*}

To evaluate the performance of different mechanisms, we use multiple metrics: success rate in anonymization, anonymization time, query processing time, size of candidate result set, successful throughput, and segment entropy against inference attacks.

\textbf{Success Rate:} An effective anonymization engine should successfully anonymize as many queries as possible, and drop as few queries as possible. Success rate measures the fraction of successfully anonymized queries divided by the number of total queries issued by the mobile users.

\textbf{Anonymization Time:} When users issue queries, they want fast answers (low temporal delay). However, an anonymization engine needs a certain amount of time to perform the anonymization. The anonymization time metric measures the average time elapsed from the query issue time until successful anonymization. From the user's perspective, it is desired that the anonymization time is as low as possible. Note that in order to ensure a fair comparison among multiple compared approaches, we measure anonymization time only on successfully anonymized queries.

\textbf{Query Processing Time:} This metric measures the processing time cost for anonymized queries. With anonymization, queries are evaluted on cloaked subgraphs instead of user's exact location. Number of border nodes and edges in the cloaked subgraph impact query processing time.

\textbf{Candidate Result Size:} This metric aims to measure the added bandwidth overhead for the communication between the anonymization engine and the LBS provider. More compact anonymized subgraphs lead to smaller candidate result sets, thus lower communication cost.

\textbf{Successful Throughput:} We use the throughput metric to evaluate the scalability of the anonymization approaches. Rate of successful throughput equals the multiplication of query execution rate (number of queries processed per second) and success rate. 

\textbf{Entropy:} We use entropy as a quantitative measure of adversarial uncertainty achieved by anonymization, where higher entropy means higher attack-resilience. Given an anonymized location $S$ for user $u$ as a subgraph consisting of multiple segments, the segment entropy of $S$ can be calculated by: 
\[
H(S) = - \sum_{s \in S} \textit{link}[u \leftarrow s] \cdot \log_2(\textit{link}[u \leftarrow s])
\]
Note that the number of segments in the generated anonymized locations may vary from anonymization algorithm to algorithm, based on their segment selection strategy. Thus, using simple entropy for different sized anonymized locations may not adequately capture the strength of protection. For this reason, we use \textit{normalized entropy} \cite{entropy_norm} defined as: $H(S)/\log_2(|S|)$. 

\setlength{\textfloatsep}{\textfloatsepsave}

\vspace{-6pt}
\subsection{Experiment Results} \label{sec:results}

\begin{figure*}[!ht]
\centering
\begin{minipage}[!b]{0.25\textwidth} 
    \includegraphics[width=\textwidth]{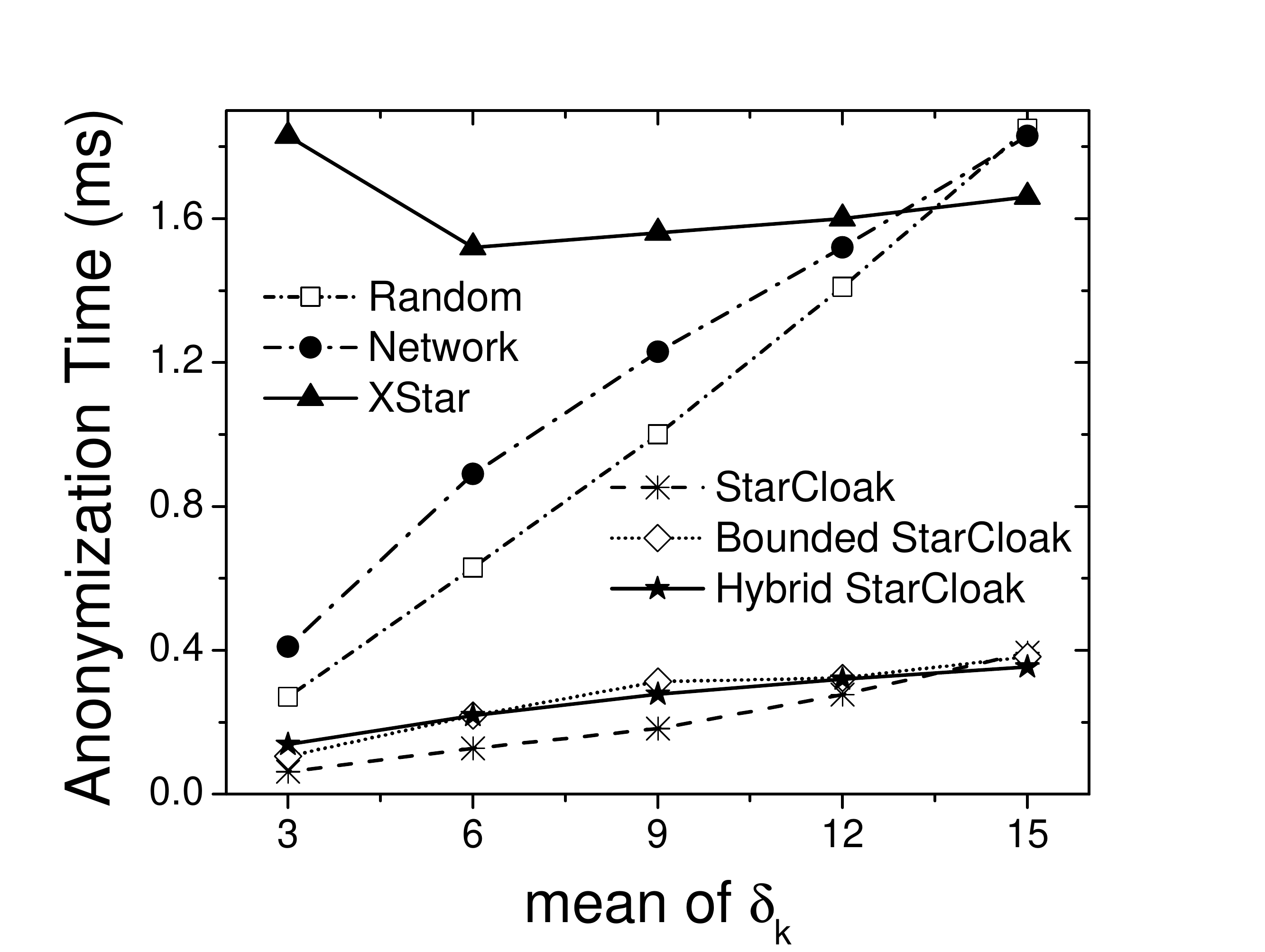}
\end{minipage}%
\hspace{-6mm}
\begin{minipage}[!b]{0.25\textwidth}
    \includegraphics[width=\textwidth]{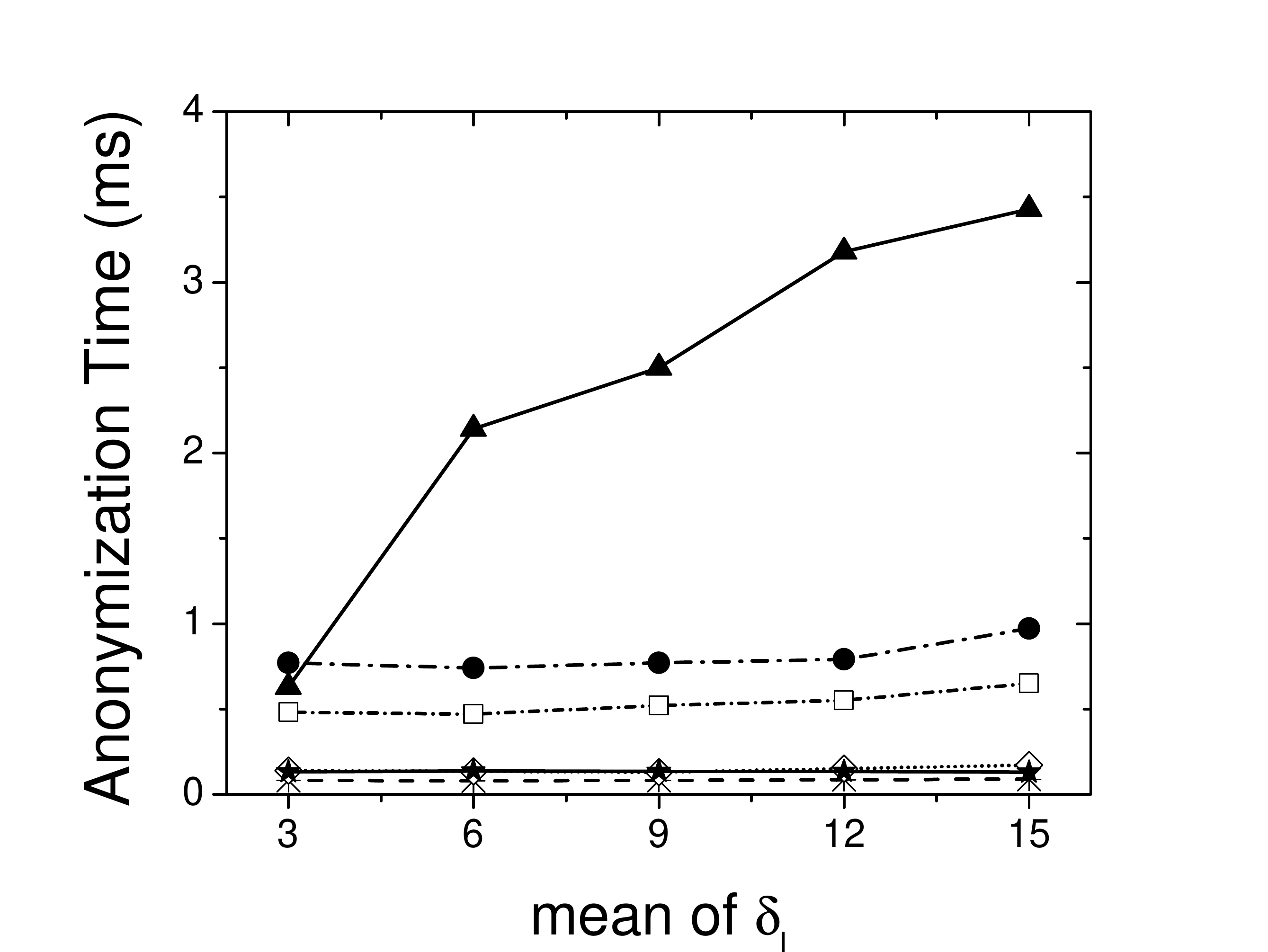}
\end{minipage}%
\hspace{-6mm}
\begin{minipage}[!b]{0.25\textwidth}
    \includegraphics[width=\textwidth]{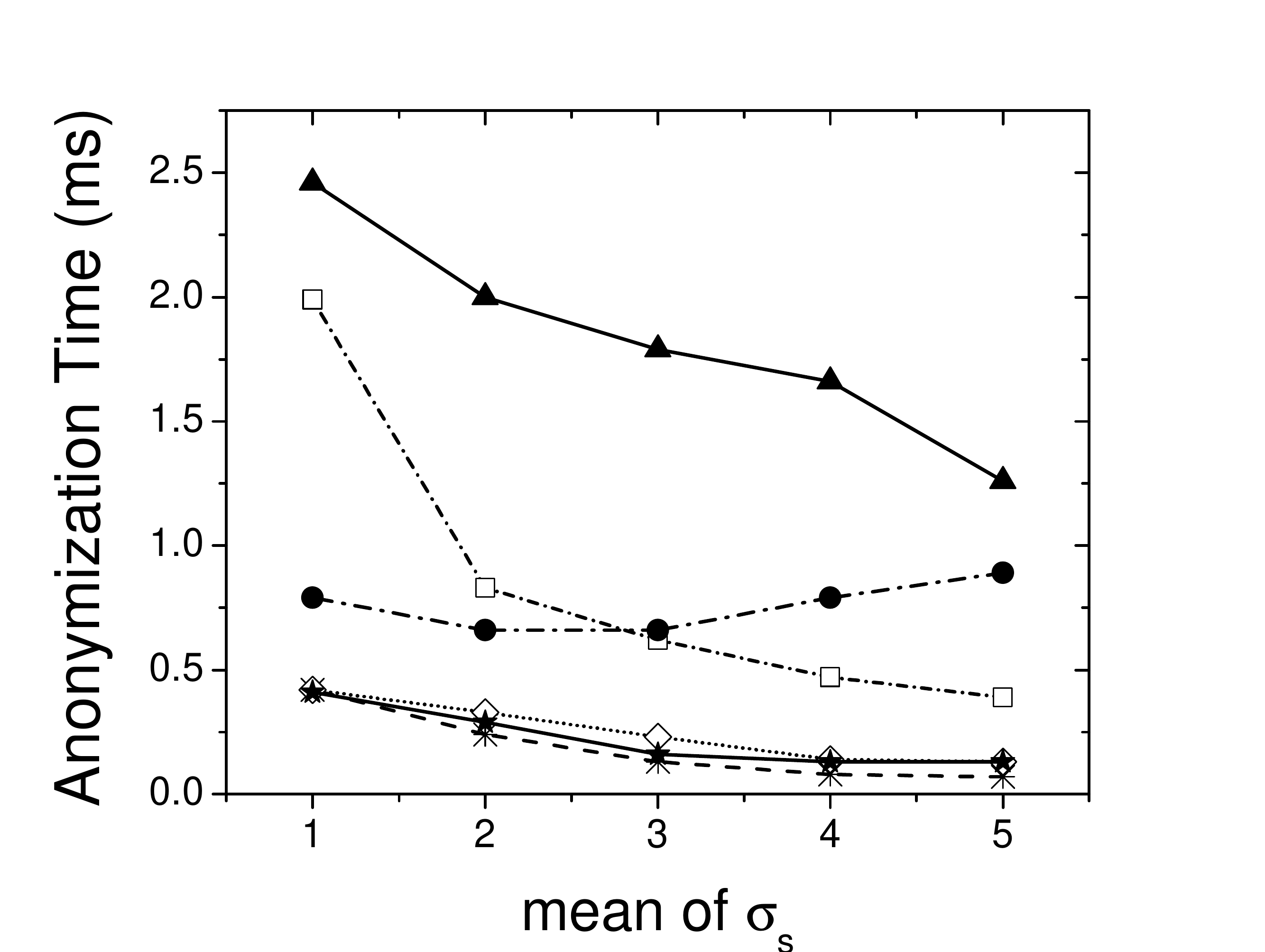}
\end{minipage}%
\hspace{-6mm}
\begin{minipage}[!b]{0.25\textwidth}
    \includegraphics[width=\textwidth]{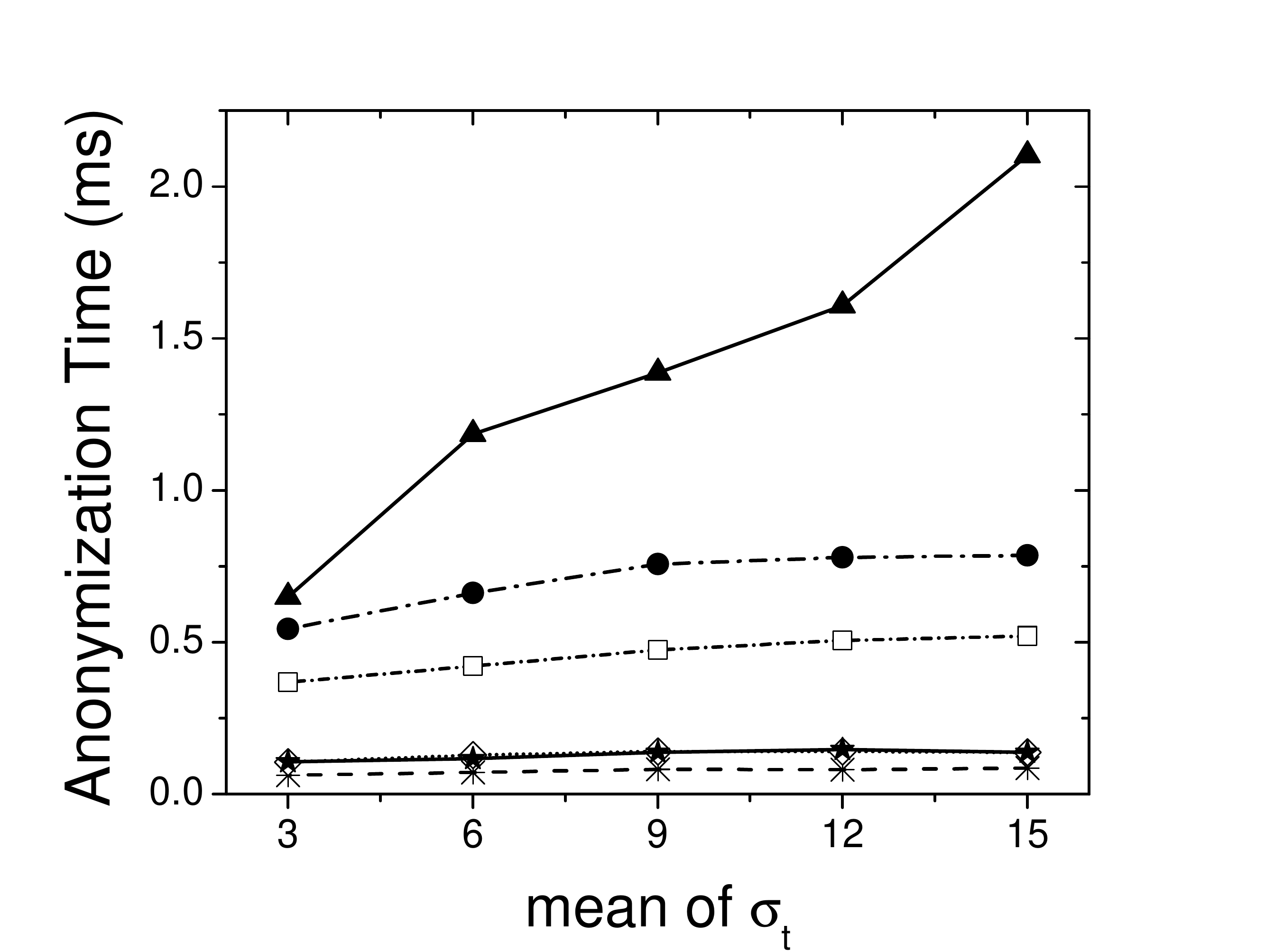}
\end{minipage}%
\\
\begin{minipage}[!b]{0.25\textwidth}
    \includegraphics[width=\textwidth]{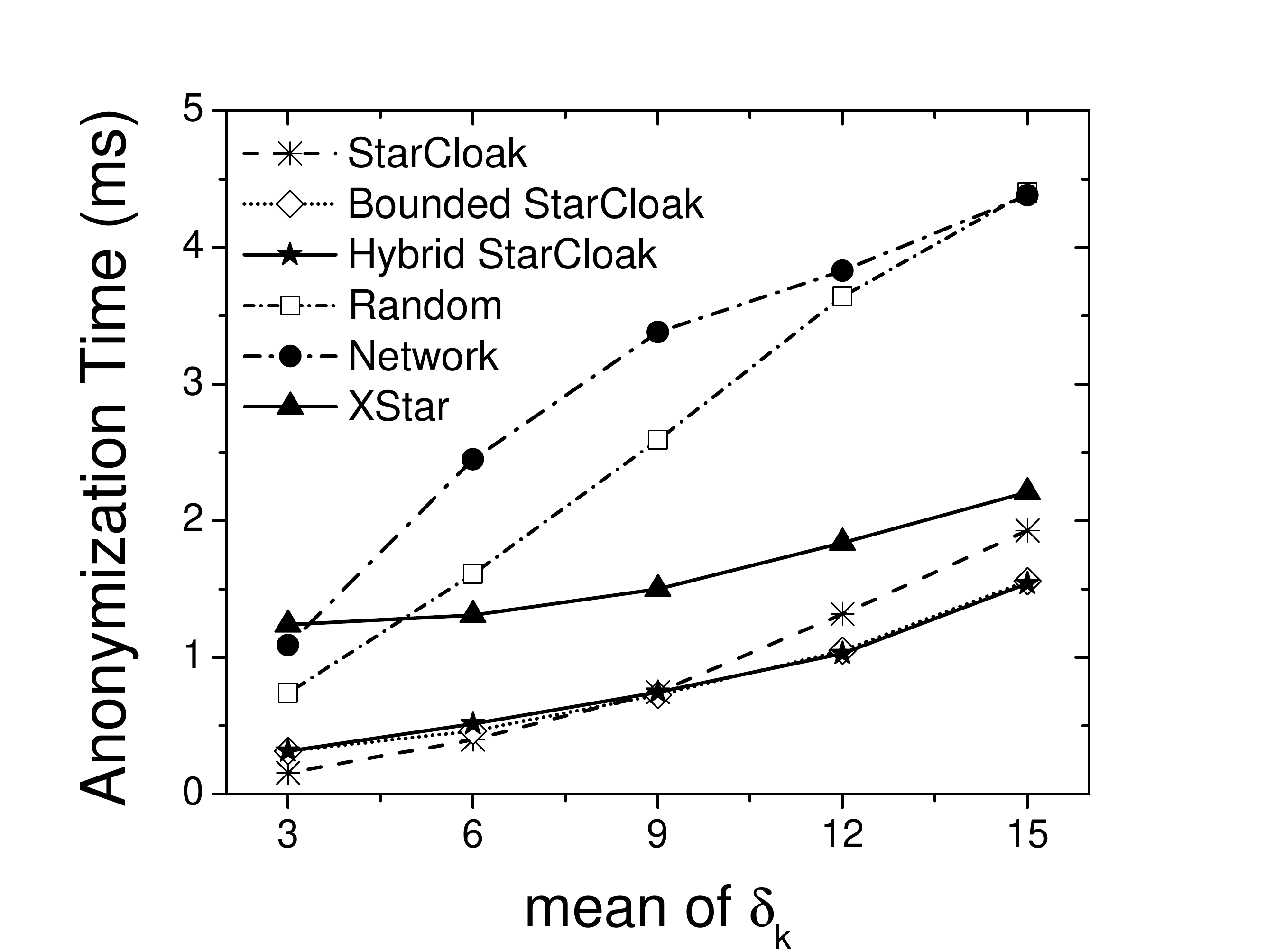}
\end{minipage}%
\hspace{-6mm}
\begin{minipage}[!b]{0.25\textwidth}
    \includegraphics[width=\textwidth]{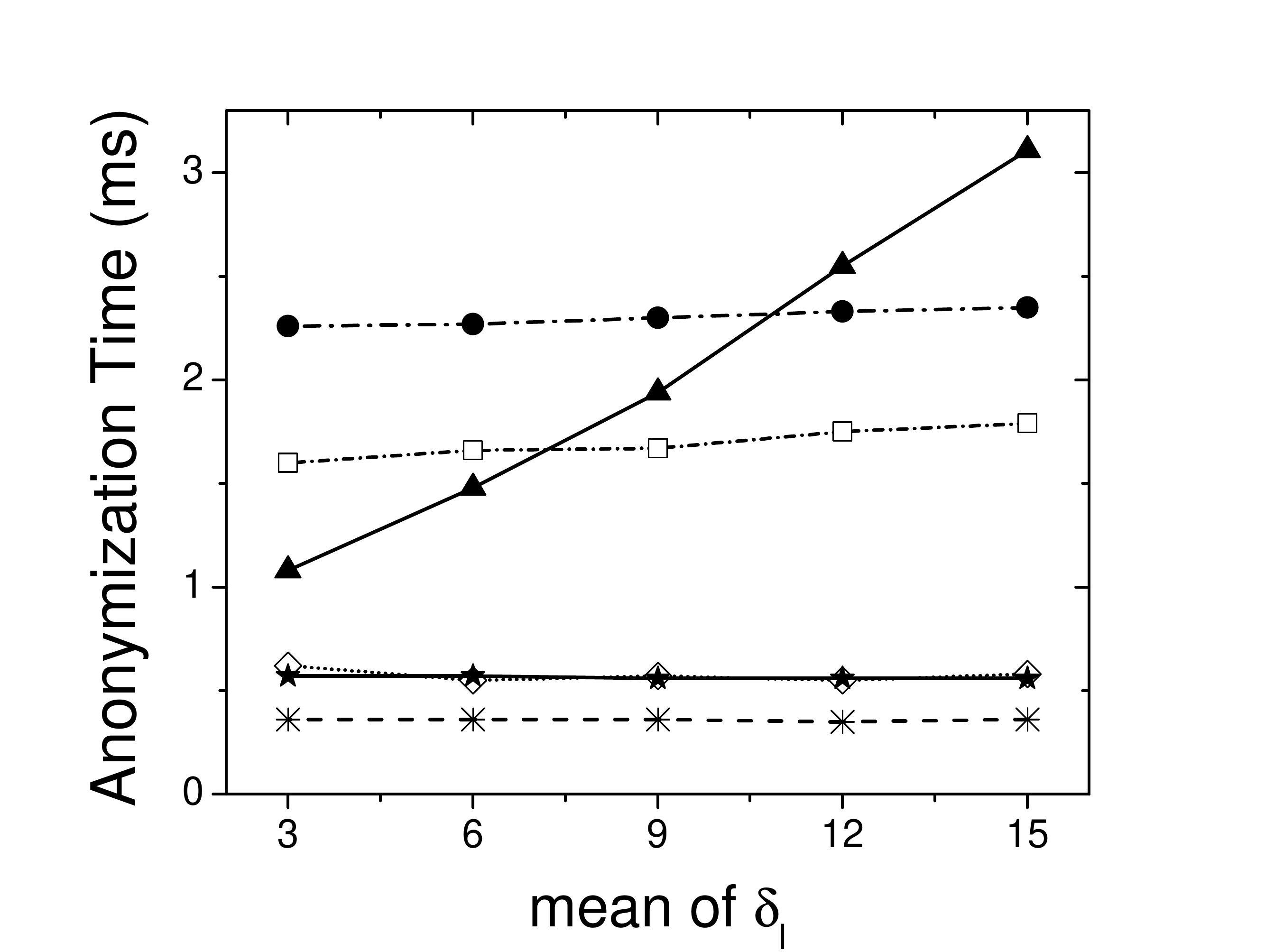}
\end{minipage}%
\hspace{-6mm}
\begin{minipage}[!b]{0.25\textwidth}
    \includegraphics[width=\textwidth]{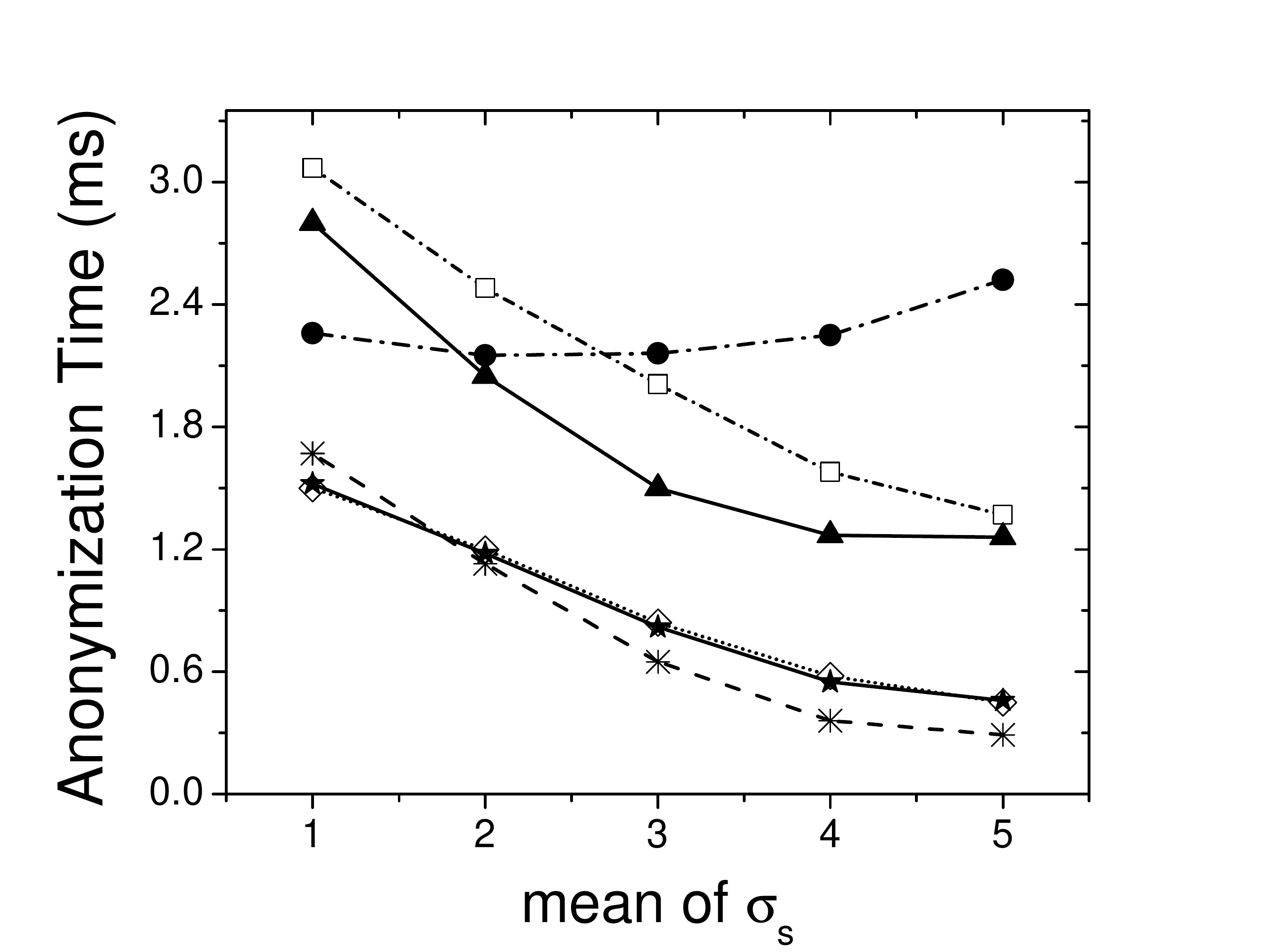}
\end{minipage}%
\hspace{-6mm}
\begin{minipage}[!b]{0.25\textwidth}
    \includegraphics[width=\textwidth]{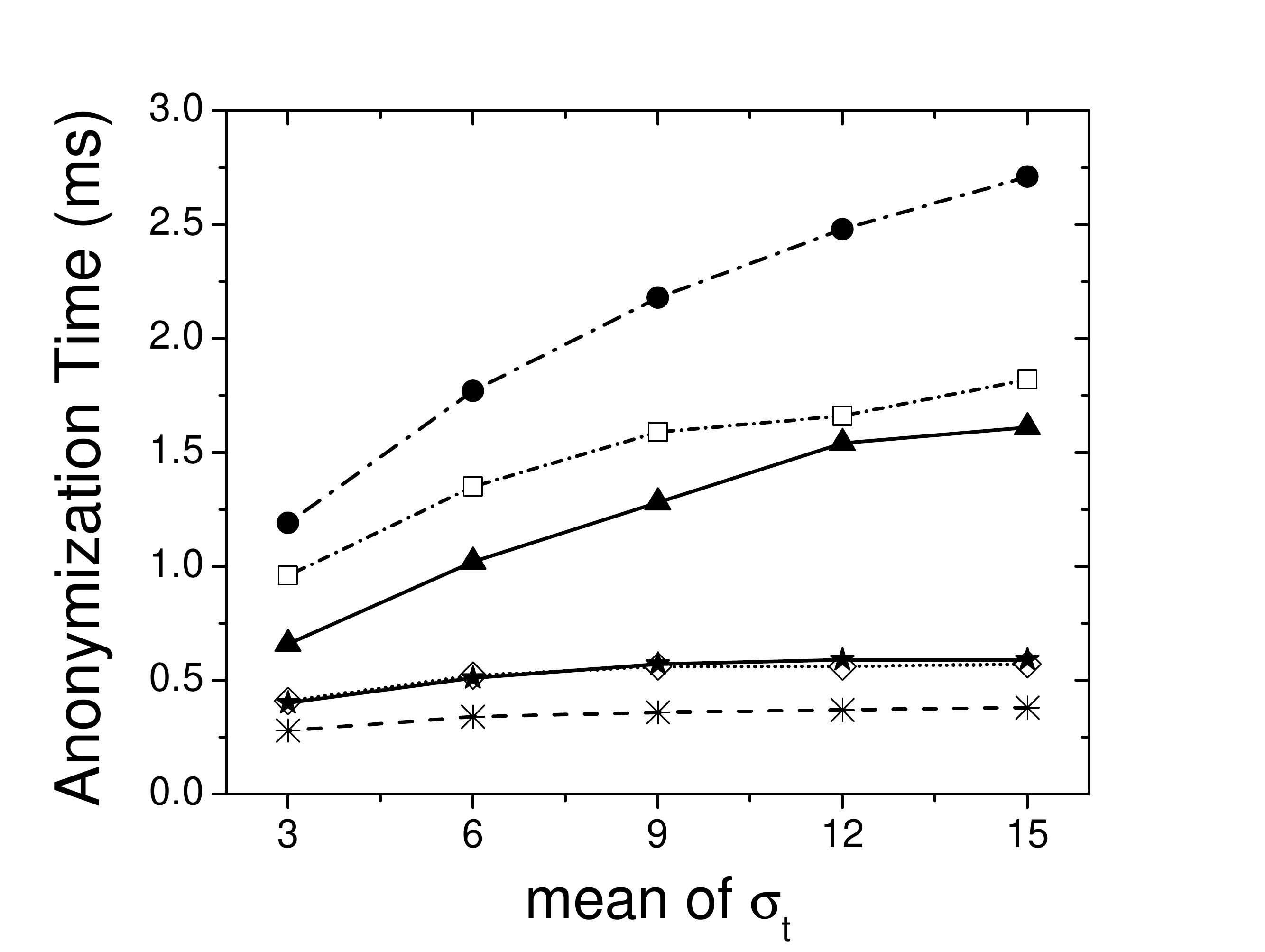}
\end{minipage}%
\vspace{-5pt}
\caption{Anonymization time for California map (four graphs in top row) and Georgia map (four graphs in bottom row)}
\label{fig:exp_anontime_ca_ga} 
\vspace{-12pt}
\end{figure*}

\begin{figure*}[!ht]
\centering
\begin{minipage}[!b]{0.25\textwidth} 
    \includegraphics[width=\textwidth]{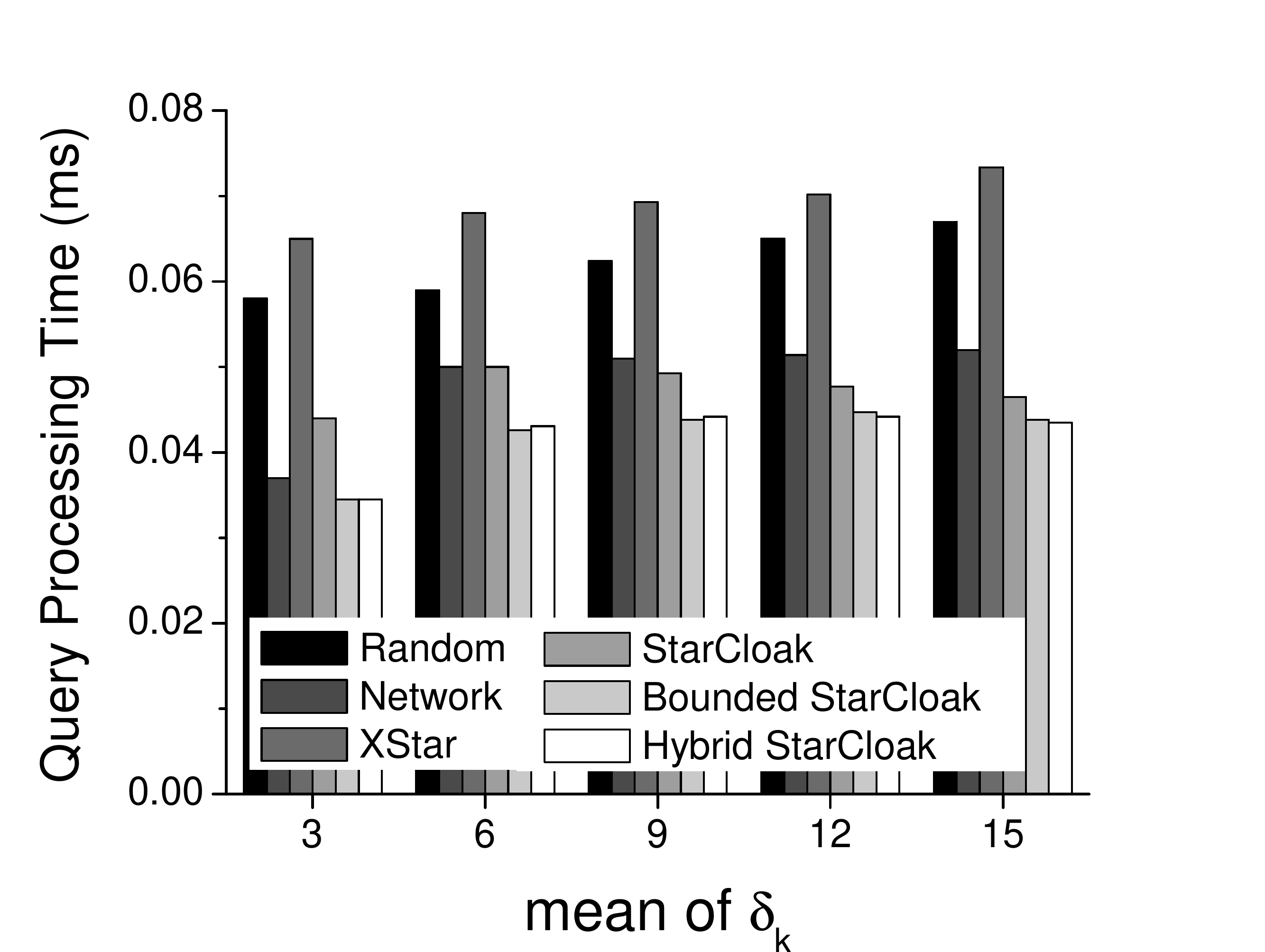}
\end{minipage}%
\hspace{-7mm}
\begin{minipage}[!b]{0.25\textwidth}
    \includegraphics[width=\textwidth]{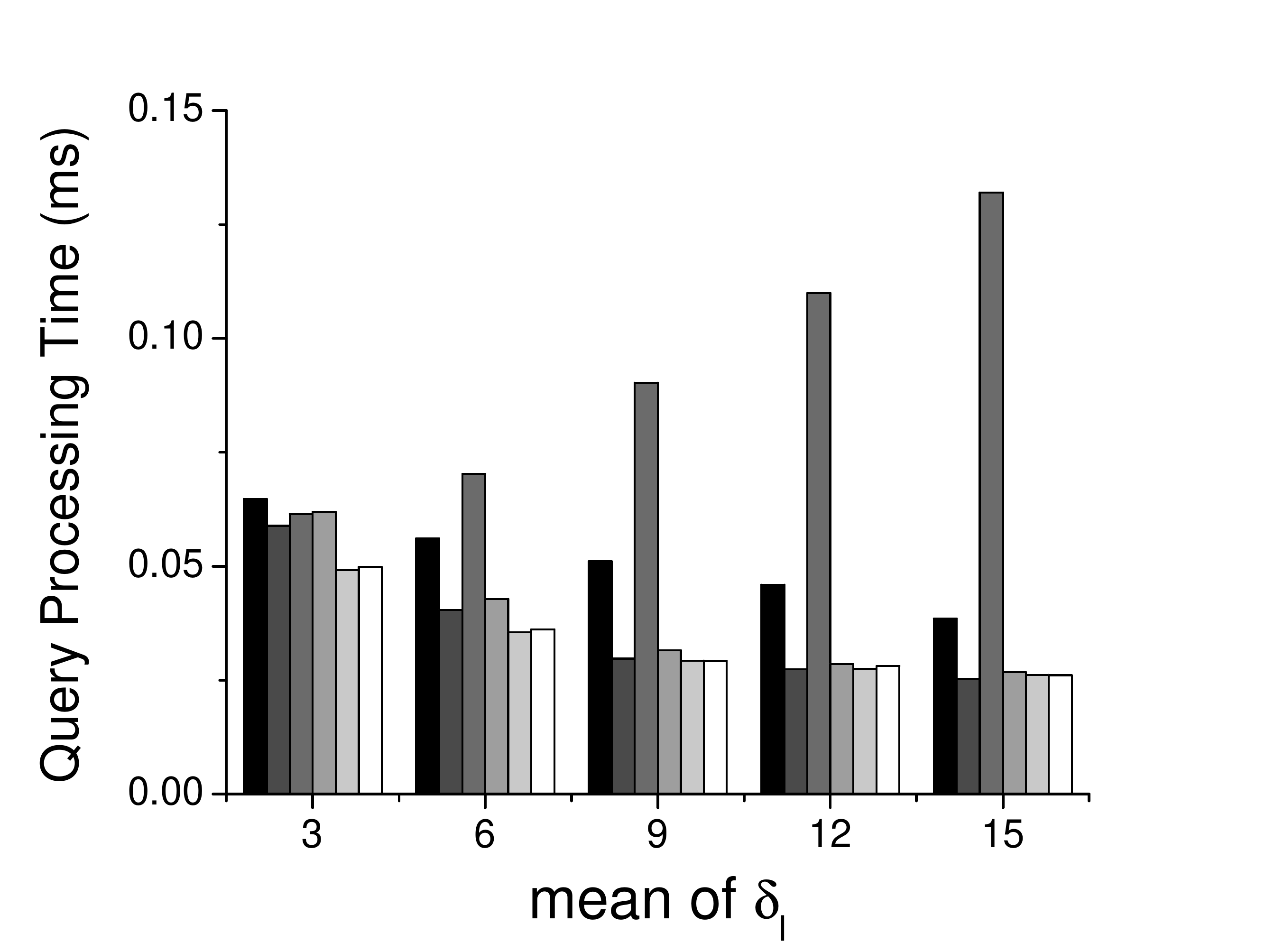}
\end{minipage}%
\hspace{-7mm}
\begin{minipage}[!b]{0.25\textwidth}
    \includegraphics[width=\textwidth]{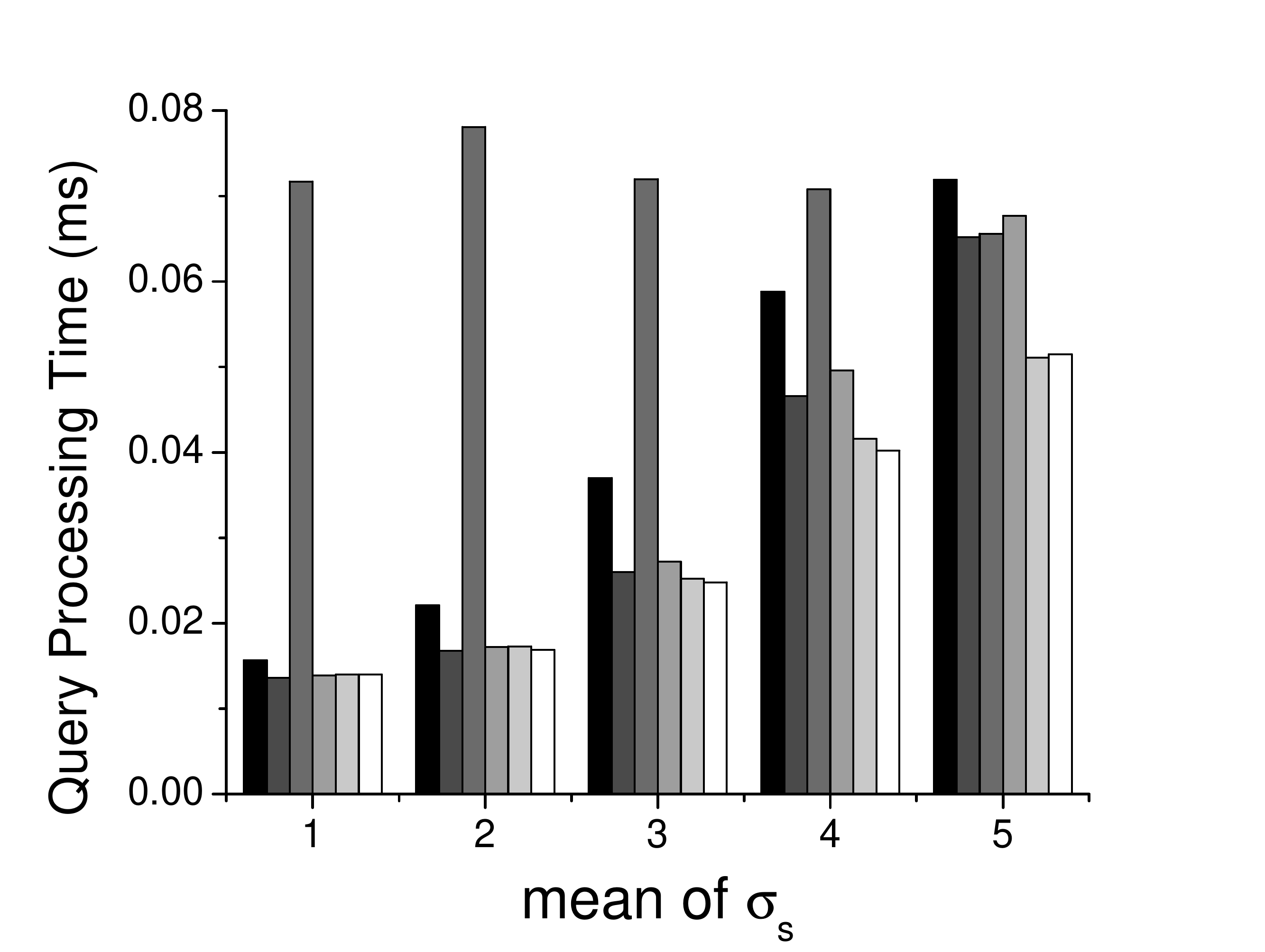}
\end{minipage}%
\hspace{-7mm}
\begin{minipage}[!b]{0.25\textwidth}
    \includegraphics[width=\textwidth]{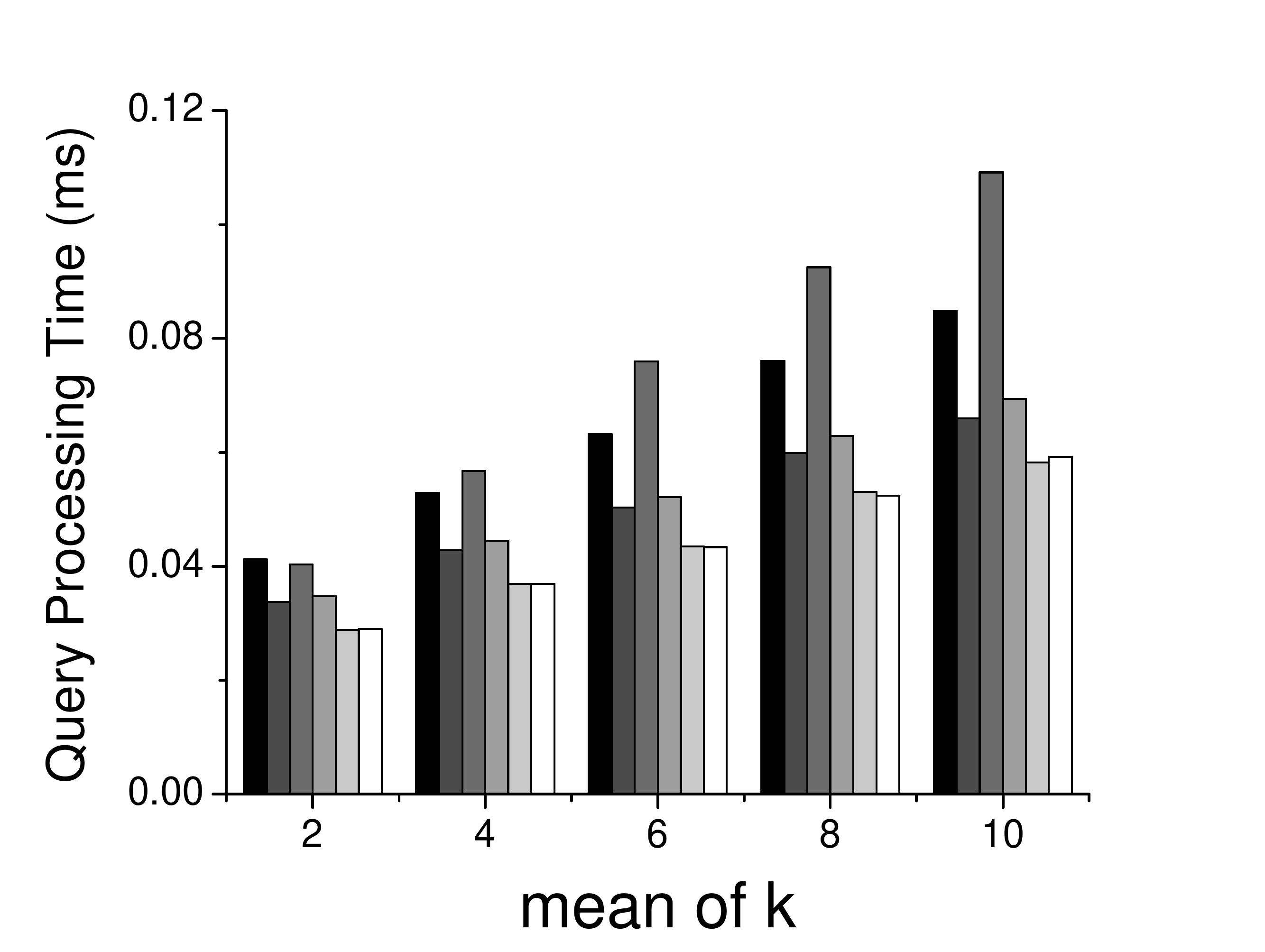}
\end{minipage}%
\vspace{-5pt}
\caption{Query processing time for California map}
\label{fig:exp_qptime_cal} 
\vspace{-10pt}
\end{figure*}

\begin{figure*}[!ht]
\centering
\begin{minipage}[!b]{0.25\textwidth}  
    \includegraphics[width=\textwidth]{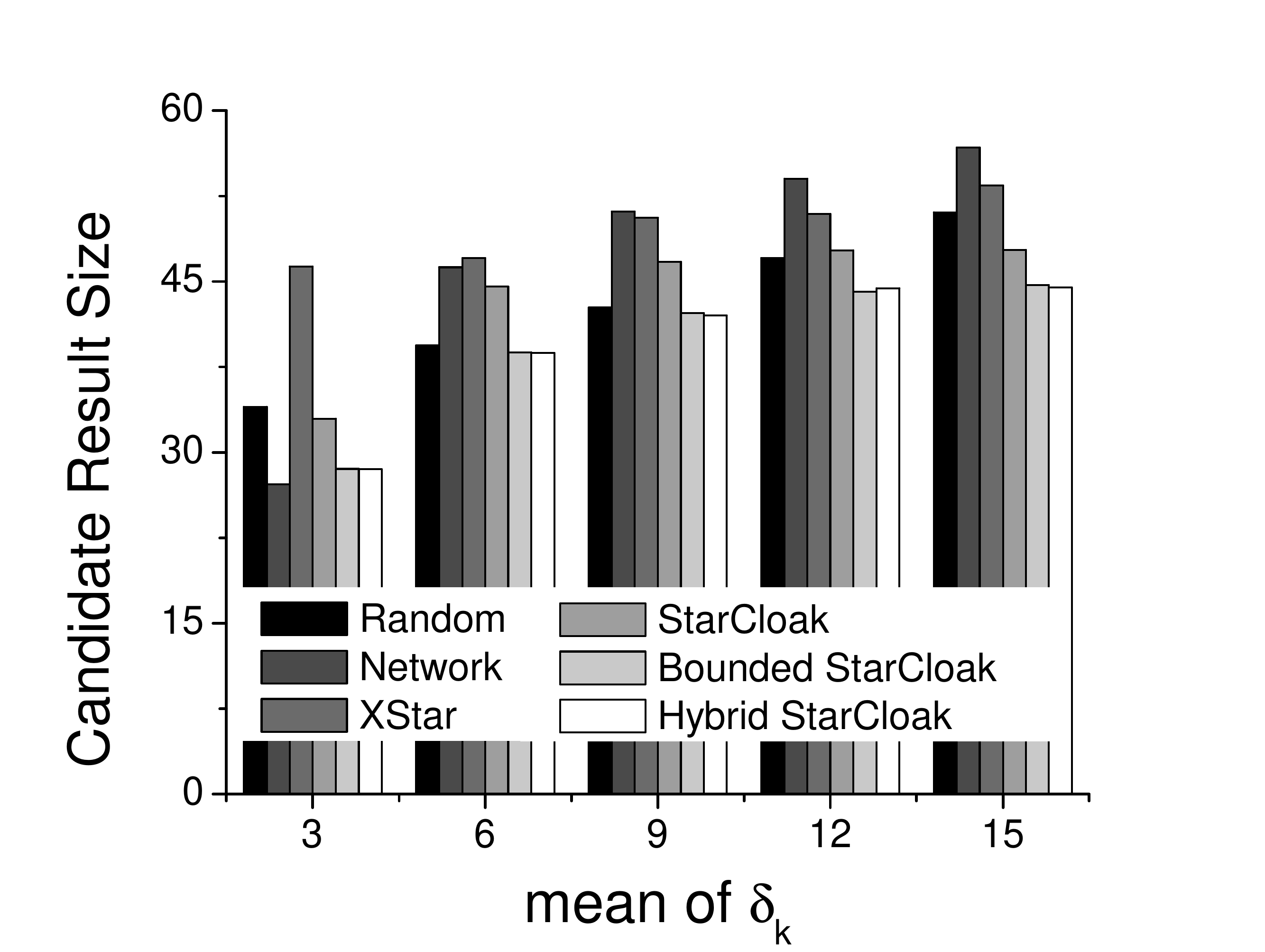}
\end{minipage}%
\hspace{-7mm}
\begin{minipage}[!b]{0.25\textwidth}
    \includegraphics[width=\textwidth]{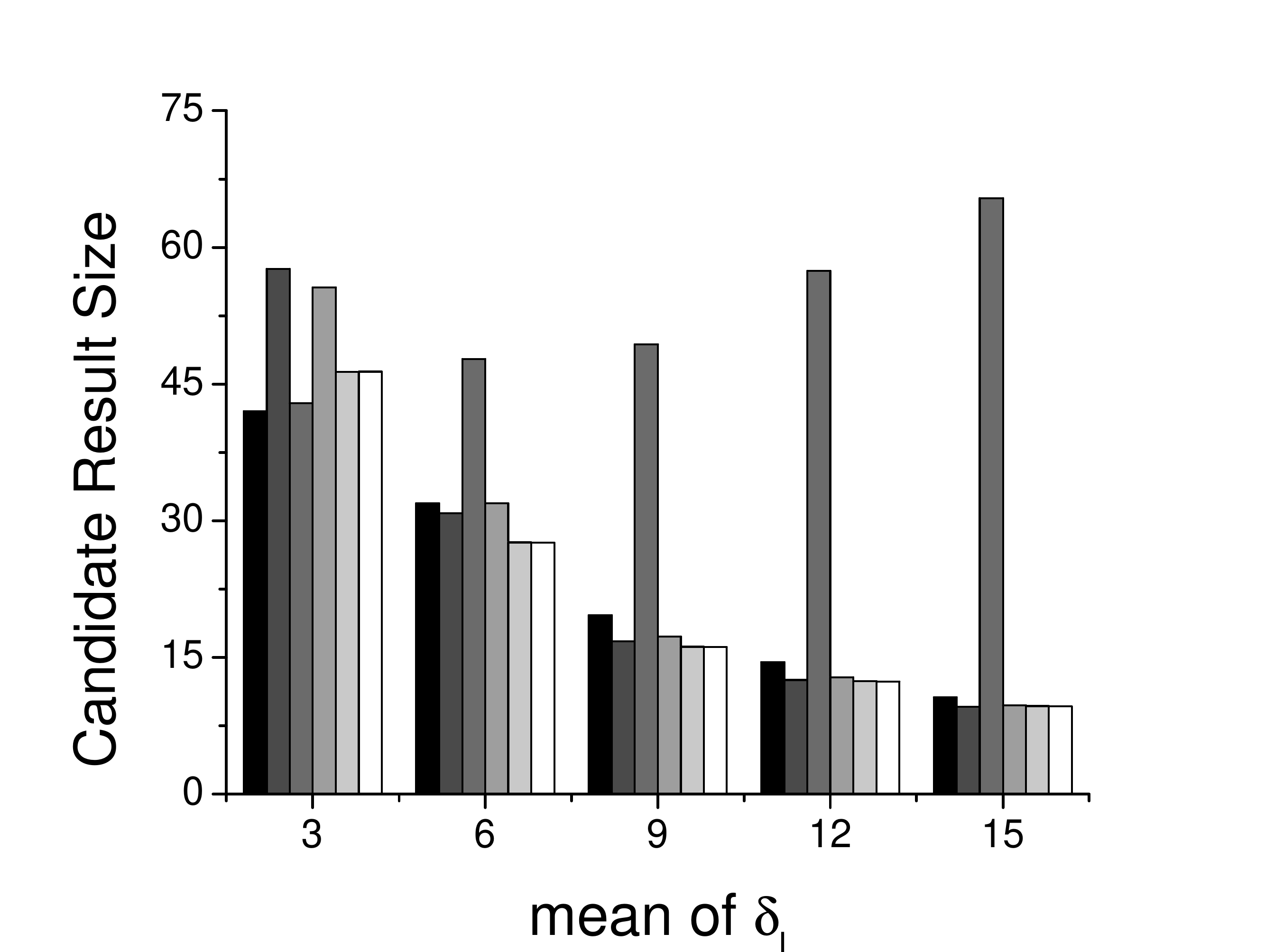}
\end{minipage}%
\hspace{-7mm}
\begin{minipage}[!b]{0.25\textwidth}
    \includegraphics[width=\textwidth]{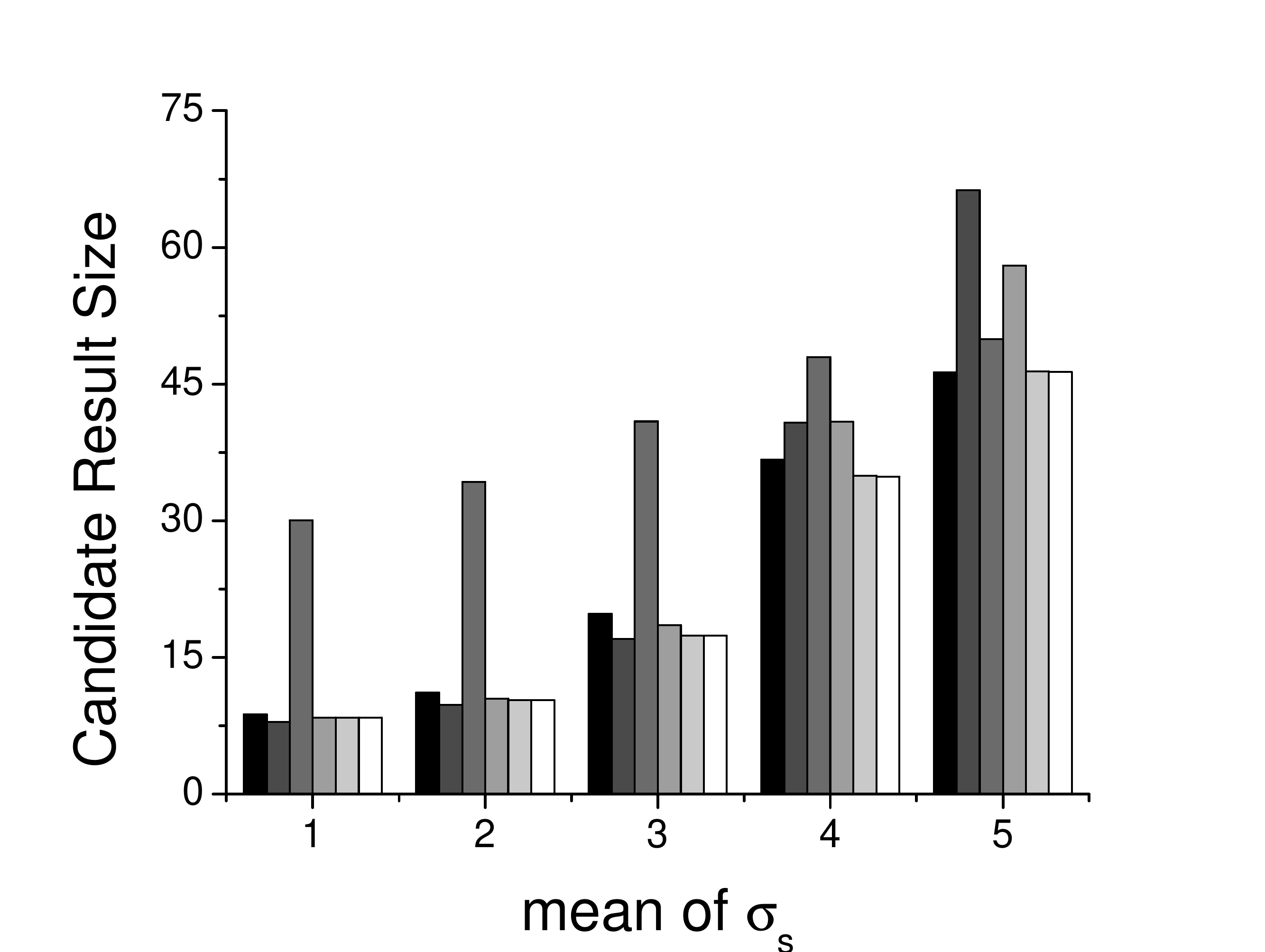}
\end{minipage}%
\hspace{-7mm}
\begin{minipage}[!b]{0.25\textwidth}
    \includegraphics[width=\textwidth]{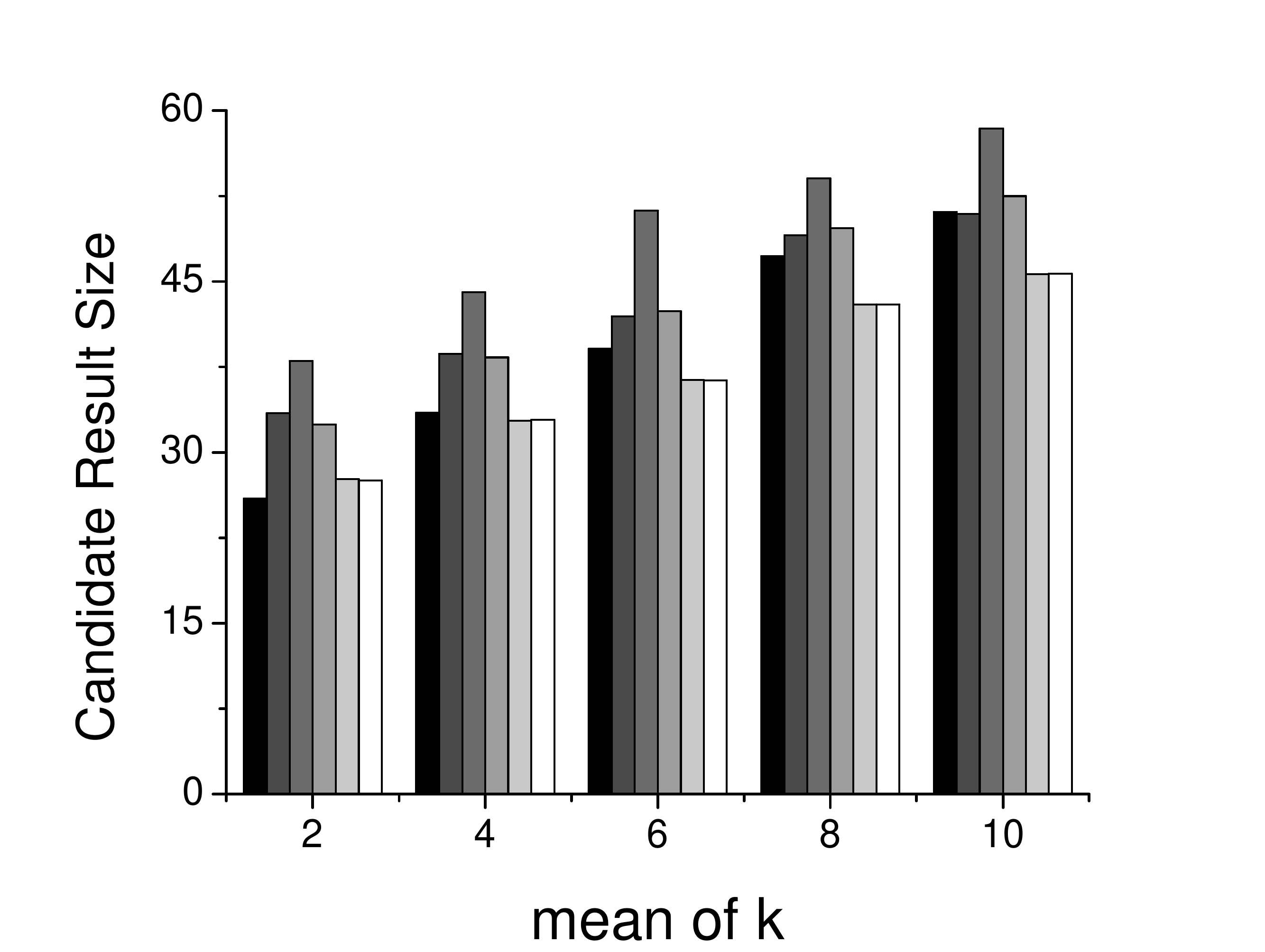}
\end{minipage}%
\vspace{-6pt}
\caption{Candidate result size for California map}
\label{fig:exp_cresult_cal} 
\vspace{-10pt}
\end{figure*}

\textbf{Results on Success Rate:} Figure \ref{fig:exp_sr_cal_ga} shows the percentage success rates of compared approaches with respect to varying $\delta_k$, $\delta_l$, $\sigma_s$, and $\sigma_t$ for California and Georgia maps. Generally, \textsc{StarCloak} and its optimized variants have high success rates because of their ability in handling different user requirements effectively. The results show that increasing the privacy requirements often decreases success rate, but the effects are different for different maps and different privacy requirements. For example, when $\delta_k$ increases, success rate decreases faster on the Georgia map compared to California. The reason for this is the query density of the maps. Keeping the number of queries constant across maps, since the Georgia map is more detailed than California, the distribution of query density on Georgia is sparser. Thus, on Georgia, the chance of finding enough queries to cloak together under the same spatial constraint is smaller, causing more queries being dropped and lowered success rate. On the other hand, for \textsc{XStar}, increasing $\delta_l$ impacts success rate on California more than Georgia, unlike \textsc{StarCloak}. The reason for this is also related to query density. \textsc{XStar} anonymizes queries on the same star together,  
whereas in \textsc{StarCloak} if there is a conflict between two queries' $l$-segment indistinguishability and $\sigma_s$ spatial tolerance, they are cloaked on different vertices of the cloaking graph. This allows \textsc{StarCloak} to maintain high success rates despite increasing $\delta_l$. 

Comparing the three \textsc{StarCloak} variants, the basic version achieves highest success rate, followed by hybrid \textsc{StarCloak} and then spatially bounded \textsc{StarCloak}. This is expected because spatially bounded \textsc{StarCloak} aims at finding compact cloak regions, whereas basic \textsc{StarCloak} allows suboptimal regions for higher success rate. Hybrid \textsc{StarCloak} achieves a trade-off between success rate and compactness of a cloak region. 
The rightmost two graphs within Figure \ref{fig:exp_sr_cal_ga} display the impact of changing spatial and temporal tolerance constraints on success rate. When users have higher tolerance, their queries are anonymized with higher success rate. We see clearly that lower spatial tolerance affects \textsc{XStar}'s success rate negatively far more than it affects \textsc{StarCloak}, once again showing \textsc{StarCloak}'s superiority.

\textbf{Results on Successful Throughput:} The throughputs of compared approaches with respect to varying $\delta_k$, $\delta_l$, $\sigma_s$ and $\sigma_t$ are shown in Figure \ref{exp:st_ca_ga} for California and Georgia maps. Note that the y-axes of these figures are in logarithmic scale. We observe that throughputs of the baseline approaches (random sampling and network expansion) are often significantly lower than \textsc{XStar} and \textsc{StarCloak}. While \textsc{XStar} and \textsc{StarCloak} maintain high throughput despite increasing $\delta_k$ (stricter privacy), the throughput of \textsc{XStar} drops when $\delta_l$ is increased. Hence, we find that \textsc{StarCloak} is much more capable of satisfying challenging $l$-segment indistinguishability requirements than compared approaches. 

With respect to varying spatial and temporal tolerances, we observe that \textsc{StarCloak} variants are capable in handling a variety of tolerance values without significant degradation in throughput. In contrast, the throughputs of the baseline approaches are often 5-6 times smaller than \textsc{StarCloak}. Furthermore, while \textsc{XStar}'s throughput is comparable to \textsc{StarCloak} when $\sigma_s$ is high, \textsc{XStar} may perform even worse than the baselines for small $\sigma_s$ (see Figure \ref{exp:st_ca_ga}). Collectively, these results show the superiority of \textsc{StarCloak} and its variants in query service and scalability compared to both \textsc{XStar} and baseline approaches, under varying privacy and utility settings.

\textbf{Results on Anonymization Time:} In Figure \ref{fig:exp_anontime_ca_ga}, we report the average anonymization times for the California and Georgia maps. The results show that \textsc{StarCloak} variants have significantly better anonymization time than compared approaches under various privacy and utility constraints. \textsc{XStar} often has the highest anonymization time on California map. Among \textsc{StarCloak} variants, hybrid \textsc{StarCloak} and spatially bounded \textsc{StarCloak} are similar, whereas basic \textsc{StarCloak} has lowest anonymization time. This is because basic \textsc{StarCloak} has no preference towards ``waiting for a better opportunity" to generate cloaked regions for incoming queries, whereas the other two variants can wait closer until the query expiration time before anonymization. 

\textbf{Results on Query Processing Time:} We measure the average query processing time of an anonymized query on the server-side and report the results in Figure \ref{fig:exp_qptime_cal}. Since each compared anonymization approach may have different success rate, in order to ensure a fair comparison, we pick the same number of anonymized locations across all approaches in this set of experiments and those experiments reported in the next subsection (which is the number of anonymized locations achieved by the approach with lowest success rate). It is expected that anonymized locations with scattered segments will cause higher query processing time. We observe that \textsc{StarCloak}'s results are often significantly better than its main competitor \textsc{XStar}. Among the three \textsc{StarCloak} variants, basic \textsc{StarCloak} has highest query processing time, whereas the hybrid and spatially bounded versions have similar processing time, because of their more compact cloaked regions. The improvement of spatially bounded \textsc{StarCloak} becomes significant particularly when $\sigma_s$ is relaxed (increased).

\textbf{Results on Candidate Result Size:} The size of the candidate result is an important measure of the added network bandwidth cost caused by anonymization. Larger the number of items returned in the candidate result set, higher the communication bandwidth cost. We measure the candidate result set size under varying $\delta_k$, $\delta_l$, $\sigma_s$, and $k$ parameters, and report the results in Figure \ref{fig:exp_cresult_cal}. Spatially bounded and hybrid \textsc{StarCloak} often provide the best results due to their compact output cloak regions. \textsc{StarCloak}'s competitors are comparable when the $\delta_k$, $\delta_l$ privacy requirements are relaxed, but as we make the privacy requirements stricter, the bandwidth cost of \textsc{XStar} in particular becomes significantly large. The increase in candidate result size caused by large $\sigma_s$ can be explained by the fact that relaxed spatial tolerance inevitably causes the \textsc{StarCloak} approaches to be more relaxed regarding the compactness of the output cloak regions, thus query candidate result sets are also more scattered and diverse. The increase in candidate result size due to increased $k$ is expected, since $k$ is the parameter controlling the number of nearest neighbors returned by the $k$-NN query. Naturally, with higher $k$, more candidates have to be returned, hence the candidate result set has larger size.

\begin{figure*}[!ht]
\centering
\begin{minipage}[!b]{0.25\textwidth}  
    \includegraphics[width=\textwidth]{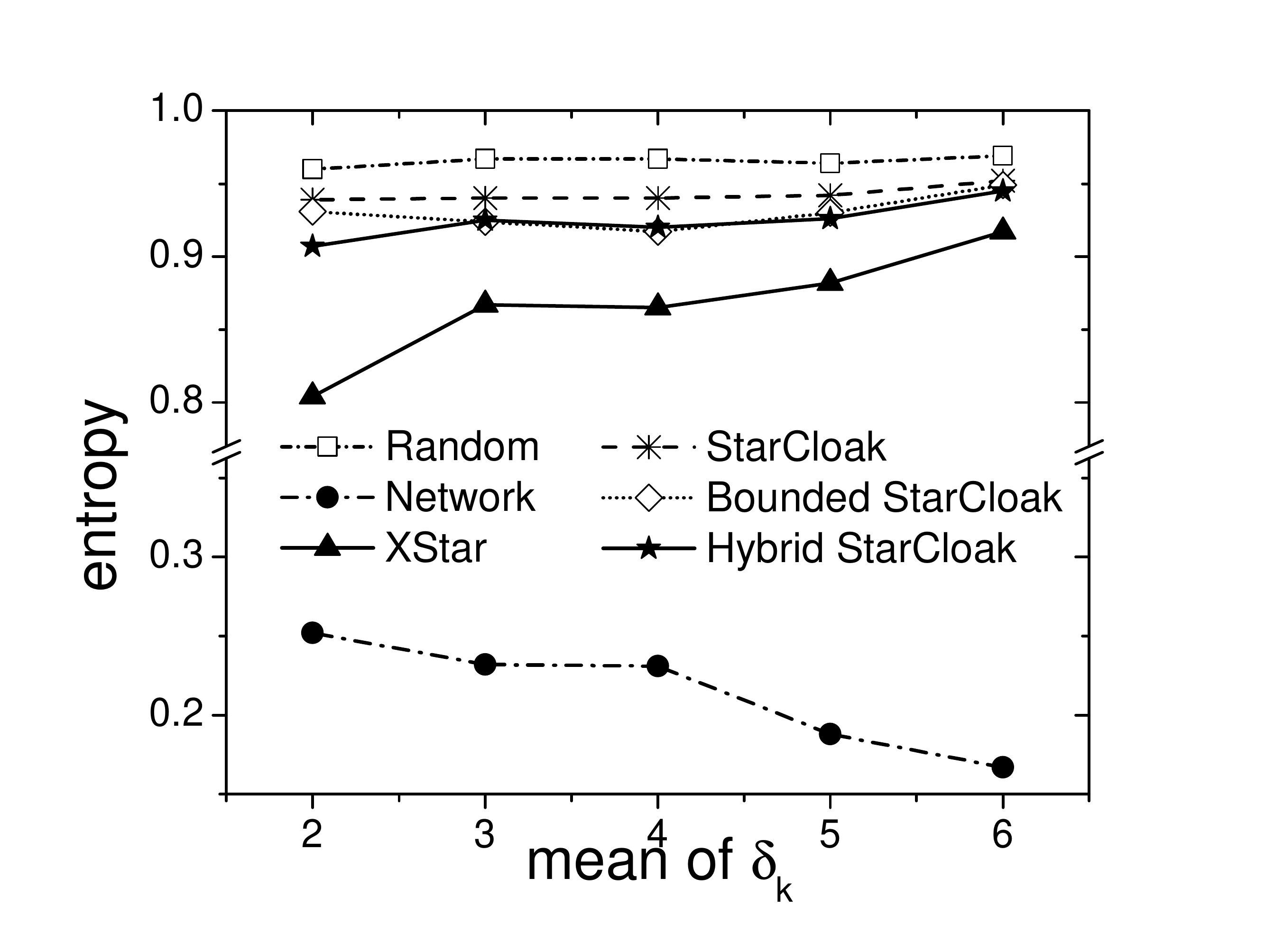}
\end{minipage}%
\hspace{-6mm}
\begin{minipage}[!b]{0.25\textwidth}
    \includegraphics[width=\textwidth]{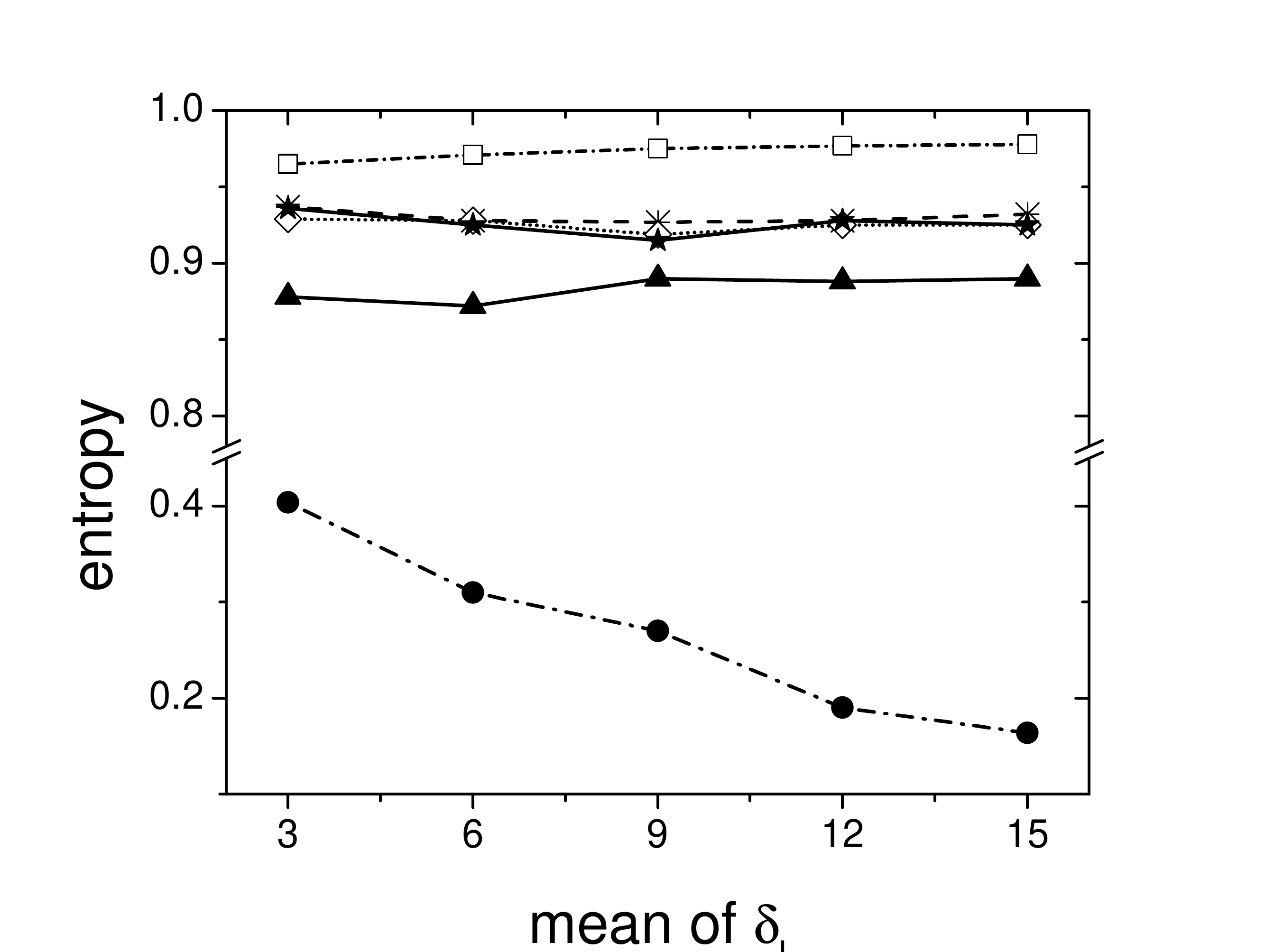}
\end{minipage}%
\hspace{-6mm}
\begin{minipage}[!b]{0.25\textwidth}
    \includegraphics[width=\textwidth]{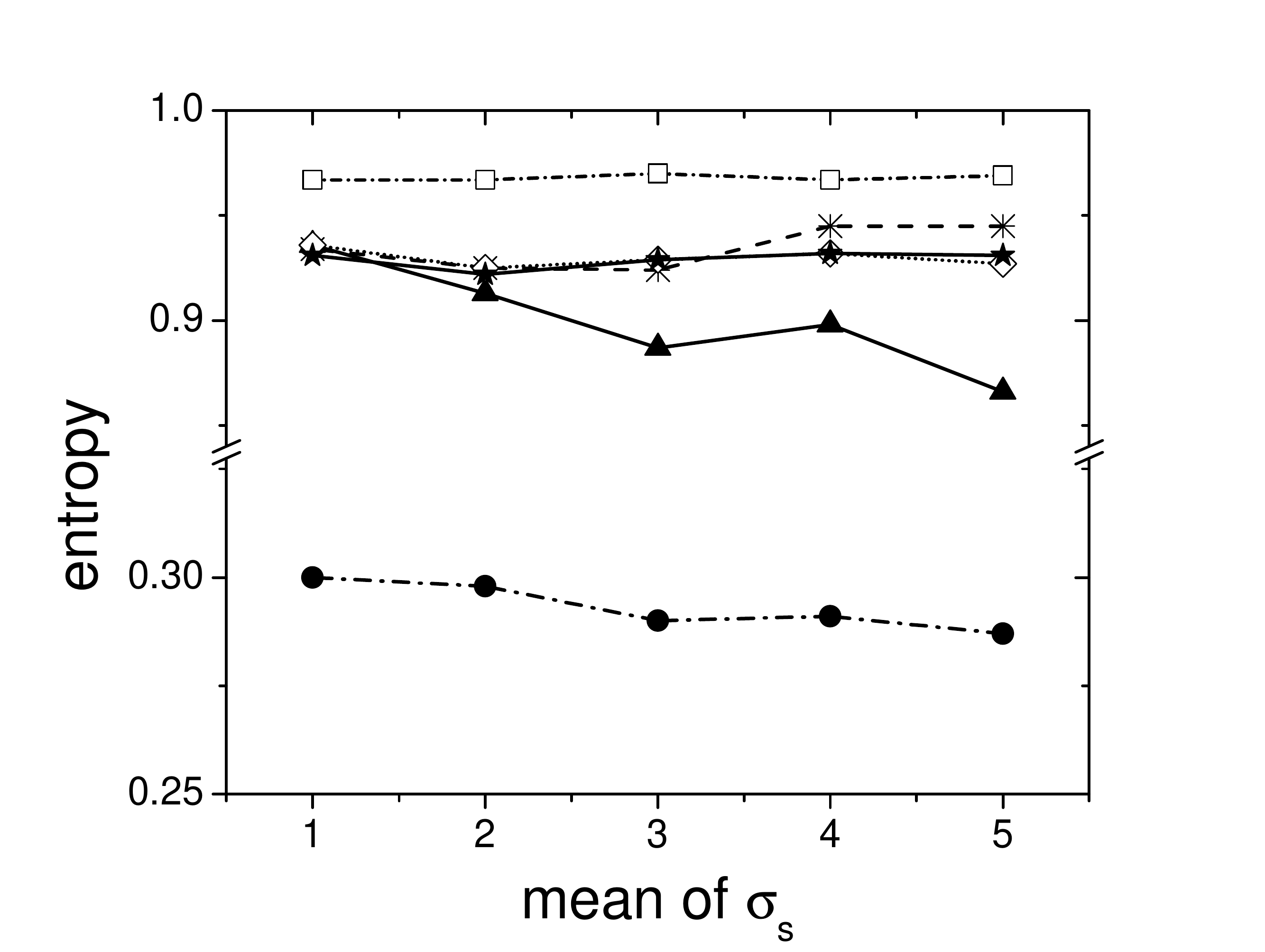}
\end{minipage}%
\hspace{-6mm}
\begin{minipage}[!b]{0.25\textwidth}
    \includegraphics[width=\textwidth]{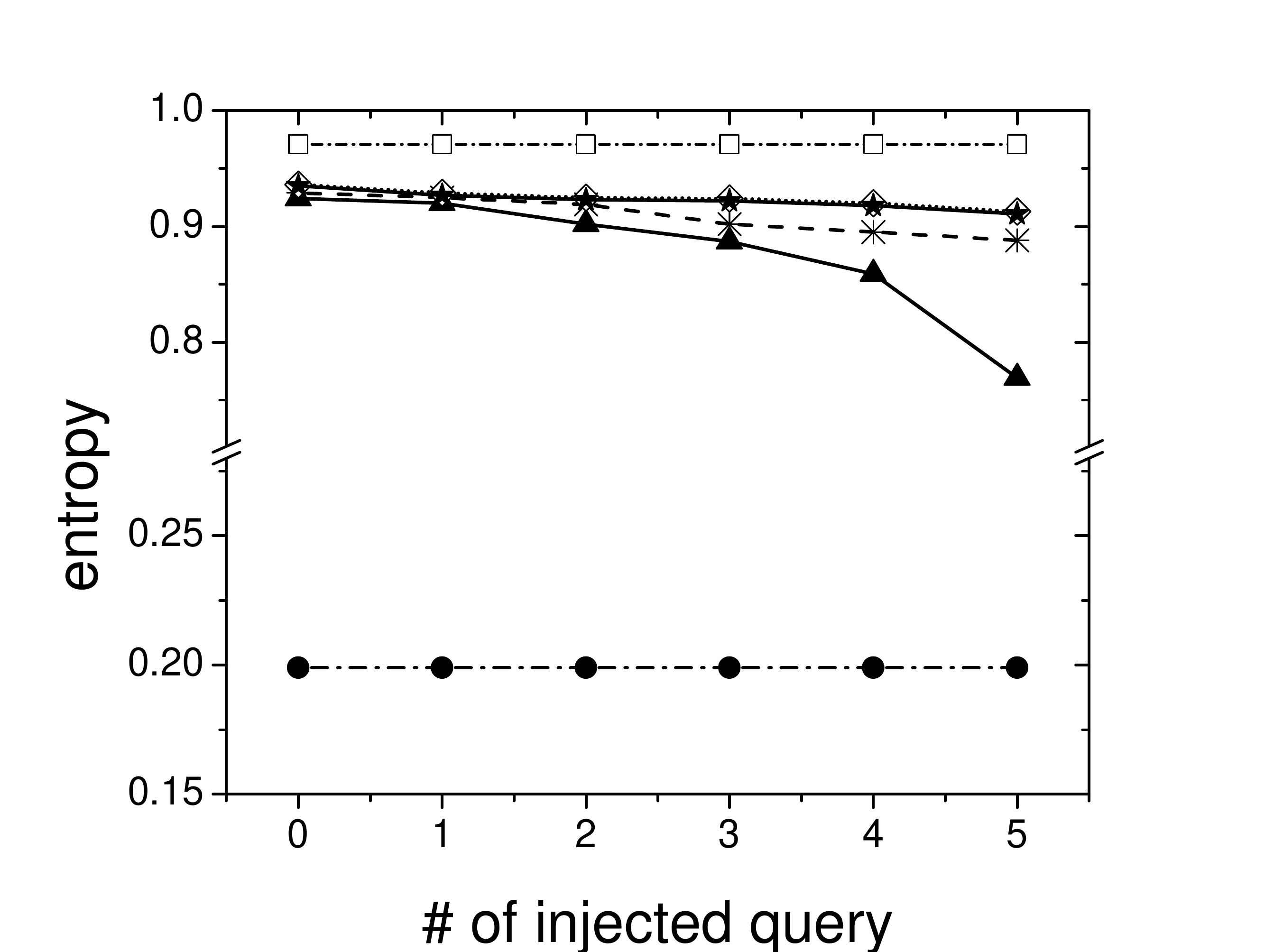}
\end{minipage}%
\\
\begin{minipage}[!b]{0.25\textwidth}
    \includegraphics[width=\textwidth]{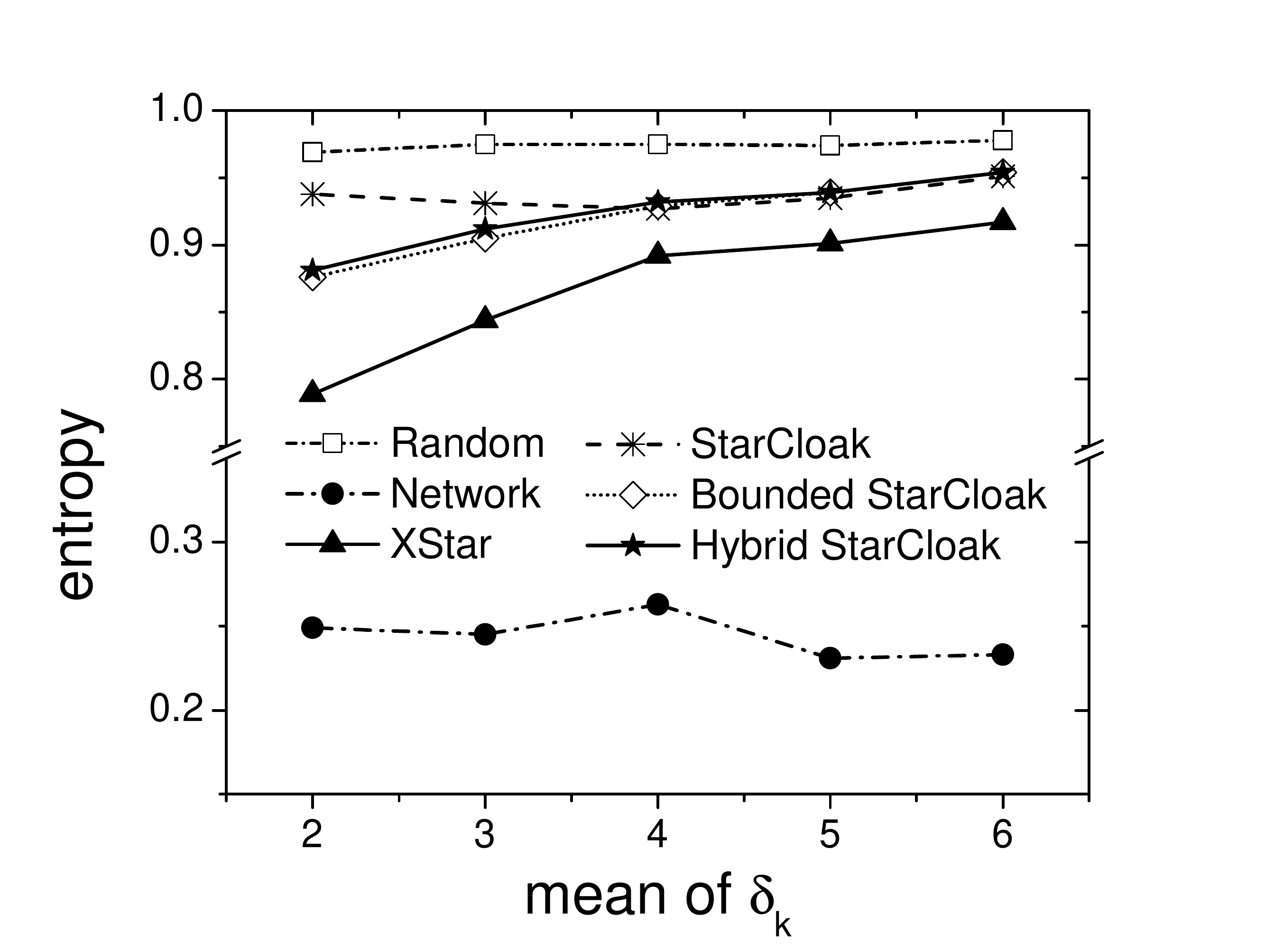}
\end{minipage}%
\hspace{-6mm}
\begin{minipage}[!b]{0.25\textwidth}
    \includegraphics[width=\textwidth]{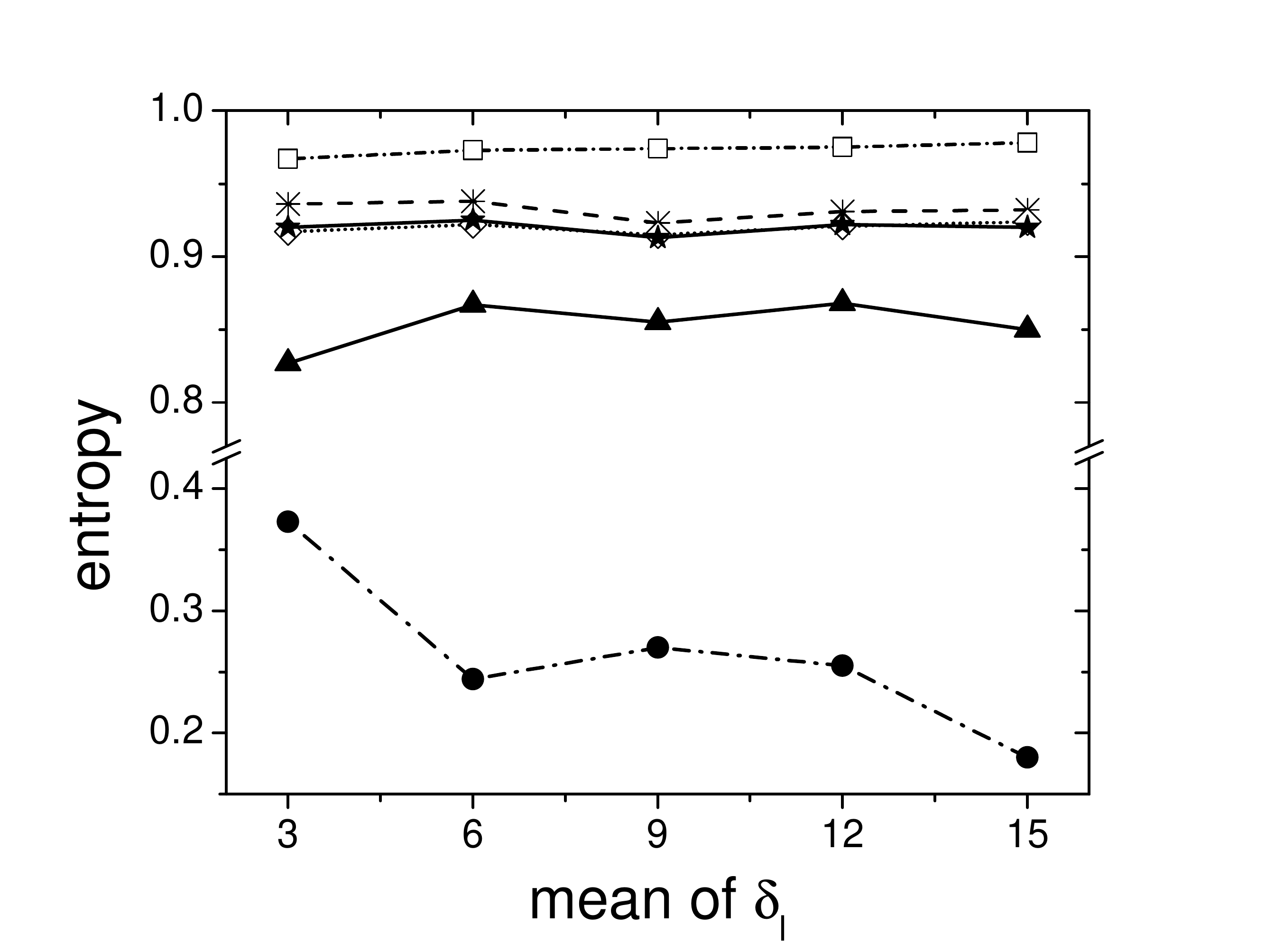}
\end{minipage}%
\hspace{-6mm}
\begin{minipage}[!b]{0.25\textwidth}
    \includegraphics[width=\textwidth]{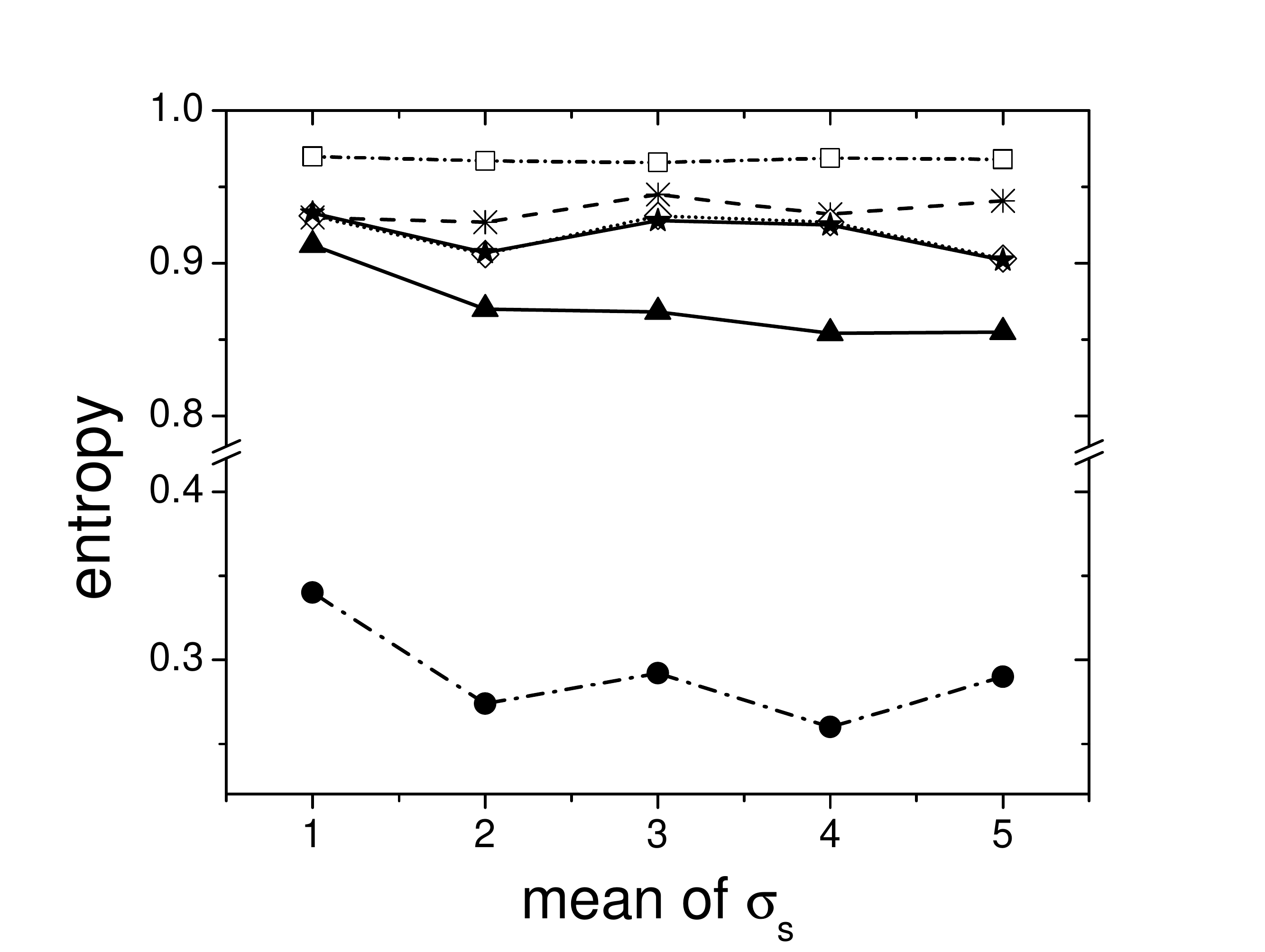}
\end{minipage}%
\hspace{-6mm}
\begin{minipage}[!b]{0.25\textwidth}
    \includegraphics[width=\textwidth]{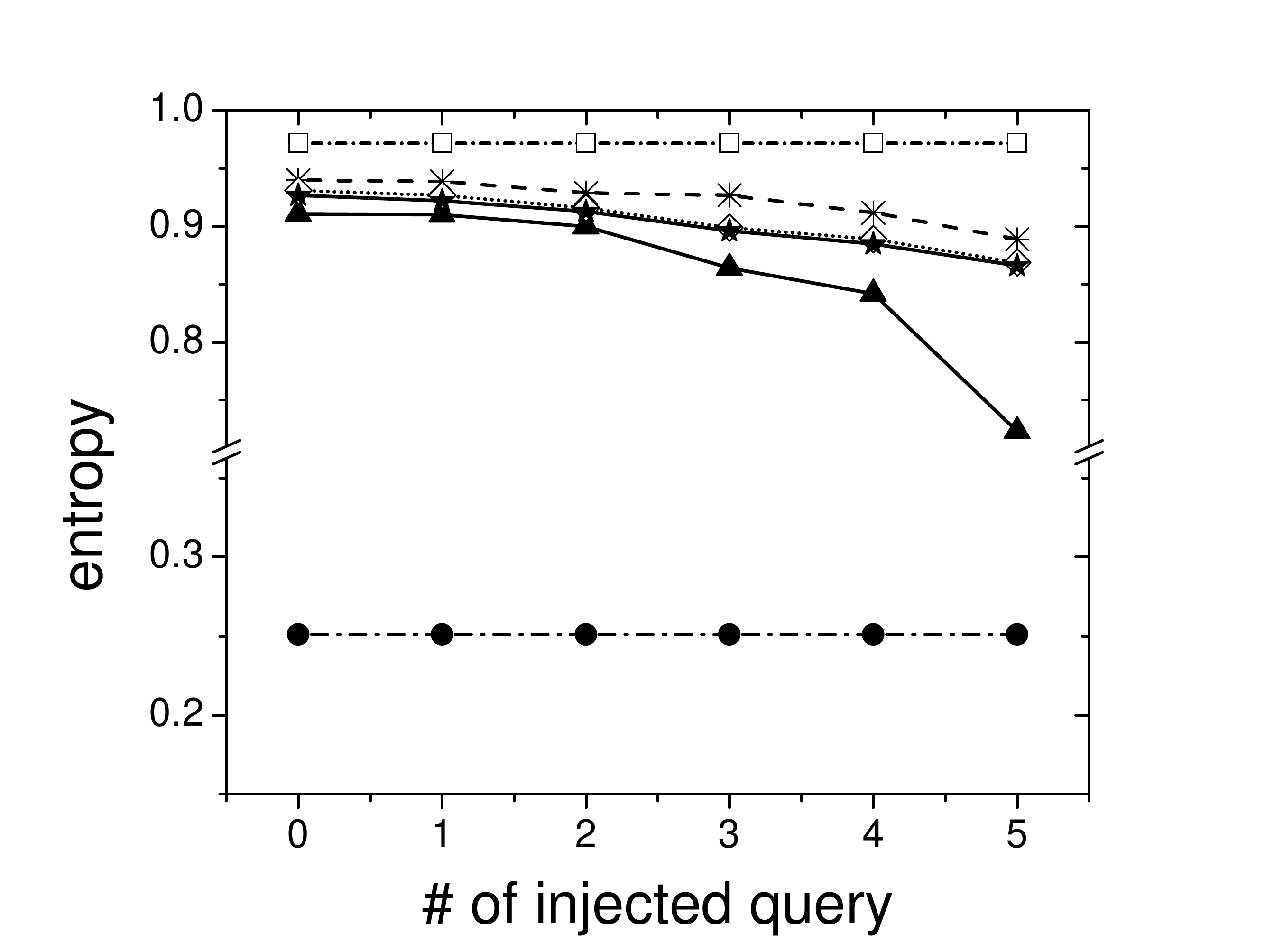}
\end{minipage}%
\vspace{-6pt}
\caption{Average entropy for California map (four graphs in top row) and Georgia map (four graphs in bottom row)}
\label{exp:en_ca_ga} 
\vspace{-10pt}
\end{figure*}

\textbf{Results on Attack-Resilience:} We use the normalized entropy metric to measure attack-resilience of compared approaches, with higher entropy meaning higher attack-resilience. The results are shown in Figure \ref{exp:en_ca_ga}. In this set of experiments, it is expected that by nature, random sampling will give highest entropy, whereas network expansion will give lowest entropy. The results in Figure \ref{exp:en_ca_ga} confirm these expectations, and show that the entropy of \textsc{StarCloak} variants and \textsc{XStar} are between random sampling and network expansion. 
Under a variety of settings, \textsc{StarCloak} has higher entropy than \textsc{XStar}. Furthermore, \textsc{StarCloak}'s entropy values are similar to random sampling, showing that it achieves near-optimal attack-resilience. As $\delta_k$ increases, since more users are cloaked together, entropy increases. The increase in entropy is more clear for spatially bounded \textsc{StarCloak} and hybrid \textsc{StarCloak} compared to basic \textsc{StarCloak}, as their output cloak regions are more compact (focused on the users' actual locations) with small $\delta_k$ in the first place. 
In the rightmost two graphs in Figure \ref{exp:en_ca_ga}, we show the impact of the number of injected queries on entropy in the query injection attack. In \textsc{XStar} and \textsc{StarCloak}, while it is generally the case that with more query injections cause a more successful attack, the vulnerability of \textsc{XStar} becomes significantly higher than \textsc{StarCloak} when 4 or more queries are injected. Unlike \textsc{XStar} and \textsc{StarCloak}, random sampling and network expansion do not consider nearby queries' locations during cloak region generation, thus their entropy remains unaffected by query injections.

\vspace{-6pt}
\section{Related Work} \label{sec:relatedwork}

Location privacy has been an active research area for more than a decade. Several location and trajectory obfuscation mechanisms have been developed to satisfy privacy notions such as $k$-anonymity, differential privacy, and geo-indistinguishability \cite{clique-cloak,andres2013geo,shokri2012protecting,bordenabe2014optimal,shokri2015privacy,gursoy2018differentially,gursoy2018utility}. However, these mechanisms operate in the Euclidean space, and do not take the road network structure under consideration. In this paper, we study location privacy protection for \textit{mobile travelers on road networks}.

Location privacy approaches on road networks can be studied under three categories: mobile permission systems, mix-zones, and location obfuscation. 
Two recent works under the permission systems category are SmarPer~\cite{olejnik2017smarper} and PrivacyZone~\cite{yigitoglu2018privacyzone}. 
Permission systems are not comparable to \textsc{StarCloak} because they either completely block location access or randomly perturb the user's location when the user is in a designated sensitive zone.  
Mix-zones were proposed to circumvent the risks of continuous location tracking on road networks. After a set of users enter a mix-zone, they change pseudonyms and exit the mix-zone such that the mapping between users' old and new pseudonyms is hidden. Among recent works under this category, MobiMix considers road network, time spent in mix-zone, and travel speed constraints to build attack-resilient mix-zones \cite{palanisamy2011mobimix,palanisamy2014attack}. Palanisamy and Liu \cite{palanisamy2014effective} further improve effectiveness and attack-resilience by studying \textit{continuous query correlation attacks} and non-rectangular mix-zones. The approach in \cite{zhang2017otibaagka} enables distribution of group secret keys in cryptographic mix-zones in the presence of malicious eavesdroppers, without relying on trusted dealers. Vaas et al.~\cite{vaas2018nowhere} propose using fictive chaff vehicles to establish attack-resilient mix-zones in areas with low traffic density. Mix-zones differ from location obfuscation and \textsc{StarCloak} in several ways. Most importantly, mix-zones do not anonymize users on demand (i.e., when user issues query to a LBS) but rather when sufficiently many users enter a mix-zone.

\textsc{StarCloak} falls under the location obfuscation category. Under this category, 
Mouratidis and Yiu \cite{mouratidis2009anonymous} provide $k$-anonymity for road network travelers under the reciprocity requirement. Chow et al.~\cite{chow:2011} support personalized privacy specifications such that a cloaked region satisfies $k$-anonymity and includes a total minimum segment length of $L$. Li and Palanisamy~\cite{li2015reversecloak} propose \textit{reversible} cloaking schemes such that anonymity levels can be reduced to accommodate multi-level privacy and selective de-anonymization. Yang et al.~\cite{yang2012path} study the orthogonal problem of \textit{path privacy}, and define the M-cut requirement to achieve path privacy. A similar path privacy problem is studied in \cite{memon2017dynamic}. Another orthogonal problem is semantic-aware and privacy-preserving sharing of sensitive locations under road network constraints \cite{yigitoglu2012privacy,li2016semantic}. In contrast, \textsc{StarCloak} does not require semantic annotation. Most closely related to our work under this category is \textsc{XStar} \cite{wang:vldb09}. We empirically compare against \textsc{XStar} and show that \textsc{StarCloak} is superior to \textsc{XStar} in several aspects.


\vspace{-6pt}
\section{Conclusion} \label{sec:conclusion}

In this paper, we proposed and evaluated \textsc{StarCloak}, a utility-aware and privacy-preserving location query system for mobile users traveling on road networks. \textsc{StarCloak} has an array of desirable features, including utility-aware and personalized location privacy protection, cost-aware star selection, and randomized star-set pruning for improved attack-resilience. The two optimized variants of \textsc{StarCloak}, namely spatially bounded \textsc{StarCloak} and hybrid \textsc{StarCloak}, improve network bandwidth usage and query processing time, with small sacrifice in success rate, throughput, and anonymization time. In comparison to \textsc{XStar}, \textsc{StarCloak} achieves reduced query processing and anonymization time, higher success rate in anonymization, and higher entropy against the considered attacks.


\ifCLASSOPTIONcaptionsoff
  \newpage
\fi

\vspace{-6pt}
\bibliographystyle{IEEEtran}
\input{references.bbl}


\vspace{-36pt}

\begin{IEEEbiographynophoto}{Emre Yigitoglu} received his PhD degree in Computer Science from Georgia Institute of Technology, his MS degree in Computer Engineering from TOBB University of Economics and Technology, Turkey, and his BS degree in Computer Science from Hacettepe University, Turkey. His research interests lie in data-intensive distributed computing systems, mobile computing, and location-based services.
\end{IEEEbiographynophoto}

\vspace{-36pt}

\begin{IEEEbiographynophoto}{Mehmet Emre Gursoy} is a PhD student in the School of Computer Science at Georgia Institute of Technology. He received his MS degree in Computer Science from University of California Los Angeles (UCLA) and his BS degree in Computer Science and Engineering from Sabanci University, Turkey. His research interests include data privacy, security, machine learning, mobile computing, and big data analytics.
\end{IEEEbiographynophoto}

\vspace{-36pt}

\begin{IEEEbiographynophoto}{Ling Liu} is a full Professor in the School of Computer Science at Georgia Institute of Technology. She is an elected IEEE Fellow, a recipient of the IEEE Computer Society Technical Achievement award in 2012, and a recipient of the best paper award from a dozen of top venues including ICDCS 2003, WWW 2004, 2005 Pat Goldberg Memorial Best Paper Award, IEEE Cloud 2012, IEEE ICWS 2013, Mobiquitous 2014, APWeb 2015, IEEE/ACM CCGrid 2015, IEEE Symposium on Big Data 2016, IEEE Edge 2017 and IEEE IoT 2017. She served as the general chair and PC chair of numerous IEEE and ACM conferences, as well as on the editorial boards of over a dozen international journals. 
\end{IEEEbiographynophoto}

\end{document}